\documentclass{aa}
\usepackage[utf8]{inputenc}
\usepackage[varg]{txfonts}
\usepackage{booktabs}
\usepackage{amssymb}
\usepackage{amsmath}
\usepackage[colorlinks=true,citecolor=blue,linkcolor=blue,urlcolor=blue]{hyperref}

\makeatletter
\renewcommand*\aa@pageof{, page \thepage{} of \pageref*{LastPage}}
\makeatother

\bibpunct{(}{)}{;}{a}{}{,} % following A&A style

\begin{document}

\title{Studying the merging cluster Abell 3266 with eROSITA}
\titlerunning{Studying A\,3266 with eROSITA}
\authorrunning{J.~S.~Sanders et~al.}

\author{
  J.~S. Sanders \inst{1}
  \and V.~Biffi \inst{2,3,4}
  \and M.~Brüggen \inst{5}
  \and E.~Bulbul \inst{1}
  \and K.~Dennerl \inst{1}
  \and K.~Dolag \inst{2}
  \and T.~Erben \inst{6}
  \and M.~Freyberg \inst{1}
  \and E.~Gatuzz \inst{1}
  \and V.~Ghirardini \inst{1}
  \and D.~N.~Hoang \inst{5}
  \and M.~Klein \inst{2}
  \and A.~Liu \inst{1}
  \and A.~Merloni \inst{1}
  \and F.~Pacaud \inst{6}
  \and M.~E.~Ramos-Ceja \inst{1}
  \and T.~H.~Reiprich \inst{6}
  \and J.~A.~ZuHone \inst{7}
  }
\institute{
  Max-Planck-Institut für extraterrestrische Physik, Gießenbachstraße 1, 85748 Garching, Germany
  \and Universitäts-Sternwarte München, Ludwig-Maximilians-Universit\"at München, Scheinerstr. 1, D-81679 Munich, Germany
  \and INAF - Osservatorio Astronomico di Trieste, via Tiepolo 11, I-34143 Trieste, Italy
  \and IFPU - Institute for Fundamental Physics of the Universe, Via Beirut 2, I-34014 Trieste, Italy
  \and Hamburger Sternwarte, University of Hamburg, Gojenbergsweg 112, 21029 Hamburg, Germany
  \and Argelander-Institut f\"ur Astronomie, Universit\"at Bonn, Auf dem H\"ugel 71, 53121 Bonn, Germany
  \and Center for Astrophysics | Harvard \& Smithsonian, 60 Garden St., MS-67, Cambridge, MA 02138, USA
}

\date{Received ---, Accepted ---}

\abstract{
  Abell 3266 is one of the X-ray brightest galaxy clusters in the sky and is a well-known merging system.
  Using the ability of the eROSITA telescope onboard \emph{SRG (Spectrum Röntgen Gamma)} to observe a wide field with a single pointing, we analysed a new observation of the cluster out to a radius of $R_{200}$.
  The X-ray images highlight sub-structures present in the cluster, including the north-east--south-west merger seen in previous \emph{ASCA}, \emph{Chandra}, and \emph{XMM-Newton} data, a merging group towards the north-west, and filamentary structures between the core and one or more groups towards the west.
  We compute spatially resolved spectroscopic maps of the thermodynamic properties of the cluster, including the metallicity.
  The merging subclusters are seen as low entropy material within the cluster.
  The filamentary structures could be the rims of a powerful outburst of an active galactic nucleus, or most likely material stripped from the western group(s) as they passed through the cluster core.
  Seen in two directions is a pressure jump at a radius of 1.1\,Mpc, which is consistent with a shock with a Mach number of $\sim1.5$--$1.7$.
  The eROSITA data confirm that the cluster is not a simple merging system, but it is made up of several subclusters which are merging or will shortly merge.
  We computed a hydrostatic mass from the eROSITA data, finding good agreement with a previous \emph{XMM-Newton} result.
  With this pointing we detect several extended sources, where we find secure associations between $z=0.36$--$1.0$ for seven of them, that is background galaxy groups and clusters, highlighting the power of eROSITA to find such systems.
}


\maketitle

\section{Introduction}
Current theories of structure formation predict structures are formed by hierarchical growth, where they are the result of the merger of smaller structures.
Clusters of galaxies are the largest objects where the resulting system is gravitationally bound.
They sit on the threshold between bound and unbound systems and are therefore ideal locations to study how this growth by merging processes happens.

Clusters consist of a number of components which are sensitive to merging processes, including the dark matter halo, the hundreds to thousands of galaxies within the cluster and their constituent gas, active galactic nuclei (AGN), and stars.
However, the one key component is the intracluster medium (ICM), the hot atmosphere which permeates the whole cluster and makes up most of the baryonic mass within the object.
This hot baryonic halo is amenable to study since, due to its multi-million K temperature, it emits X-rays via bremsstrahlung and related processes, with an emissivity which is sensitive to density variations and a spectrum which can be fitted to measure the temperature via its shape and metallicity from emission lines  \citep[e.g.][]{Bohringer10}.

Cluster mergers are some of the most energetic events in the universe.
A great deal of energy needs to be thermalised during a merger and much of that energy goes into the ICM.
The ICM reveals the presence of shock heated material and sloshing processes \citep[e.g.][]{MarkevitchCFShock07,ZuHoneRoediger16}, low entropy material brought into a cluster, stripping of sub-halos, and turbulence \citep{Sanders10_A1835,ZhuravlevaFluct14,Hitomi16}.
It is also the reservoir for many of the metals produced by stellar processes in the galaxies within the cluster.
The metallicity distribution is affected by transport processes within the ICM, for example caused by AGN feedback or mixing \citep[e.g.][]{RebuscoDiff05}, acting as a tracer of gas motions.
The hot atmosphere also provides important information about the unseen dark matter halo in a cluster.
When a cluster is in a relaxed state and has a known geometry, the ICM pressure and density profiles can be used to measure the mass profile under the assumption of hydrostatic equilibrium.
Merging processes, however, can break these assumptions by introducing non-virialised material, additional pressure contributions such as turbulence, or disturbing the 3D geometry \citep[e.g.][]{Lau09,Biffi16}.

Abell 3266 is an X-ray bright galaxy cluster (2-10 keV flux of $
5.9\times10^{-11}$~erg~cm$^{-2}$~s$^{-1}$; \citealt{Edge90}; 0.1-2.4 keV flux of $
5.8\times10^{-11}$~erg~cm$^{-2}$~s$^{-1}$; \citealt{Reiprich02}) at a redshift of 0.0589  \citep{StrubleRood99}, which has been the target of many generations of X-ray telescopes.
The cluster is hot ($\sim 8$~keV) and massive.
\cite{Ettori19} analysed new \emph{XMM-Newton} observations out to the virial radius of the cluster and obtain using a hydrostatic analysis, masses of $M_{500}=8.8\times10^{14}$~M$_{\odot}$ and $M_{200}=15\times10^{14}$~M$_{\odot}$, and radii of $R_{500}=1.43$~Mpc and $R_{200}=2.33$~Mpc, assuming an NFW mass model \citep{NFW96}.

\emph{ASCA} observations of the cluster provided evidence for a merger, including detecting a temperature variation along the merger axis \citep{Henriksen00}.
\emph{Chandra} observations of the cluster found a cooler filamentary region centred on the central cD galaxy aligned along the merger axis running north-east (NE) \citep{Henriksen02}.
\emph{XMM-Newton} data of the cluster were analysed by both \cite{Sauvageot05} and \cite{Finoguenov06}.
\cite{Finoguenov06} termed this cooler, denser material from the core to the north-west (NW) as the low entropy gas, which they estimate has a mass of $1.3\times 10^{13}$~M$_{\odot}$.
They found it has a high metallicity and favour a scenario where it is stripped material falling along the plane of the sky to the south-west (SW).
They suggested that this results from a merger with a mass ratio of 1:10.
\cite{Sauvageot05} put forward a similar picture that the low entropy material is due to a merger on a direction close to the plane of the sky (otherwise projection effects would wash it out).
There is a hotter region west of the low entropy material, which Rankine-Hugoniot jump conditions suggest that if it is a shock, it is a weak one ($M\sim 1.2$).
They also studied numerical simulations finding two scenarios for the merger.
The subcluster could have merged from the NE and is now exiting to the SW, after passing the core 0.15--0.20 Gyr ago.
Alternatively, it could have entered from the SW, passing the core 0.8 Gyr ago and is now nearing turnaround.

\cite{Dehghan17} conducted a detailed structural analysis of A\,3266, using over 1300 spectroscopic redshifts and find that the cluster can be decomposed into six groups and filaments to the north of the cluster, in addition to a cluster core which can be split into two components.
The core has a velocity dispersion of $\sim 1460$~km~s$^{-1}$, while the dispersion of the whole cluster is $\sim 1340$~km~s$^{-1}$, although these are not due to proper virialised motions.
They find the dynamics indicate that the merger with the material to the NE has not yet reached core passage.
Their conclusion is that there is not a simple NE-SW merger, but there is a range of continuous dynamical interactions taking place.

The eROSITA X-ray telescope \citep{Predehl21} onboard the \emph{SRG} observatory is an excellent instrument with which to study mergers of nearby galaxy clusters.
It has the advantage of a wide ($\sim 1$~deg) field of view, good soft response and stable X-ray background \citep{Freyberg20}.
A\,3266 was the target of a calibration observation for the eROSITA mission during its calibration and performance verification (CalPV) phase \citep{Dennerl20}.
In this paper we describe an analysis of this observation in order to further study the merging processes within A\,3266.

We assume $H_0 = 70$~km~s$^{-1}$~kpc$^{-1}$, $\Omega_\Lambda=0.7$ and $\Omega_\mathrm{m}=0.3$.
1 arcsec on the sky at the redshift of A\,3266 corresponds to a distance of 1.139 kpc.
We use the relative Solar abundance ratios of \cite{Asplund09}, noting that
the obtained abundances are around 50\% larger than the commonly used \cite{AndersGrevesse89} values if the measurements are dominated by iron.
Coordinates are given in the J2000 system.
Unless otherwise indicated, uncertainties are given at the $1\sigma$ level.
All images are aligned with north upwards and east to the left.

\section{Data analysis}

\begin{figure}
  \centering
  \includegraphics[width=\columnwidth]{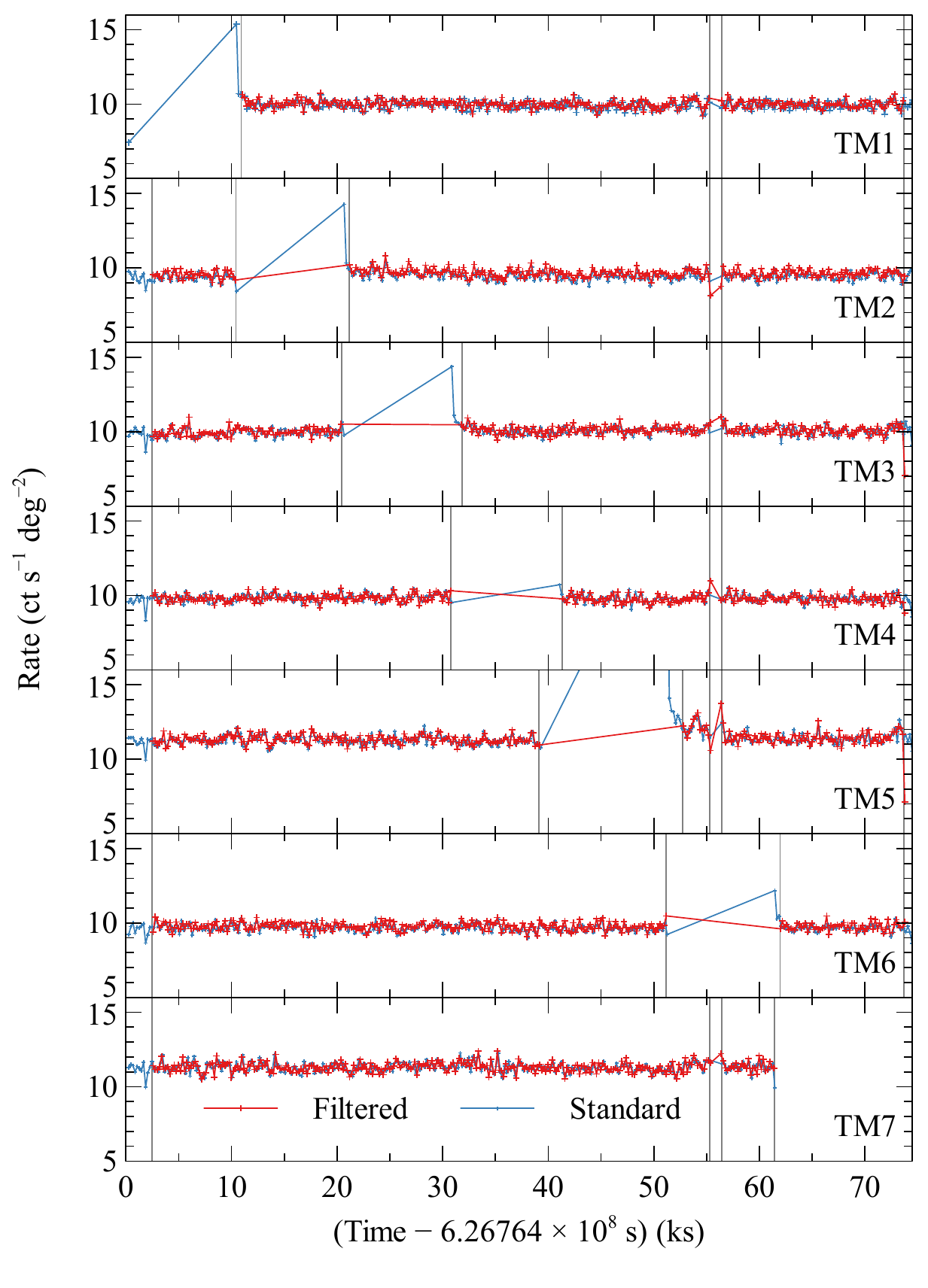}
  \caption{
    Lightcurves of the telescope modules in 200s bins, with standard GTI filtering and after applying an extra GTI selection.
    The energy range is 0.2 to 10 keV.
    The vertical lines show the start and stop times of the GTIs.
    Times are relative to a modified Julian date of 51543.875.
  }
  \label{fig:lightcurves}
\end{figure}

\begin{table}
  \caption{
    eROSITA telescope modules used in the observation.
  }
  \begin{tabular}{lll}
  \hline\hline
  Module & Good time (ks) & Minimum energy (keV) \\ \hline
TM1 & 61.61 & 0.3 \\
TM2 & 59.34 & 0.3 \\
TM3 & 58.63 & 0.3 \\
TM4 & 59.52 & 0.6 \\
TM5 & 56.49 & 0.4 \\
TM6 & 60.40 & 0.3 \\
TM7 & 57.81 & 0.3 \\ \hline
Average & 59.11 \\ \hline
  \end{tabular}
  \tablefoot{
  Listed are the total length of the good time intervals and the minimum energy of events used in the analysis.
  }
\label{tab:tms}
\end{table}

\subsection{X-ray data}
A\,3266 was selected as a calibration target for eROSITA.
It was observed on 2019-11-11 for a total of 86~ks (observation ID 700154).
We took as input processing version 001 of the eROSITA A\,3266 dataset and used version \texttt{eSASSusers\_201009} of the eSASS analysis software \citep{Brunner21}.

The seven telescope modules (TMs) of eROSITA lie in three different orientations \citep{Meidinger20}.
There is an almost circular region on the sky at any one time where there is signal from all seven TMs.
Normally counts are discarded from outside this region, because the PSF increases at large radius and the effective area is also rapidly declining.
As this cluster fills the field of view we instead retained these detector corners, by applying a \texttt{FLAG} filter of \texttt{0xC0007000}, when combining the event files of each TM to make a merged event file.
A \texttt{PATTERN} filter of 15 was used to include single, double, triple, and quadruple events.

\begin{figure*}
  \centering
  \includegraphics[width=\textwidth]{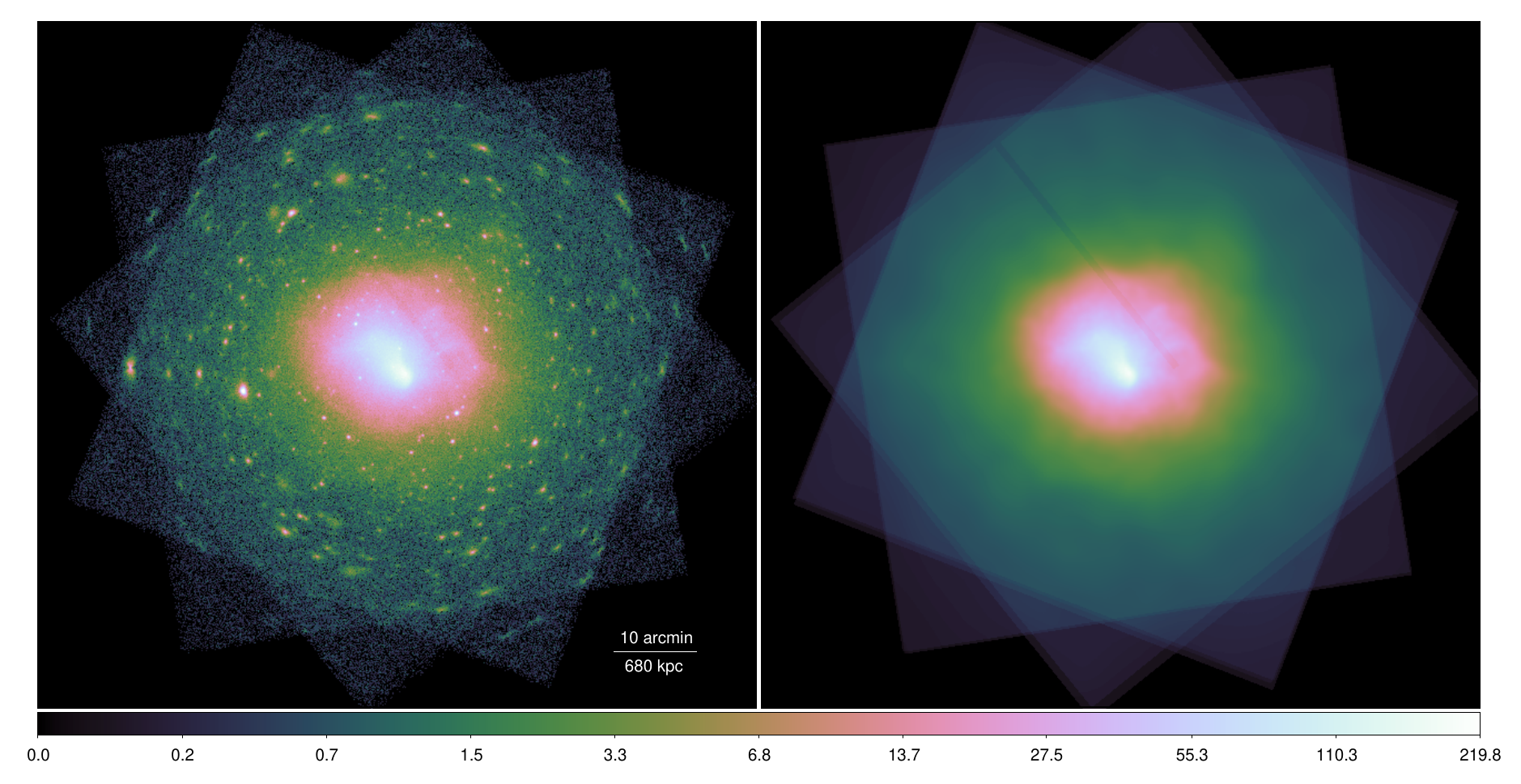}
  \caption{
    X-ray count image of cluster (left) and generated background map for source detection (right).
    The images are in the 0.3 to 2.3 keV band using the standard CubeHelix colour scheme \citep{GreenCubehelix11}.
    Shown are the number of counts per 4 arcsec pixel.
    The strip running towards the north east is the excluded bad pixel region on TM2.
  }
  \label{fig:cts_bkgmap}
\end{figure*}

Figure~\ref{fig:lightcurves} shows the lightcurve for each TM created using the eSASS \texttt{flaregti} task.
Each TM did not observe the sky for $\sim 10$~ks periods during the observation for calibration purposes and there was also a short period around 56~ks where there was no data from any TM.
At the start and end of the observation there was also a period where the attitude of the spacecraft was slewing to or from the target.
To ensure accurate exposure times and to avoid the unstable attitude periods we applied an extra good time interval (GTI) filtering, shown as the vertical bars in Fig.~\ref{fig:lightcurves}.
We also excluded the period immediately after each closed filter period due to temporary increases in the background count rate.
We do not see any evidence for any other flaring during the observation and therefore did not exclude any further time intervals.
The total remaining good time intervals after filtering in each TM are listed in Table \ref{tab:tms}, with a combined effective exposure of 59.1~ks.

TMs 5 and 7 are susceptible to excess soft emission which is believed to be optical light leaking into these cameras, dependent on the orientation of the spacecraft relative to the sun \citep{Predehl21}.
Due to this light leak we used a higher minimum energy of 0.4 keV for TM5.
We do not see evidence for the light leak at the time of observation for TM7, so we use our standard minimum energy of 0.3 keV.

TM4 shows some bright columns between 2019-10-28T10:29:25Z and 2019-11-12T17:11:05Z.
These excess events are only visible below 0.6 keV.
Therefore we increased the minimum energy to 0.6 keV for TM4.

\subsection{Images and exposure maps}
Images of the observation were made using two different pixelisation schemes.
For the source detection we used the standard sky coordinate binning of a factor of 80 (4 arcsec), for which the source detection software has been optimised.
For imaging and spectral maps, we use images with a bin factor of 40 (2 arcsec).
For our imaging analysis we created images in four bands: 0.3--0.8, 0.8--1.3, 1.3--2.3, and 0.3--2.3 keV, adjusting the lower energy threshold for each TM as discussed above.
The counts image with 4 arcsec binning is shown in Fig.~\ref{fig:cts_bkgmap}.

Exposure maps, with and without vignetting, were created using the eSASS task \texttt{expmap}.
These maps have the units of time, giving the exposure relative to an on-axis observation at the energy given.
Conventionally for eROSITA, the exposure map of each TM is the exposure time divided by 7, so that the summed exposure maps gives the average exposure of all the TMs.
When creating maps over an energy range including below 0.6 keV we applied an empirical correction factor to the exposure maps for TMs 5 and 4 due to their higher lower-energy threshold.
To do this, we calculated the exposure-corrected count rate over the inner 9.3 arcmin of the cluster for each TM.
The exposure map of TM5 was then scaled so that its count rate matched that of TM7 in the same energy band.
TMs 5 and 7 do not have the on-chip filter shared by the other TMs and so have a different effective area curve at low energies.
Similarly we decreased the exposure of TM4 to match the average rates in TMs 1, 2, 3, and 6.

TM2 had a bad column in this observation which was not properly handled by the task \texttt{expmap}.
The bad column, which can be seen as missing X-ray events, is not at the correct location in the exposure map.
This problem was fixed in newer versions of \texttt{expmap}, but not in version \texttt{eSASSusers\_201009}.
Therefore for TM2, we manually masked the region containing this column and the offset column from images of the data and exposure maps.

\subsection{Source detection}
\label{sect:src_detn}
A bright cluster with a complex morphology in the field presents a problem for standard source detection.
We therefore used a non-standard method to create a background map for source detection which includes the majority of the extended cluster emission.
Source detection optimised for point sources in this observation is also done by Liu et al.~(in prep.).

Firstly, we created an initial set of regions to mask point sources.
This was used to make an input mask image, identifying point source regions and regions outside the field of view (identified from the exposure map).
The program \texttt{accumulate\_counts}, supplied with the Contour Binning software \citep{SandersBin06}, was used to calculate the radius around each pixel (including those within point sources) which contains at least 1024 counts when point sources were masked.
This `scale map' was then input into the same program, to apply Gaussian smoothing using this radius as the per-pixel 2D Gaussian $\sigma$, to both the count and exposure map images.
Dividing the smoothed count image and smoothed exposure map images produced a smoothed exposure-corrected image.
We then multiplied this by the unsmoothed exposure, to make a background map in count units suitable for point source detection (the unsmoothed exposure was appropriate here, to include sharp edges in exposure associated with bad pixels or the edges of the detector).

To do the source detection we used the eSASS task \texttt{erbox} to make an initial box detection using the background map above.
We then took this list of sources as the input to the \texttt{ermldet} maximum likelihood detection task.
For this tool we used a maximum fitting extent (\texttt{extmax}) of 80 pixels, a maximum number of simultaneous sources to fit of 4 (\texttt{nmaxfit}), a maximum number of new sources when source splitting of 3 (\texttt{nmulsou}), a minimum extended source likelihood of 8 (\texttt{extlikemin}), and a minimum detection likelihood of 8 (\texttt{likemin}).
We also used photon based detection, which does source detection based on individual photons rather than using image pixels.

The output source list was used to construct a new set of point source mask regions and a mask image.
This was done by calculating for each point source an ellipse where the surface brightness of the source at the ellipse edge is a maximum of 10 per cent of the background surface brightness.
As input, the calculation used ellipses fitted to contours for the set of PSF images taken as a function of off-axis angle in the eROSITA calibration database (file \texttt{tm1\_2dpsf\_190219v05}; \citealt{Dennerl20}), at an energy of 0.93 keV.
In addition to the point sources, there are sources within the observation detected as being extended.
We added these manually to the mask, except for source X8 (see below), which appears local to the cluster.

With our new mask, we repeated the creation of the background map and detection of the sources, to reduce the dependence of our results on the manual input source mask.
The final background map is shown in Fig.~\ref{fig:cts_bkgmap}.

\begin{figure}
  \centering
  \includegraphics[width=\columnwidth]{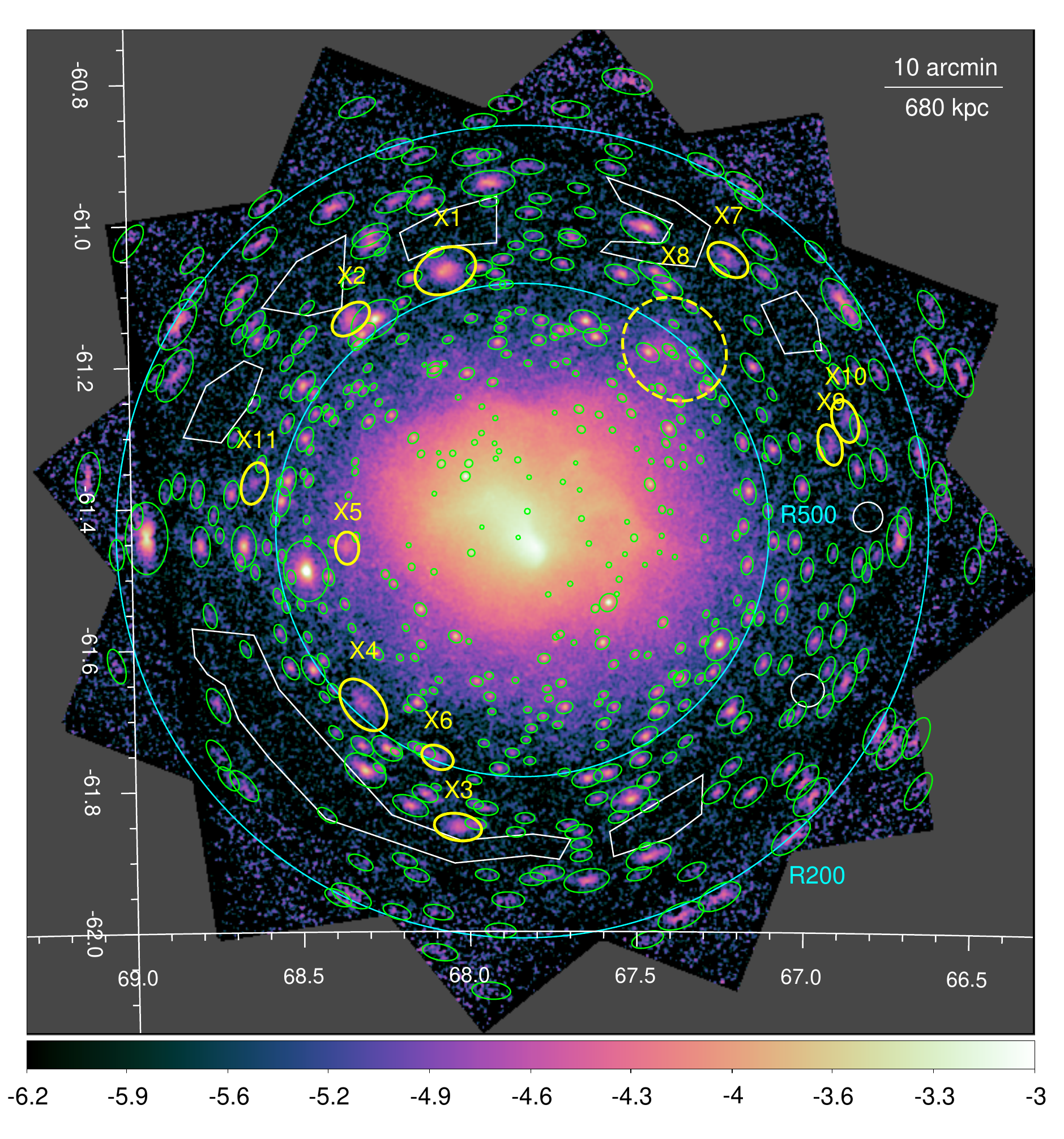}
  \caption{
    Detected X-ray sources and other regions.
    The X-ray image has been exposure corrected and shows the $\log_{10}$ count rate per 4 arcsec pixel after background subtraction (Section \ref{sect:back}).
    The large cyan circles show the radii $R_{200}$ and $R_{500}$, taken from \cite{Ettori19}.
    The green ellipses are the detected point sources.
    The yellow ellipses are the examined extended sources.
    The extended sources were masked when examining the cluster emission, except for X8 which is local to the system.
    The white regions are background regions chosen to overlap with all detectors.
  }
  \label{fig:regions}
\end{figure}

In Fig.~\ref{fig:regions}, the detected point source and extended source regions, are shown.
These extended sources are labelled as X1 to X11, the largest of which is X8, which appears local to the cluster.
We discuss these extended sources later in Section \ref{sect:ext}.
In addition, we highlight the radii of $R_{500}=1.43$~Mpc and $R_{200}=2.33$~Mpc obtained from \emph{XMM-Newton} data assuming hydrostatic equilibrium \citep{Ettori19}.
It can be seen that although $R_{200}$ is within the field of view of the observation, there is little area not covered by the cluster.

\subsection{Background modelling}
\label{sect:back}
The field of view of the eROSITA observation only just contains the radius $R_{200}$, and so there is no source free region which can be used to fit a background model (the vignetting in the corners of the TMs has not been calibrated fully).
Therefore we used data from the first all-sky eRASS survey (eRASS1) to obtain a model for the astrophysical X-ray background (XRB).
This was then combined with a fit to spectra extracted from the outskirts of the observation to compute the non-X-ray background (NXB) component.

We note that the backgrounds in this section are not the same as used for the source detection.
The source detection background included the extended emission from the cluster, while this is a background which only includes non-cluster components.

\begin{figure}
  \includegraphics[width=\columnwidth]{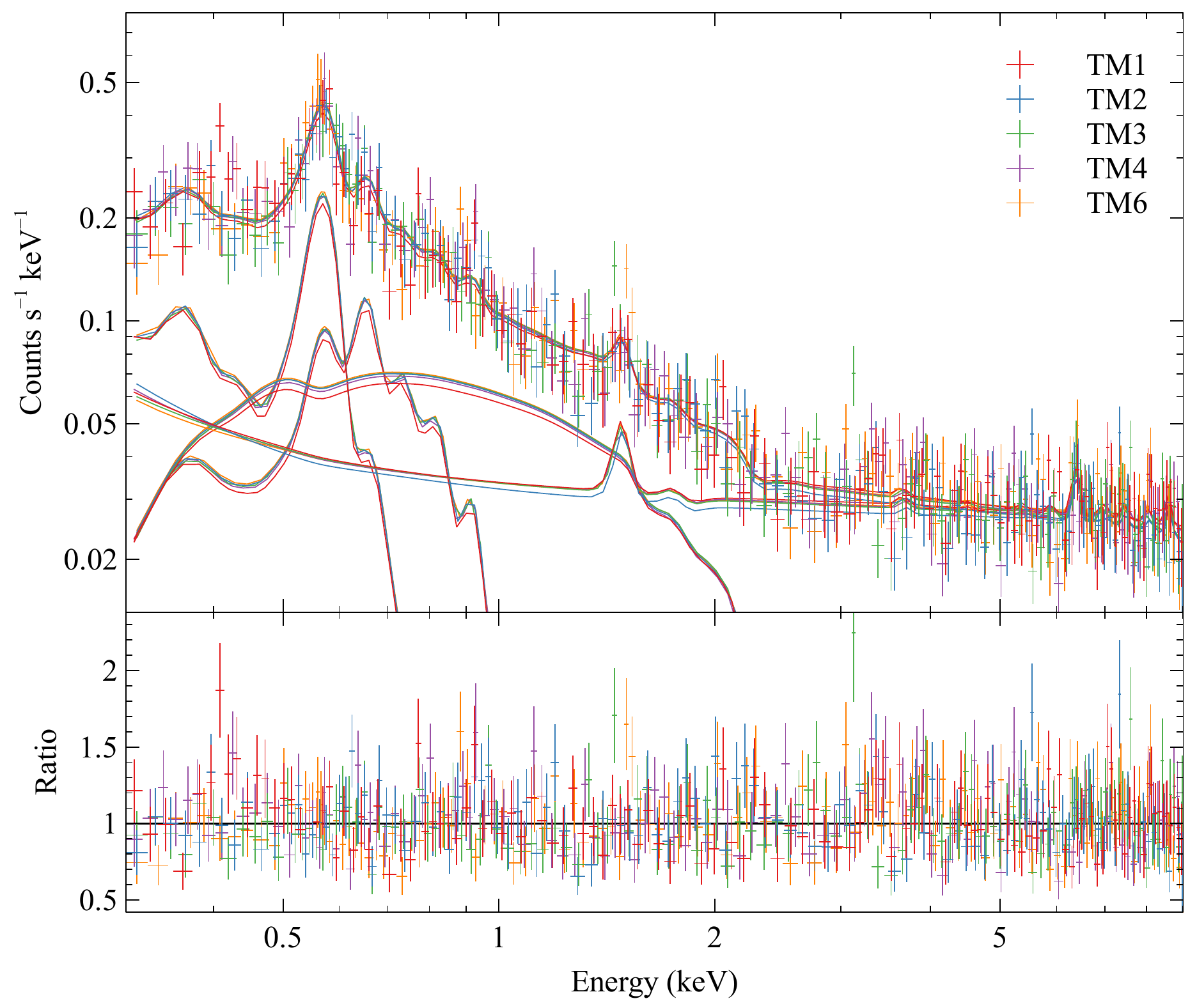}
  \caption{
    Best fitting model to the background spectrum extracted from eRASS1 survey, also showing the data and the residuals.
    The data were rebinned to a minimum signal to noise ratio of 5 for display purposes.
    Also shown are the model components, consisting of the NXB contribution, the two soft thermal components, and the AGN contribution.
  }
  \label{fig:back}
\end{figure}

For the XRB determination, we extracted a spectrum from an annulus between 35 and 50 arcmin radius around the cluster, excluding detected point sources.
The annulus was chosen to avoid nearby extended objects and the cluster itself.
The typical exposure within this region was around 770s.
Sources were detected in the 0.3 to 2.3 keV band within a $10.7$ by $10.7$~deg box centred on the cluster.
An initial \texttt{erbox} box source detection was run to provide a list of sources to mask, \texttt{erbackmap} was used to make an initial background map, \texttt{erbox} was run again with a minimum source likelihood of 6 to create an input list of sources, and then \texttt{erbackmap} was run again with this new list of masked sources to make a final background map.
With the input list of box-detected sources, \texttt{emldet} was run using the parameters \texttt{likemin=8}, \texttt{nmaxfit=4}, \texttt{nmulsou=3}, and \texttt{extlikemin=8}, in photon detection mode.
Sources were excluded from the annulus by removing a circle out to where the source surface brightness was 10\% of the sky surface brightness.

Figure~\ref{fig:back} shows the resulting spectrum for the 0.946~deg$^2$ region on the sky.
The \texttt{srctool} task produces an exposure which is the total amount of time the source area is visible on the sky after accounting for dead time, but adjusts the auxiliary response file (ARF) to account for vignetting variation of X-rays and for the fact that eROSITA only observes part of this region at any one time.

The eRASS1 spectra for the TMs were fitted by a model for the XRB and non-X-ray (NXB) backgrounds.
The fit was made simultaneously between 0.3 and 9.0 keV by minimising the C-statistic in \texttt{Xspec} (\citealt{ArnaudXspec}; version 12.11.1).
We excluded TM5 and TM7 here and for the majority of our spectral analysis because they currently have a more uncertain energy calibration.
The XRB model consisted of an absorbed and unabsorbed \texttt{apec} thermal model (\citealt{SmithApec01}; version 3.0.9), plus an absorbed powerlaw with the photon index fixed to be 1.45 \citep{Cappelluti17} to account for AGNs.
For the photoelectic absorption we used the \texttt{tbabs} model \citep{Wilms00} with the hydrogen column density fixed to a value of $2.26 \times 10^{20}$~cm$^{-2}$, which was derived from the HI4PI survey \citep{HI4PI16} value of $2.13 \times 10^{20}$~cm$^{-2}$, with a correction for missing molecular hydrogen added following the prescription of \cite{Willingale13}.

The NXB was modelled using fits to filter-wheel-closed (FWC) data from each of the cameras, with models consisting of a broken powerlaw with an exponential cut-off at high energies, plus Gaussian lines to model emission from fluorescent emission from the K$\alpha$ lines of Al, Co, Ti, Cr, Mn, Fe, Co, Ni, Cu, and Zn, and the K$\beta$ lines of Ni, Cu, and Zn.
We used the best fitting parameters of these models when applied to the FWC data, but allowing a free overall normalisation.

Given the XRB background model from eRASS1 we then obtained an observation-specific NXB model by fitting spectra in the outskirts of the pointed cluster observation (see regions in Fig.~\ref{fig:regions}) using the fixed XRB model combined with NXB components which were allowed to vary.
In addition we also added another absorbed \texttt{apec} component to account for any residual cluster emission in these background regions (which is not included in our background when applied to the cluster).
The normalisation of the NXB component was around 5\% higher for this pointed observation compared to the eRASS1 fit.
We fitted the data using all the TMs to calculate imaging background for each TM, and excluding TMs 5 and 7 to create a spectroscopic background model.
The minimum energies for the TMs listed in Table \ref{tab:tms} were used in these fits.

The background images were then calculated in each band by taking the count rate in each pixel from the best fitting model for the background region.
The XRB and NXB components were extrapolated over the sky by scaling using the normalised, relative to the background region, exposure (for the XRB) and non-vignetted exposure maps (for the NXB).
These two components were summed to give a total number of background counts in each band in each pixel.

\begin{figure*}
  \centering
  \includegraphics[width=\textwidth]{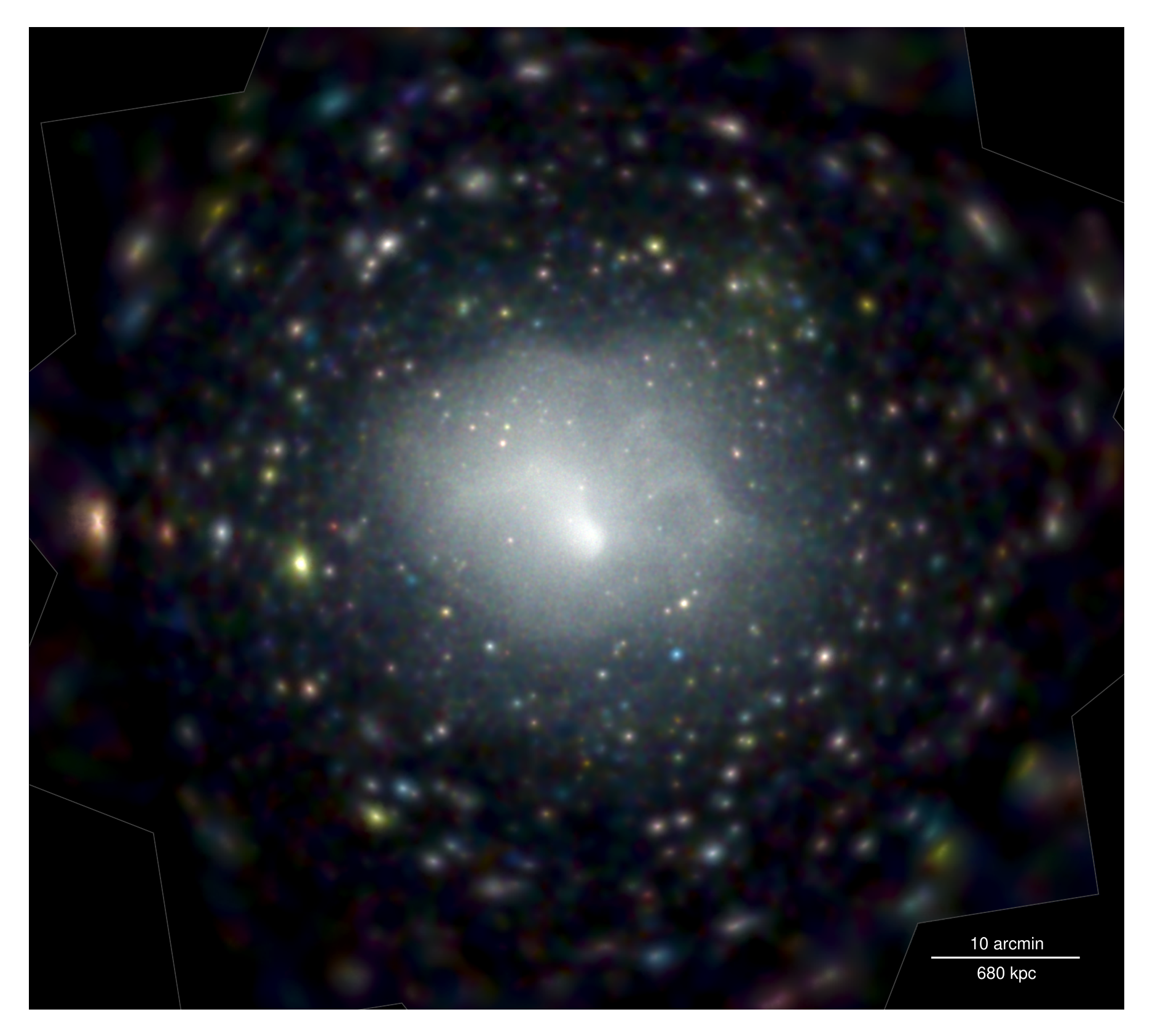}
  \caption{
    RGB image of A\,3266. The bands used for the red, green and blue channels were 0.3--0.8, 0.8--1.3, and 1.3--2.3 keV, respectively.
    We applied location-dependent Gaussian smoothing to the count, exposure and background maps.
    The $\sigma$ of the Gaussian function was taken from the radius of a top hat function which contains a minimum of 36 counts in the 0.3 to 2.3 keV band.
  }
  \label{fig:rgb}
\end{figure*}

\begin{figure*}
  \centering
  \includegraphics[width=\columnwidth]{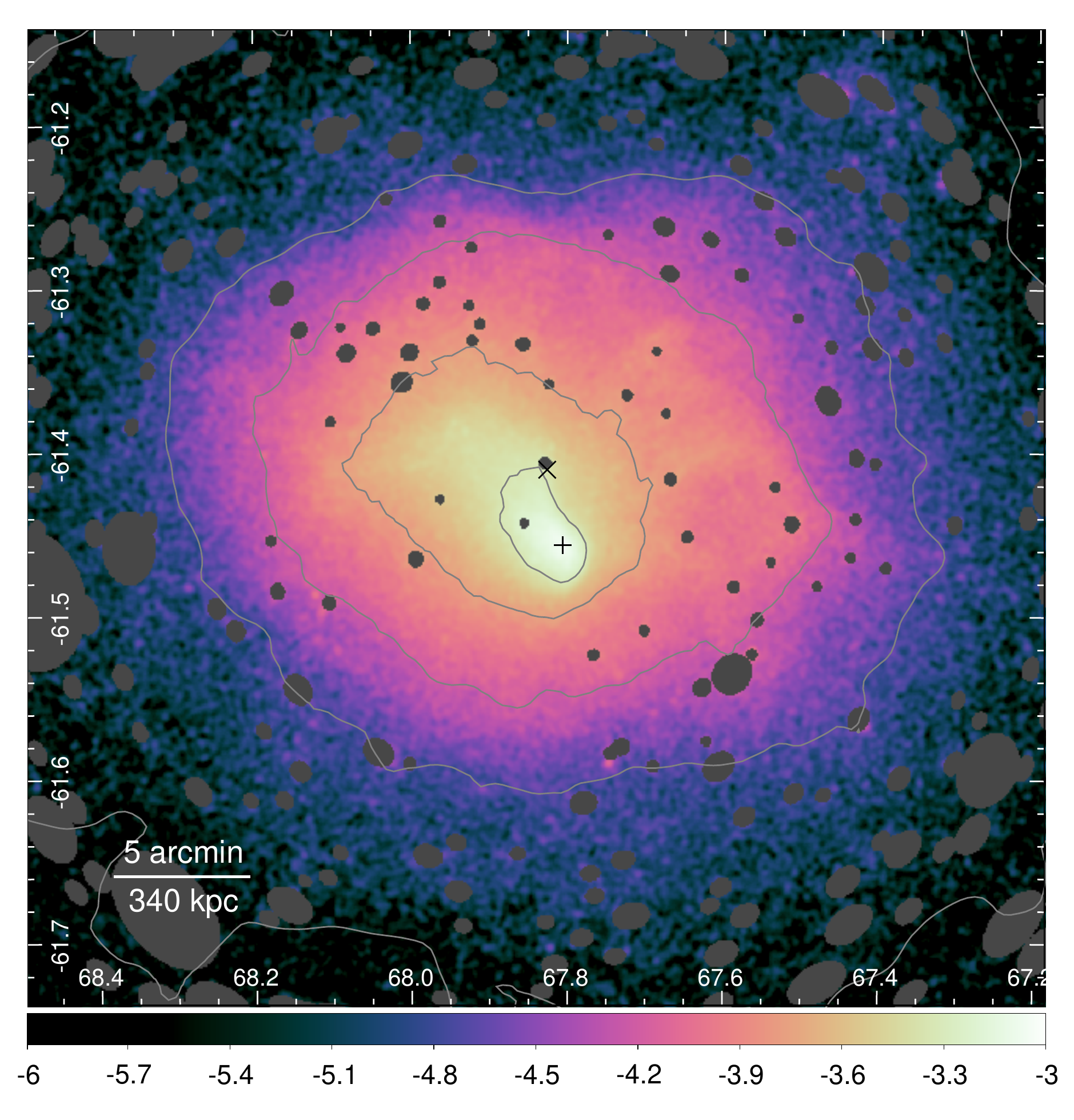} \\
  \includegraphics[width=\columnwidth]{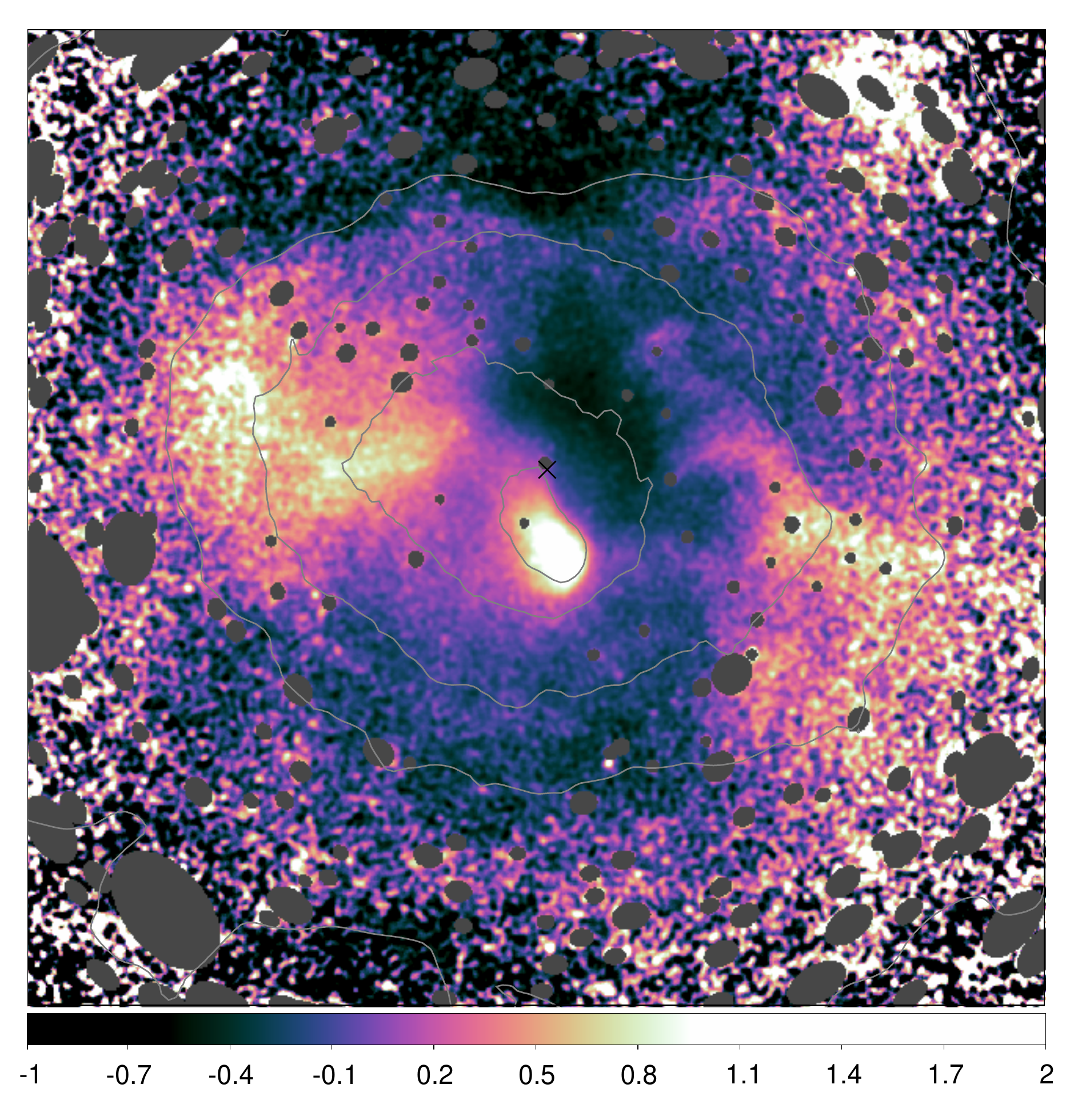}
  \includegraphics[width=\columnwidth]{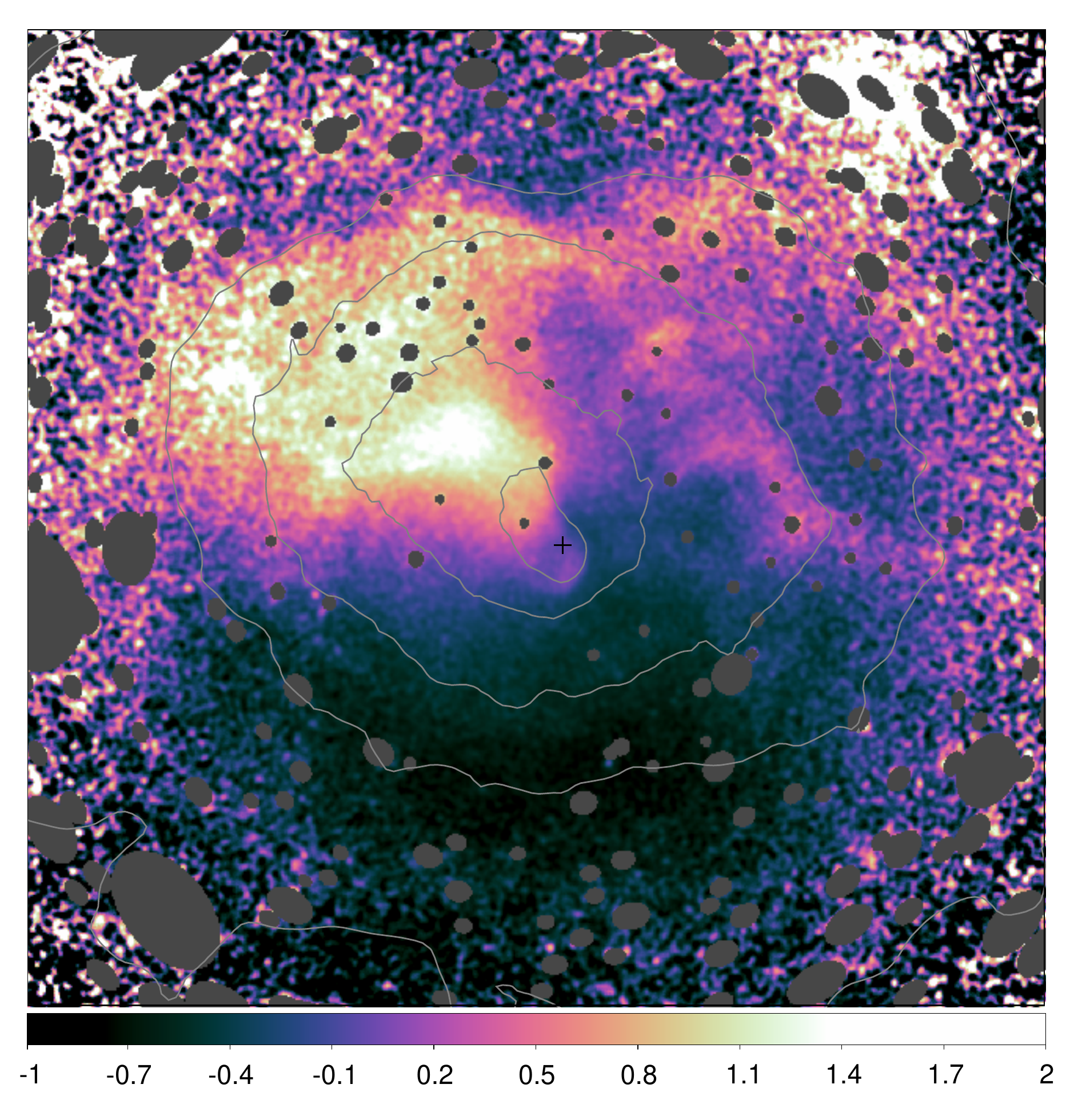}
  \caption{
    X-ray image (top panel) and fractional residuals (bottom panels).
    The X-ray image was created in the 0.2 to 2.3 keV band and is exposure corrected by the effective exposure (relative to on-axis; units $\log_{10}$~counts~s$^{-1}$~pixel$^{-1}$).
    The image was smoothed by a Gaussian of $\sigma=6$~arcsec (3 pixels).
    Point sources were masked before smoothing.
    Contours are at 5 logarithmic levels between $0.00004$ and $0.0019$ counts~s$^{-1}$~pixel$^{-1}$.
    The residual maps shows the fractional difference from the average at each radius, ignoring masked regions.
    Two different centres are used: $(\alpha=67.8759^{\circ},\delta=-61.4214^{\circ})$, marked $\times$, chosen to lie at the centre of the 4th contour level (bottom left panel), and $(\alpha=67.8060^{\circ},\delta=-61.4570^\circ)$, marked $+$, which is the X-ray centroid within 12 arcmin radius (bottom right panel) and lies very close to the centre used by \cite{Ghirardini19}.
    When showing the fractional differences later in the paper we will use the hand-chosen centre.
  }
  \label{fig:img_resid}
\end{figure*}

\subsection{Background-subtracted images}
\label{sect:bgsub}
Shown in Fig.~\ref{fig:regions} is the background-subtracted exposure-corrected X-ray image of the cluster.
We also split the cluster counts into three energy bands containing similar numbers of counts: 0.3--0.8, 0.8--1.3, and 1.3--2.3 keV, to make an RGB image (Fig.~\ref{fig:rgb}).
In detail, taking the 0.3 to 2.3 keV image we computed for each pixel the radius containing a minimum of 36 counts.
Each pixel was then convolved with a separate Gaussian, taking the above radius as its $\sigma$.
This smoothing was applied to the image data, background model and exposure map in each energy band.
Exposure-corrected background-subtracted images were then computed from these smoothed images.

The complex disturbed morphology of the intracluster medium can be seen.
There is a bright dense core with sharp edges to the west (W).
Surrounding the core of the cluster is a roughly elliptical region aligned along the NE to SW direction with an edge of around 700 to 800 kpc (`inner edge').
Beyond this emission can be seen out to a radius of $R_{500} \sim 1.4$~Mpc, although it is more visible along the SW direction.
There appears to be an edge in the X-ray profile at this radius (`outer edge').
To the NW there is extension of emission, which we call the NW structure, which appears greener in the RGB colour scale, likely because it is cooler than the main cluster.

To show the features in more detail, Fig.~\ref{fig:img_resid} shows an X-ray image of the cluster with the point sources masked, compared to residual maps showing the fractional deviation to a smooth model.
The smooth model was constructed by calculating the average surface brightness of the unmasked regions as a function of radius from a central position.
We chose two different centres for this analysis: a centre based on the centroid of the inner 12 arcmin after iteration, and another chosen by hand to lie at the centre of one contour level.

A bright central core can be seen, with a fan shaped structure extending from it towards the NE.
The central core appears to have a sharp edge to the W, except there is also an elongated bright region which extends westwards to a bright clump in the residual map (the `bridge').
There are also other filamentary structures extending inwards from this bright clump.
These filamentary structures have the appearance of the bright cooler material which surrounds bubbles of radio emitting plasma in some clusters \citep[e.g.][]{Fabian12}, although they are only seen to one side of the cluster.
To the far north-west (NW) we see a bright region of emission.
This is the extended source X8 mentioned previously.

\begin{figure}
  \includegraphics[width=\columnwidth]{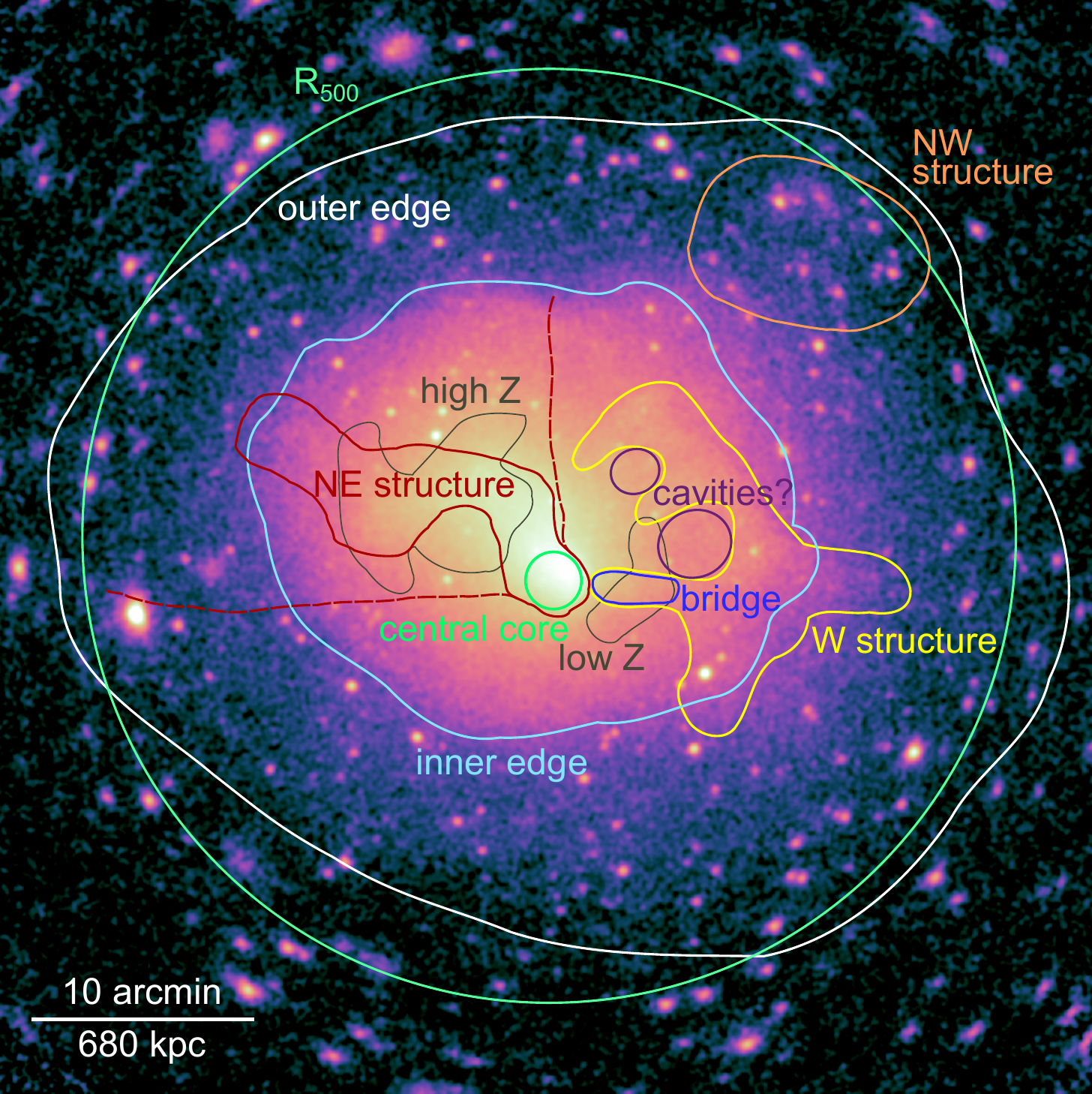}
  \caption{
    Image labelling some of the features discussed.
    The image is exposure corrected and background subtracted in the 0.3 to 2.3 keV band, and smoothed by a Gaussian with $\sigma=6$~arcsec.
  }
  \label{fig:schematic}
\end{figure}

To make our discussion clearer, we show in Fig.~\ref{fig:schematic} a labelled diagram of some of the features present.
Seen are the central core, the fan shaped structure to the NE, the filamentary structure to the W, the bridge connecting the core to the W structure, the inner and outer edges, and the NW structure.

\begin{figure*}
  \centering
  \includegraphics[width=\textwidth]{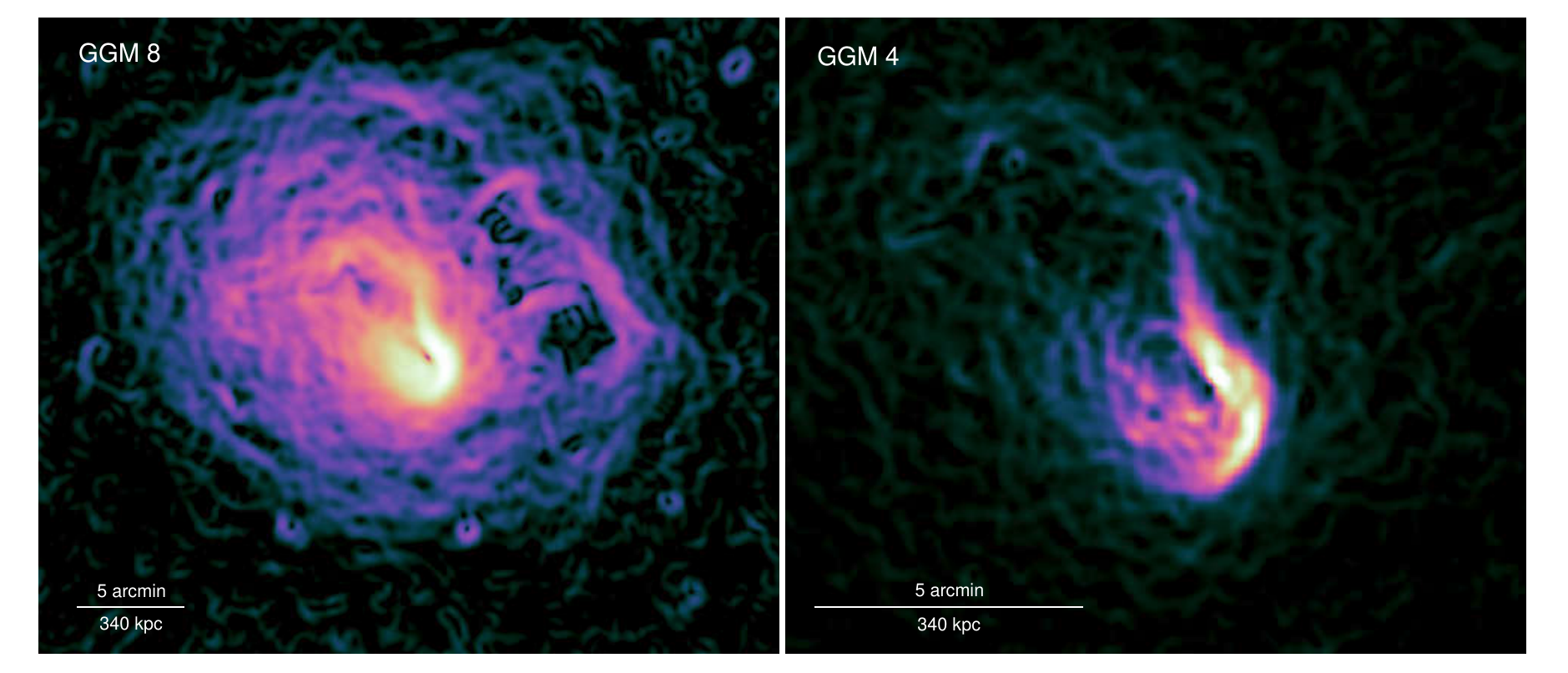}
  \includegraphics[width=\columnwidth]{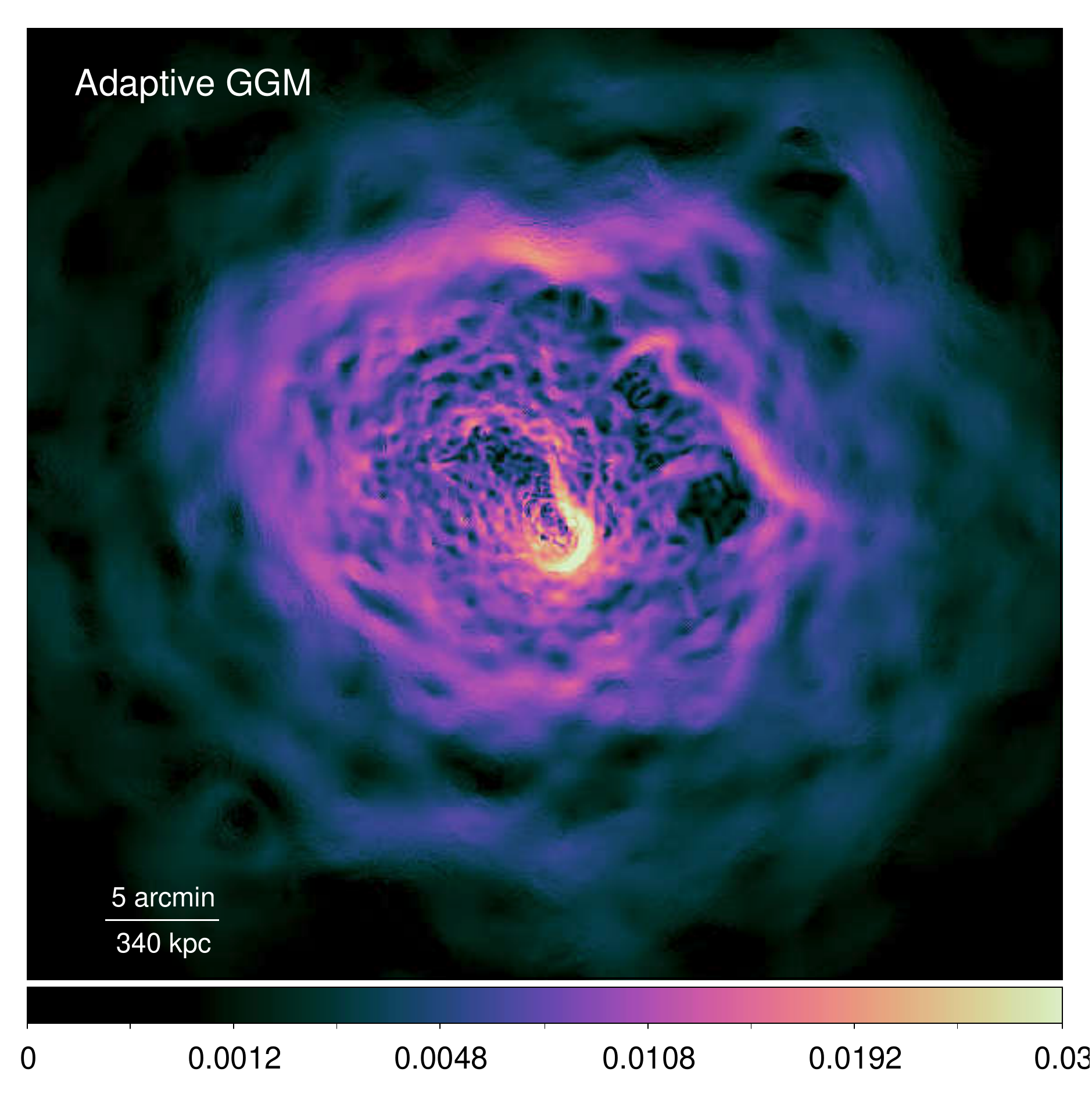}
  \caption{
    Edge-filtered X-ray images.
    In the top panels the X-ray 0.3 to 2.3 keV image was filtered with the Gaussian Gradient Magnitude (GGM) filter on scales of 8 pixels (16 arcsec; top left panel) and 4 pixels (8 arcsec; top right panel).
    Before filtering, point source regions were cosmetically filled using the values of random neighbouring pixels.
    The bottom panel has been processed by an adaptive GGM filter (using a signal to noise ratio of 32), where the scale shows the gradient as the log$_{10}$ change per 4 arcsec pixel.
  }
  \label{fig:ggm}
\end{figure*}

These filamentary structures and the central core can be seen in more detail by applying filtering to the X-ray image.
In Fig.~\ref{fig:ggm} is shown data filtered using the Gaussian Gradient Magnitude filter \citep[GGM;][]{SandersCent16,SandersGGM16}.
The effect of the filter is to compute the gradient magnitude of the X-ray image when convolved with a Gaussian of a particular size, here with scales of 16 and 4 arcsec.
The GGM filtered images are sensitive to the edges in the X-ray data.

Standard GGM filtering becomes noisier in regions where the count rate is low, for example in the outskirts of galaxy cluster.
Therefore, we also show the results from a new method to apply an adaptively smoothed GGM filter.
To do this, the adaptively smoothed X-ray image was taken from Section \ref{sect:src_detn} (computed by convolving the image and exposure map by a Gaussian with a $\sigma$ given by a radius containing 1024 counts).
The log$_{10}$ value of this map was then taken and the gradient calculated, by taking the difference between neighbouring pixels along the two axes and adding them in quadrature.
This method is described in more detail in Appendix \ref{appen:ggm}.
The filtered images clearly show the sharp edge to the west of the nucleus, the filamentary structure further out to the west and the 700--800 kpc elliptical surface brightness edge surrounding the core (the inner edge).

\begin{figure}
  \includegraphics[width=\columnwidth]{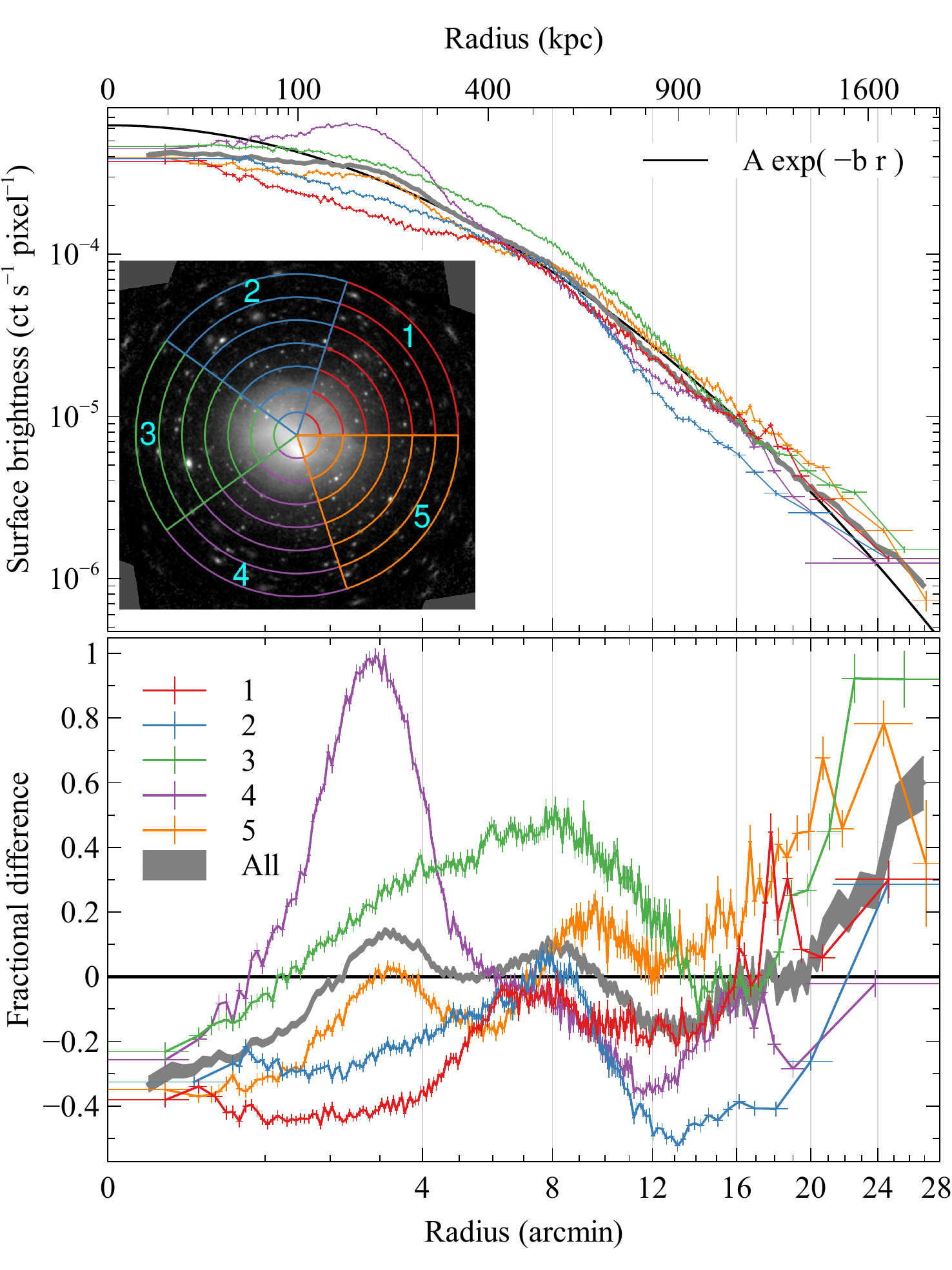}
  \caption{
    Surface brightness profiles in five different sectors between 0.3 and 2.3 keV using 2 arcsec pixels, after masking sources.
    The profiles were binned to have less than a 4\% uncertainty in each bin.
    The top panel shows the profiles, while the bottom panel shows the residuals to a smooth model.
    The smooth model (shown in a top panel) has the functional form $6.25 \times 10^{-4} \exp [ -0.260 \, ( r/\textrm{arcmin} ) ]$~ct~s$^{-1}$~pixel$^{-1}$.
    The radial axis uses square-root scaling where the vertical lines mark the same radii shown in the inset cluster image.
  }
  \label{fig:sbprof}
\end{figure}

The edges in surface brightness can be more clearly seen in radial profiles.
Figure~\ref{fig:sbprof} shows profiles along five different sectors and for the whole cluster.
The plot also shows the residuals to a smooth scale-free model.
In the centre, the bright central core can be seen in sector 4 (SE).
In sectors 1, 2, and 3 towards the north, the profile appears smooth in the centre, with breaks in slope between 6 and 8 arcmin radius where the profiles appear to steepen.
This is the radius of the inner edge.
There are further breaks in the slopes between 12 and 19 arcmin radius in all sectors where they flatten out again.
Towards the edge there is further evidence for steepening (the outer edge).

\subsection{Central spectrum}
\label{sect:central}

\begin{figure}
  \centering
  \includegraphics[width=\columnwidth]{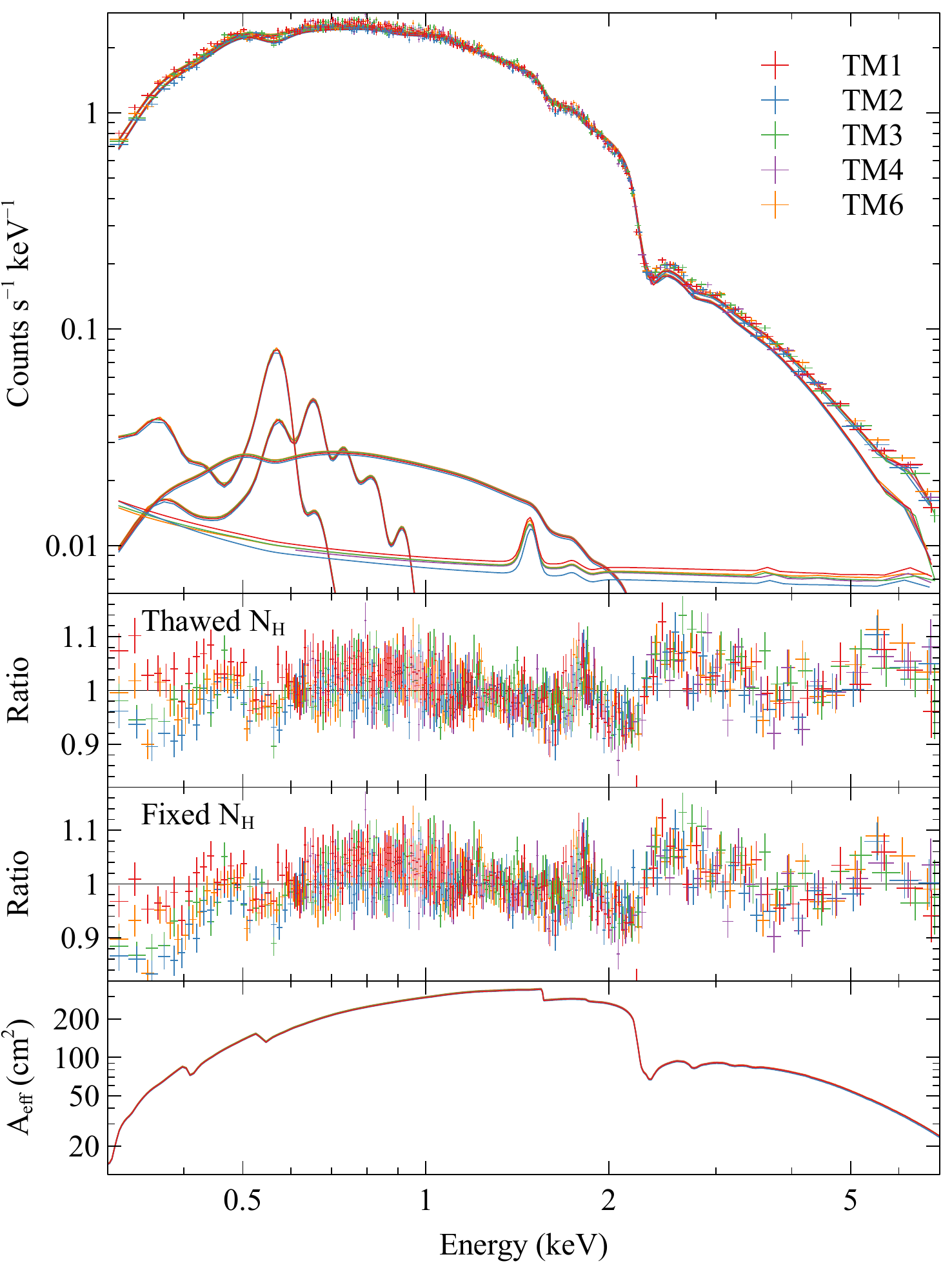}
  \caption{
    Spectrum of the central 6 arcmin radius around the cluster centre.
    The data were rebinned to have a signal to noise ratio of 30 in each spectral bin for display purposes.
    In the top panel are the data, total model, cluster model, XRB components, and NXB in each TM.
    The second panel shows the residuals after thawing the absorbing column density parameter.
    The third panel shows the residuals with the column density fixed to be the corrected HI4PI value.
    The fourth panel shows the average effective area of the TMs as a function of energy over the extraction region based on the source model.
  }
  \label{fig:inner_spec}
\end{figure}

\begin{table*}
  \caption{
    Spectral fitting results from the central 6 arcmin radius.
  }
  \centering
  \begin{tabular}{lcccc}
  \hline\hline
    Type & $N_\mathrm{H}$ ($10^{20}$ cm$^{-2}$) & $k_\mathrm{B}T$ (keV) & $Z$ (Z$_{\odot}$) & Normalisation (cm$^{-5}$ deg$^{-2}$) \\ \hline
    Fixed $N_\mathrm{H}$ & $2.26$ & $8.39 \pm 0.07$ & $0.44 \pm 0.02$ & $1.167 \pm 0.003$ \\
    Free $N_\mathrm{H}$ & $3.18 \pm 0.05$ & $7.46 \pm 0.07$ & $0.30 \pm 0.02$ & $1.197 \pm 0.003$ \\
%    Fixed $N_\mathrm{H}$, singles only & $2.26$ & $9.25 \pm 0.19$ & $0.60 \pm 0.06$ & $1.041 \pm 0.007$ \\
%    Thawed $N_\mathrm{H}$, singles only & $3.12 \pm 0.08$ & $7.91 \pm 0.17$ & $0.32 \pm 0.05$ & $1.073 \pm 0.007$ \\
    Survey, fixed $N_\mathrm{H}$ & $2.26$ & $6.7 \pm 0.5$ & $0.7 \pm 0.2$ & $1.04 \pm 0.03$ \\
    Survey, free $N_\mathrm{H}$ & $2.8 \pm 0.5$ & $6.3 \pm 0.6$ & $0.6 \pm 0.2$ & $1.07 \pm 0.04$ \\
    Survey, fixed $Z$ & $3.2 \pm 0.4$ & $5.8 \pm 0.5$ & $0.30$ & $1.11 \pm 0.02$ \\
    Survey, fixed $N_\mathrm{H}$, $Z$, $k_\mathrm{B}T$ & $3.18$ & $7.46$ & $0.30$ & $1.14 \pm 0.01$ \\ \hline
  \end{tabular}
  \tablefoot{
      The results for the pointed observation described here, and for an early version of the eRASS1 survey, are shown.
    Normalisations are given per square degree of extraction area.
  }
  \label{tab:central_fit}
\end{table*}

As a first spectral analysis, we fitted the spectrum from the inner 6 arcmin radius after excluding point source regions.
The same centre was used as \cite{Ghirardini19} for their \emph{XMM-Newton} analysis $(\alpha=67.8434^\circ, \delta=-61.4297^\circ)$.
In this analysis we supplied a spatial model, the adaptively smoothed image from Section \ref{sect:src_detn}, to the \texttt{srctool} task to better model the effect of the vignetting of the source on the ARF.
When modelling the XRB in the same spectrum we instead used an ARF assuming a flat spatial model.
As with all the analyses in this paper, we disable the correction for the telescope PSF in \texttt{srctool}, as the PSF is smaller than our typical extraction region.
We fitted the spectrum for the region with an \texttt{tbabs} absorbed \texttt{apec} model to account for the cluster emission, the XRB components from the eRASS1 best-fitting model and the NXB components for the region in the outskirts of the cluster.
To allow for variation in the background, we allowed the normalisations of the XRB and NXB components to vary with a Gaussian prior of 5\%.

The spectra from each TM was fitted between the minimum range in Table \ref{tab:tms} to 7.0 keV, excluding TM5 and TM7.
The C-statistic plus prior was minimised to fit the spectra.
Figure~\ref{fig:inner_spec} shows the data, best fitting model and model sub-components.
As there were clear residuals at low energies, we also fitted the data allowing the absorption to vary.
The plot shows the residuals for the varying and fixed absorption cases.
The best fitting parameters for the two models are listed in Table \ref{tab:central_fit}.
It can be seen that the data prefer a temperature around 12\% higher if the photoelectric absorption is frozen at our Galactic value rather than fitted, while the metallicity is around 50\% higher than the obtained with a fitted column density.

There are clear residuals (Fig.~\ref{fig:inner_spec}) at the 10\% level over the entire energy band for both versions of the fit.
These are seen in particular at the edges in the effective area curves.
The residuals are likely due to residual calibration uncertainties of the telescopes.
The higher value of the best-fitting absorption may also be due to uncertainties in the calibration, as the other residuals in the spectral fit have a similar magnitude.
For hot clusters like A\,3266, it is difficult to disentangle the absorption, metallicity and temperature with the softer response of the eROSITA telescopes.

The cluster has also been observed by the eRASS1 survey.
As it is relatively bright, we can make comparisons against the pointed observation.
If the vignetting model is inaccurate, this would lead to differences between the pointed and survey fits.
We note that the eRASS1 data have not been processed with the final eRASS1 version of the eSASS software or calibration, but only with a similar one to the pointed observation and therefore this comparison is preliminary.
Table \ref{tab:central_fit} shows the results of spectral fitting for free and fixed absorption values.
It can be seen that the temperature found from the survey is lower than found with the pointed observation and the metallicity is higher.
The absorption, if free, is roughly consistent with, and lies between, the HI4PI  value and the fitted value obtained from the pointed observation.
If the metallicity is fixed to a more realistic value ($0.3$~Z$_\odot$), then the temperature drops lower still and the absorption increases to what was found in the pointed observation.
If the temperature, metallicity and absorption are forced to be the best fitting values from the pointed observation, the normalisation agrees to within 5\%.

It should be noted, however, as seen previously by \emph{XMM-Newton} or in Section~\ref{sect:mapping}, the cluster has temperature sub-structure within the central region.
Our temperature maps later show a range of $6.6$ to $9.1$~keV within 6 arcmin radius ($1\sigma$ percentiles).
Components at different temperatures will preferentially be detected by different instruments depending on the sensitivity of their telescopes to different X-ray energies \citep[e.g.][]{Reiprich13}.
eROSITA in a survey has on average less effective area at high energies than it does on-axis in a pointing.
Therefore the temperature differences seen between the pointed and survey data could be due to this effect, or it could be due to calibration uncertainties.
Further work is required to understand this better.

\subsection{Spectroscopic profiles}

\begin{figure}
  \centering
  \includegraphics[width=\columnwidth]{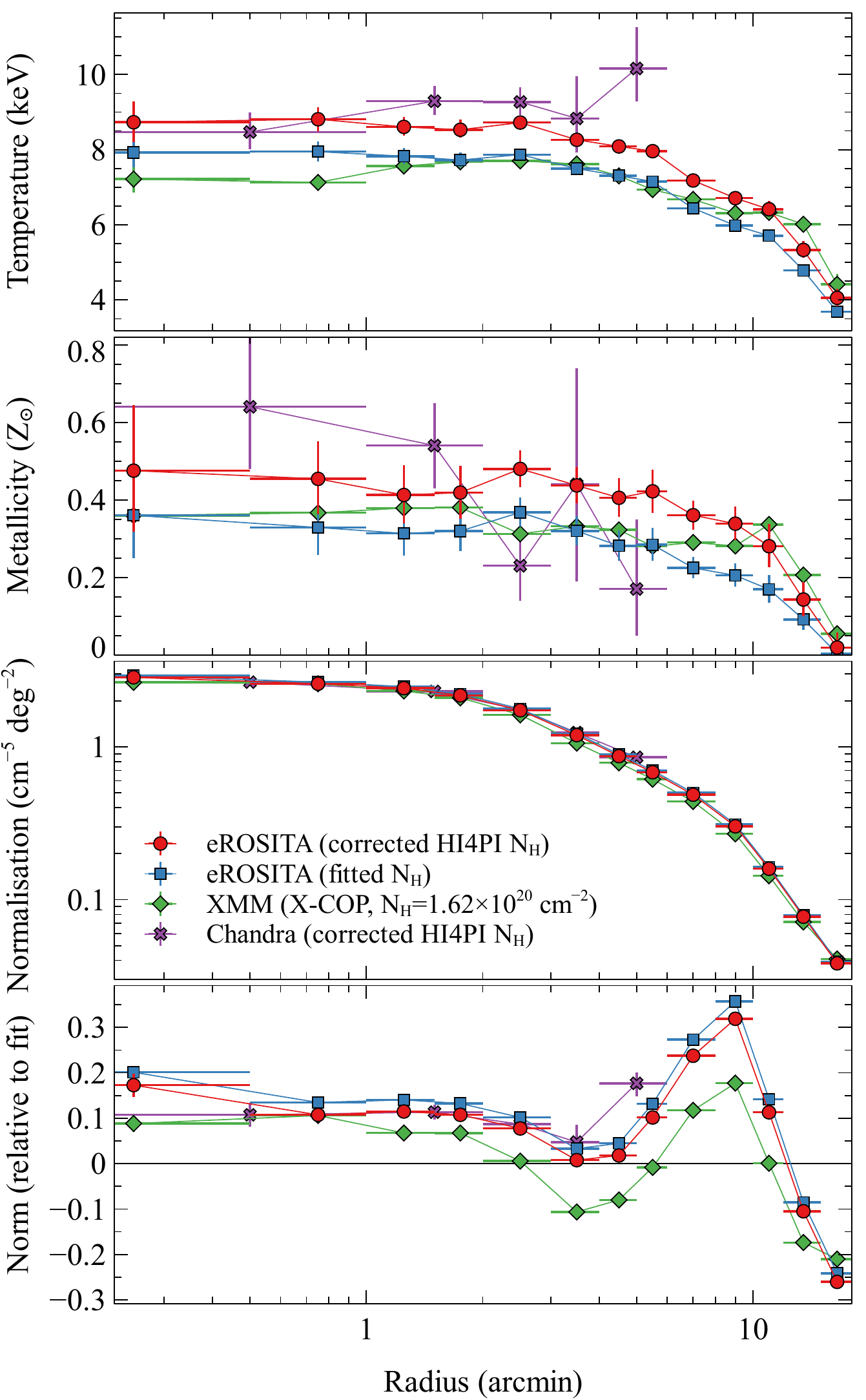}
  \caption{
    Comparison between eROSITA, \emph{XMM-Newton}, and \emph{Chandra} spectroscopic profiles for annular regions, showing the temperature, metallicity, normalisation, and relative normalisation.
    The \emph{XMM-Newton} results are the from the X-COP analysis of \cite{Ghirardini19}, scaling the metallicity profiles from \cite{AndersGrevesse89} to \cite{Asplund09} solar abundance ratios by multiplying by a factor of $1.48$ to account for the relative H/Fe ratio.
    The eROSITA profiles use the same centre and annuli as the \emph{XMM-Newton} profiles, while the \emph{Chandra} profiles use wider annuli.
    Normalisations are the \texttt{Xspec} normalisation factor per square degree.
    The relative normalisation shows the fractional difference to a $\beta$-model fitted to the \emph{XMM-Newton} data.
  }
  \label{fig:compar_profiles}
\end{figure}

A\,3266 has been the subject of observations by other X-ray telescopes and is therefore a good target to use to compare instrumental calibrations.
Figure~\ref{fig:compar_profiles} compares profiles obtained by spectrally fitting eROSITA, \emph{XMM-Newton}, and \emph{Chandra} data.
The \emph{XMM-Newton} results were taken from \cite{Ghirardini19} which fits both EPIC-pn and EPIC-MOS data.
We used the same cluster centre and annular regions as this analysis, extracting the spectra using \texttt{srctool}, excluding point sources.
Two sets of ARFs were created: one assuming the cluster emission is distributed like the adaptively smoothed X-ray emission (Fig.~\ref{fig:regions}) and was applied to the cluster model, while the second assumed a flat spatial distribution and was applied to the XRB model.
The two XRB and NXB background components from Section \ref{sect:back} were fitted to the data from each annulus in addition to an \texttt{apec} model absorbed by a \texttt{tbabs} model to account for the cluster emission.
We fitted the data from the different TMs simultaneously from our standard minimum energies to 7 keV.

We repeated the fits using two different absorption values.
The first was the corrected HI4PI value ($2.26 \times 10^{20}$~cm$^{-2}$) and the second an average value obtained by allowing the absorption to be free in these radial profile fits ($3.15 \times 10^{20}$~cm$^{-2}$).
This second value is close to that obtained from the inner 6 arcmin ($3.18 \times 10^{20}$~cm$^{-2}$).
We did not see evidence for absorption variation in these profiles inside 8 arcmin radius (beyond this the measurement uncertainties become larger).
The second value is what we use in the remainder of the spectral fits in this paper.

The \emph{Chandra} observation IDs 899 and 7687 were used for the \emph{Chandra} analysis.
Data reduction was performed using the software {\tt CIAO v4.12} \citep{Fruscione06}, with the latest release of the \emph{Chandra} Calibration Database at the time of writing ({\tt CALDB v4.9}).
Time intervals with a high background level were filtered out by performing a 3$\sigma$ clipping of the background level on the light curve in the 2.3--7.3~keV band, and binned with a time interval of 200~s. The cleaned exposure times were 28.0~ks and 4.5~ks for obsIDs 899 and 7687, respectively.

Unresolved sources within the ICM were identified with {\tt wavdetect}, checked visually, and eventually removed. The ARF and redistribution matrix file (RMF) for each observation were extracted with the commands {\tt mkarf} and {\tt mkacisrmf}, respectively. The background spectra were extracted from the `blank sky' files, and processed using the {\tt blanksky} script (we used the default options with {\tt weight\_method} `particle' and {\tt bkgparams=[energy=9000:12000]}).

Galactic hydrogen absorption was described by the model {\tt tbabs}, where the Galactic column density was fixed at our HI4PI-corrected value of $2.26\times10^{20}~{\rm cm}^{-2}$.
To ensure flat azimuthal coverage, only obsID 899 was used within 3 arcmin radius and only obsID 7687 outside that radius.
The ICM spectrum in the 0.5--7.0~keV band was fitted with the {\tt apec} thermal plasma emission model, where the redshift was fixed at 0.0589, and the temperature, abundance, and normalisation were thawed as free parameters, minimising the C-statistic.

The eROSITA temperature profiles lie somewhere between the \emph{XMM-Newton} and \emph{Chandra} results, depending on what value is assumed as the Galactic absorption in the spectral fits.
If the absorption is fixed to the corrected HI4PI value, then the eROSITA temperatures are similar to the \emph{Chandra} values.
If the absorption is fixed to be our best fitting value, then the temperature profile is around 1 keV hotter in the very centre, but matches \emph{XMM-Newton} out to around 7\,arcmin radius, but then becomes cooler.
The temperature and metallicity are inversely correlated.
We obtain similar metallicities to \emph{XMM-Newton} over most of the cluster ($\sim 0.3$~Z$_\odot$) with the higher absorption value.
The uncertainties on the \emph{Chandra} metallicities make comparison more difficult, but are suggestive of higher metallicities in the very centre, but matching \emph{XMM-Newton} and high-absorption eROSITA further out.
We note that the \emph{Chandra} values are also somewhat dependent on the assumed absorption.
If our best fitting eROSITA absorption is used rather than the corrected HI4PI value, the \emph{Chandra} temperatures increase by $\sim 1$\,keV and the metallicites decrease by $\sim 0.05$\,Z$_\odot$.
The \emph{Chandra} data themselves prefer an even larger best fitting column density of $\sim 4.9\times 10^{20}$\,cm$^{-2}$, depending on radius.

The normalisation profiles for the eROSITA and \emph{Chandra} fits are around 10\% larger than \emph{XMM-Newton} over most of the radial range.
This difference could be due to a lower absorption value ($1.62 \times 10^{20}$\,cm$^{-2}$) used for the \emph{XMM-Newton} X-COP analysis.

If the emitting plasma in a cluster is not isothermal, it is expected that non-identical X-ray telescopes would produce different best fitting temperatures.
Depending on the relative amount of effective area as a function of energy, telescopes will preferentially detect hotter or colder material.
A\,3266 contains spatial variation in temperature (see Section \ref{sect:mapping}) which is not simply radial, making comparison difficult.
There are also intrinsic calibration differences between X-ray telescopes.
\cite{Schellenberger15} found that hotter galaxy clusters produce higher
temperatures in \emph{Chandra} than \emph{XMM-Newton}, even after accounting for multi-temperature gas or sensitivity in different bands.

The discrepancy between the temperatures obtained from the different telescopes in this hot cluster may be due to calibration differences or could be due to multiphase gas.
Future work to disentangle could include measuring temperatures with narrower energy bands (to reduce the effect of calibration differences) or selecting isothermal regions within the cluster for comparison (although multi-temperature material can be present along the line of sight and may be difficult to detect).

\begin{figure*}
  \centering
  \includegraphics[width=0.48\textwidth]{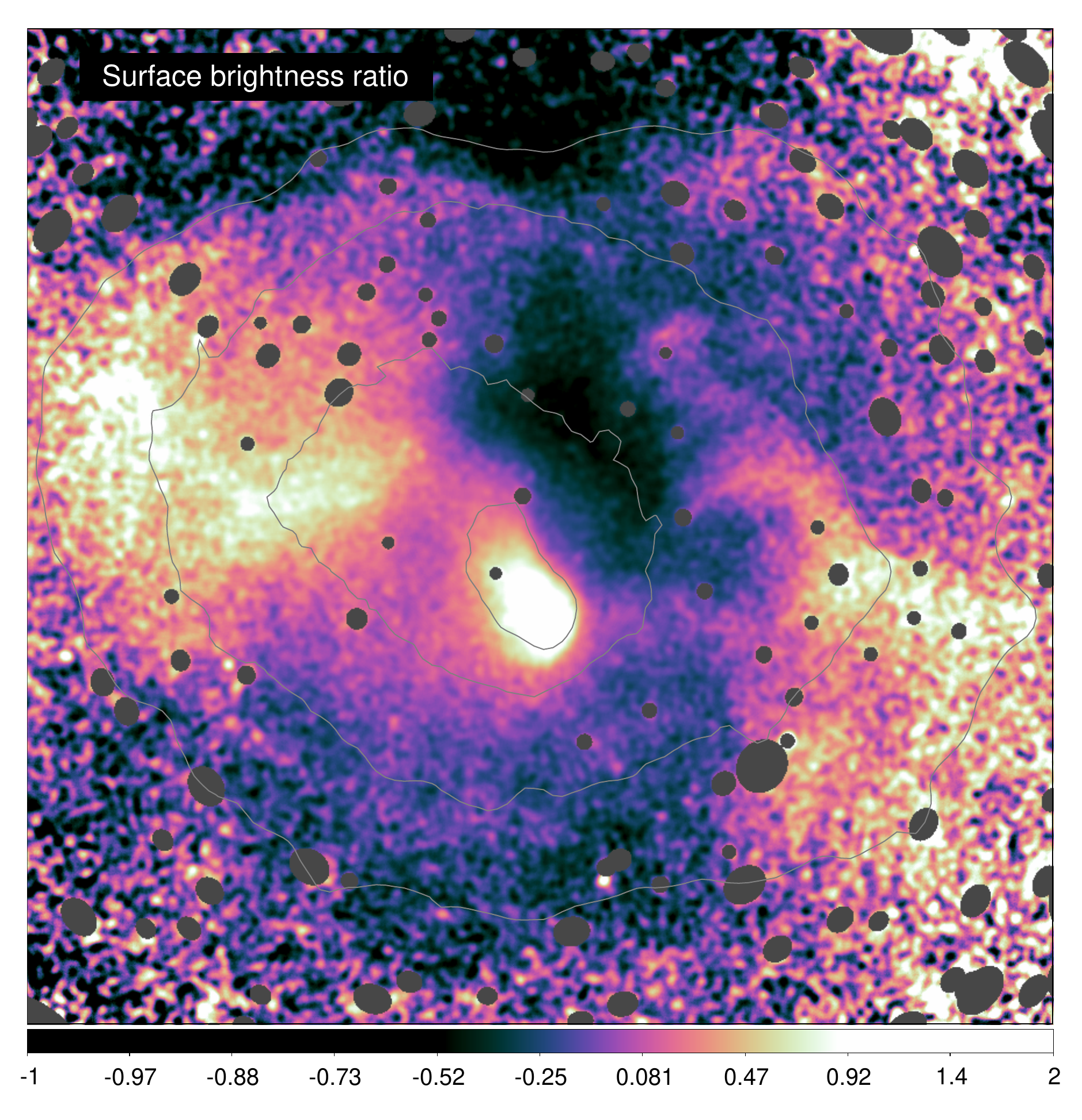}
  \includegraphics[width=0.48\textwidth]{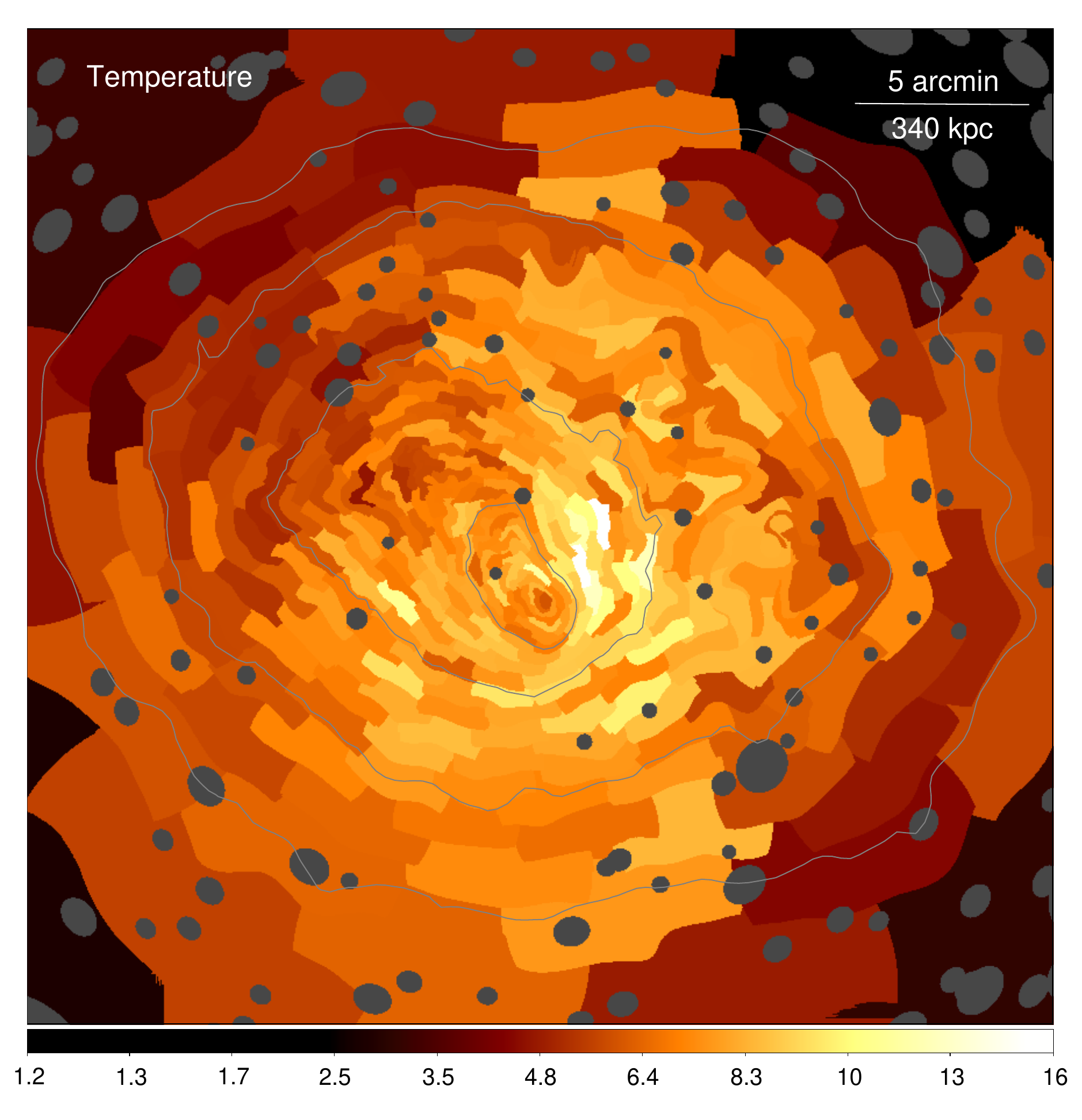}
  \includegraphics[width=0.48\textwidth]{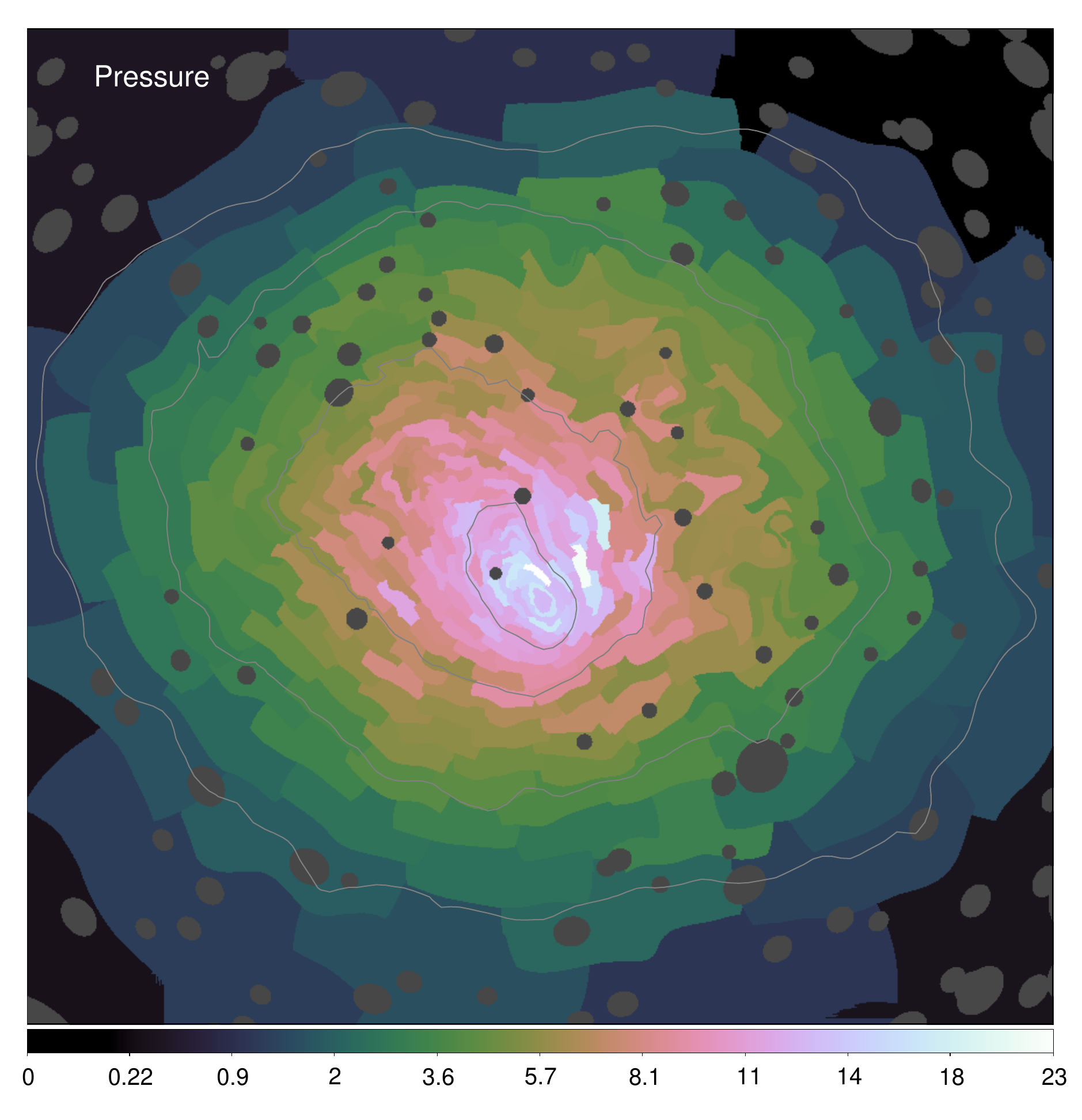}
  \includegraphics[width=0.48\textwidth]{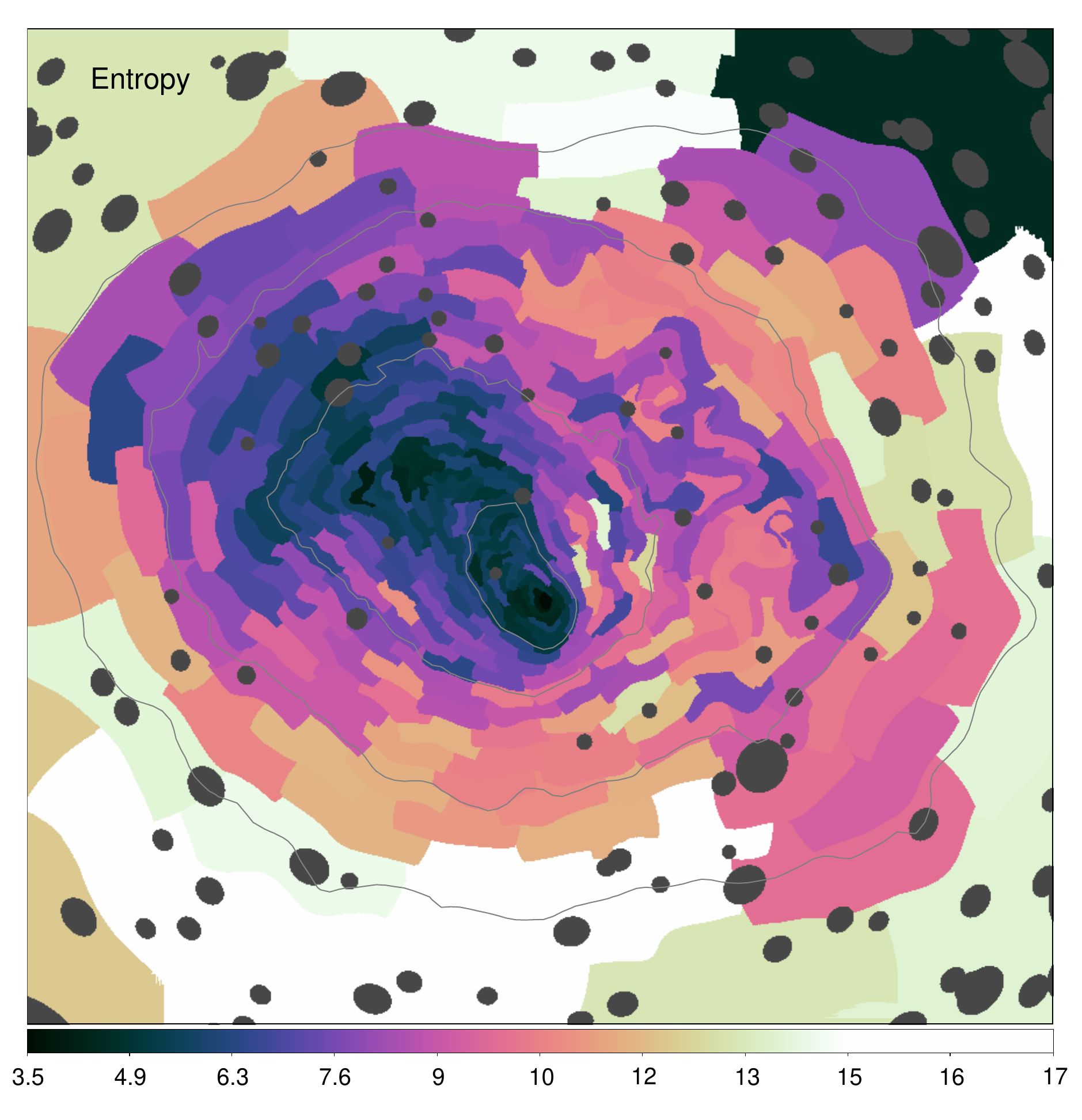}
  \caption{
    Spatially resolved spectroscopic maps created using the contour binning method.
    (Top left) Ratio of surface brightness to average at radius, for comparison, smoothed by a Gaussian with $\sigma=6$~arcsec.
    (Top right) Temperature, in keV.
    (Bottom left) Pseudo-pressure (keV~cm$^{-5}$~deg$^{-2}$).
    (Bottom right) Pseudo-entropy (keV~cm$^{5/3}$~deg$^{2/3}$).
    The maps were create by spectrally fitting regions with a signal to noise ratio of 75.
    Median statistical temperature uncertainties are around 12\% (1 keV) for each region.
  }
  \label{fig:contbin_maps}
\end{figure*}

\begin{figure*}
  \centering
  \includegraphics[width=0.48\textwidth]{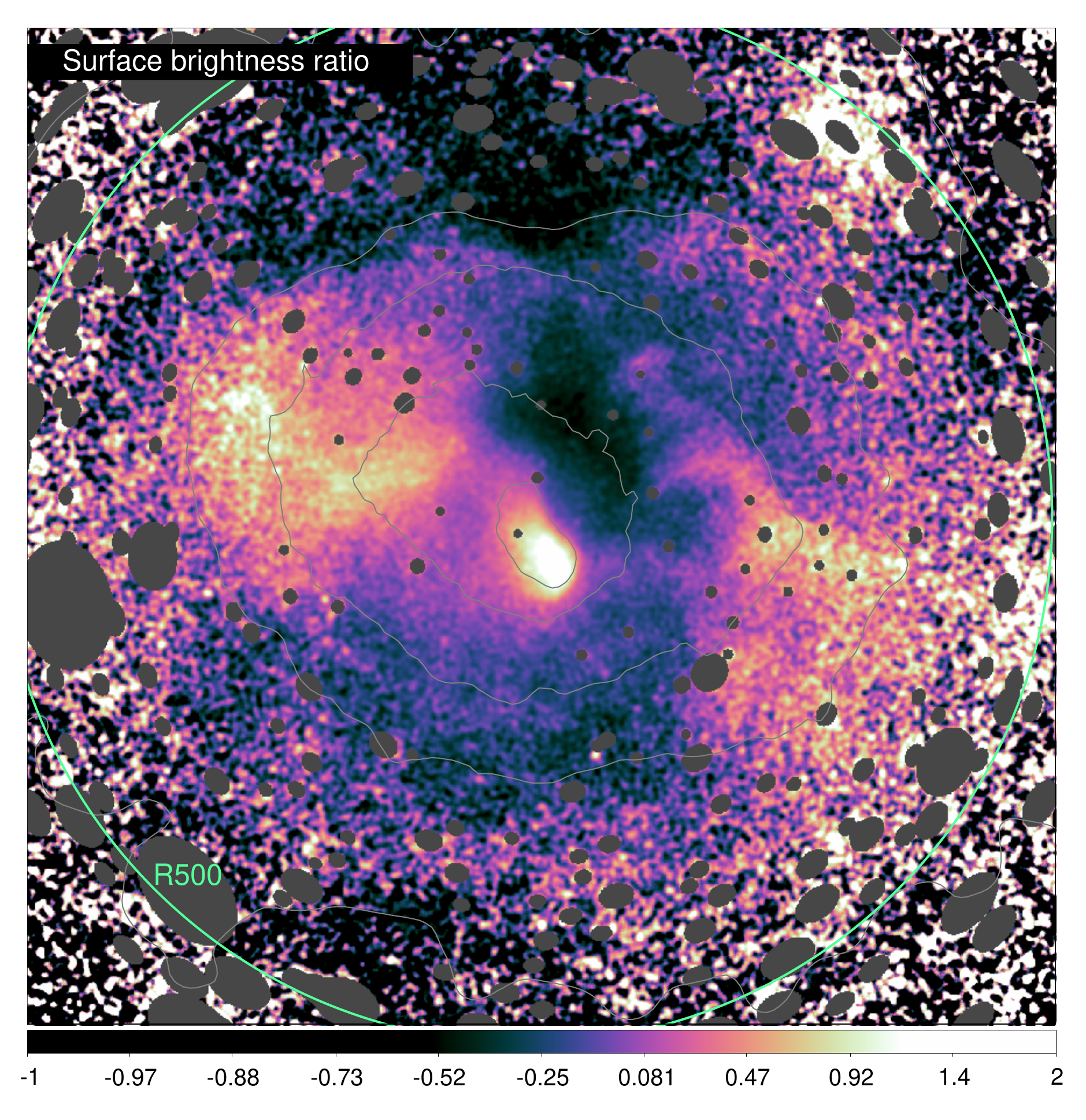}
  \includegraphics[width=0.48\textwidth]{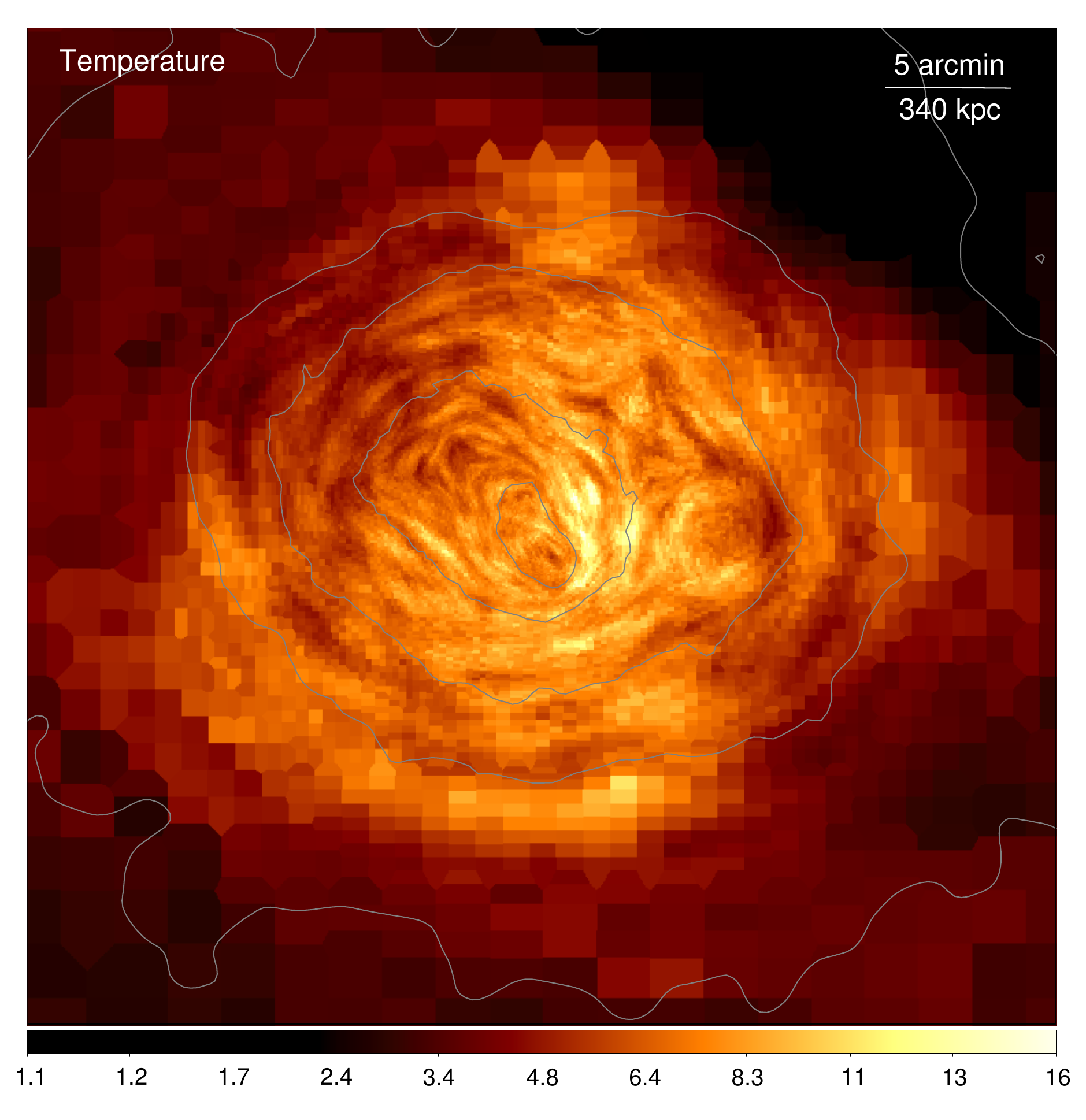}\\
  \includegraphics[width=0.48\textwidth]{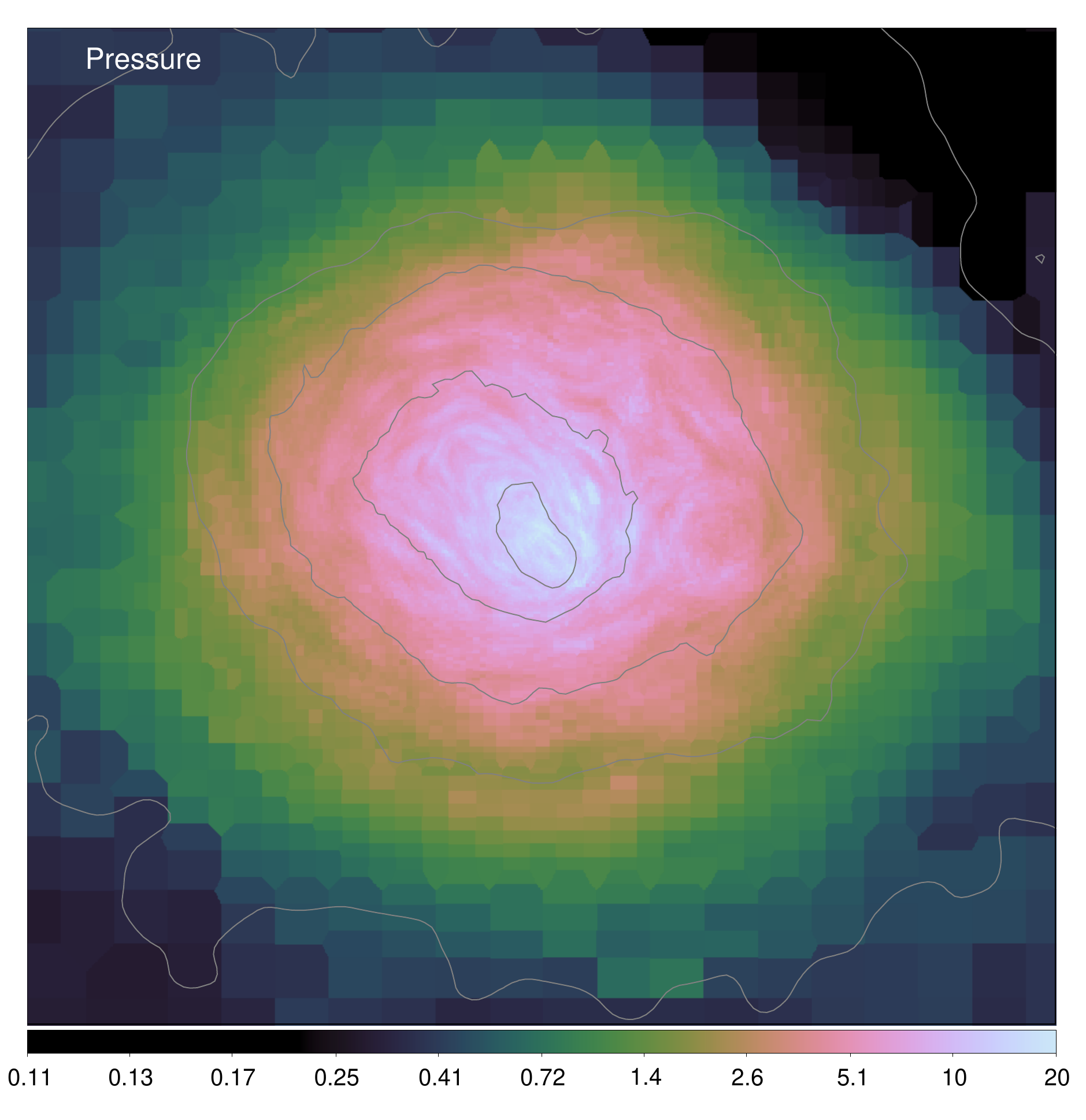}
  \includegraphics[width=0.48\textwidth]{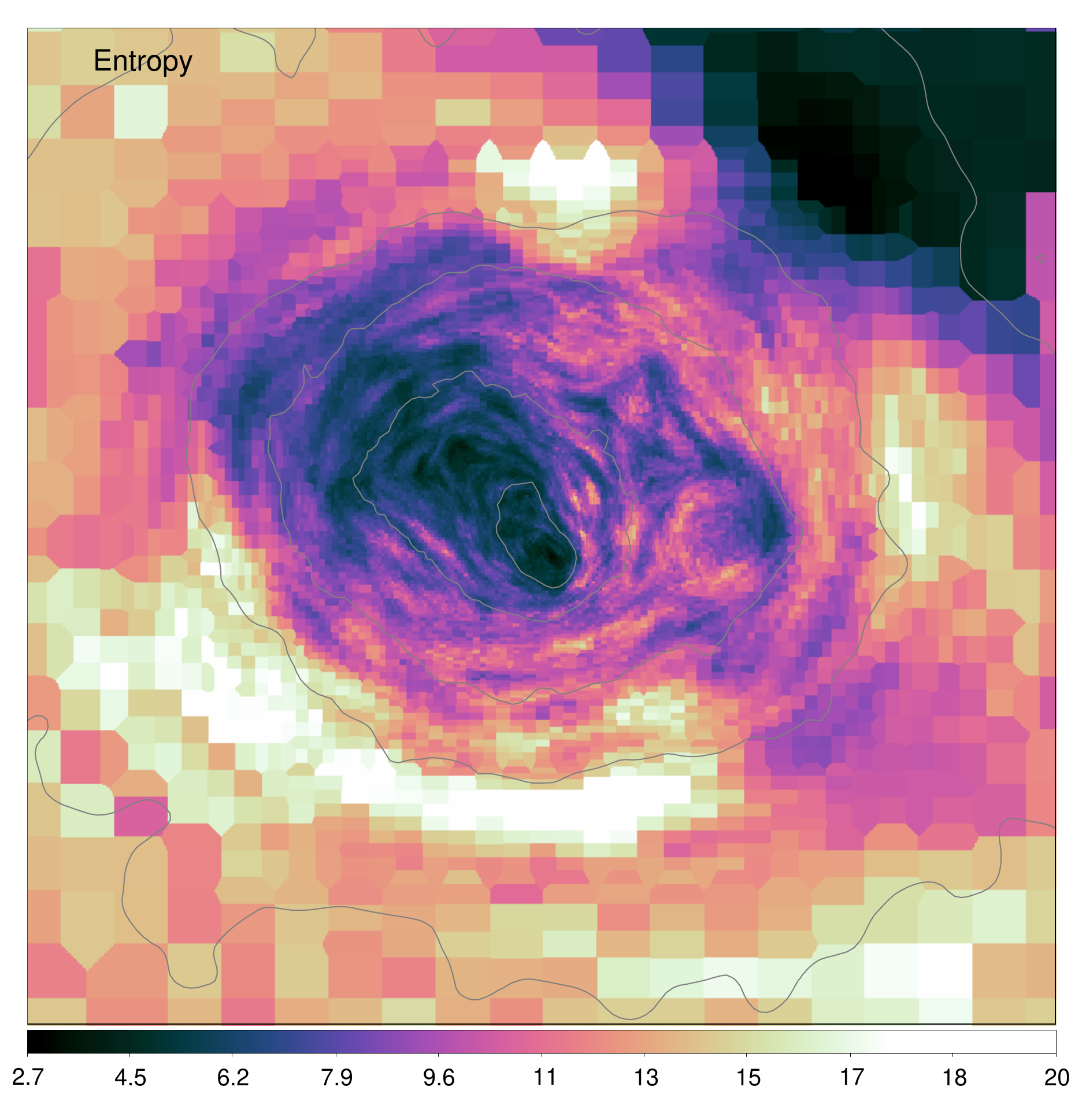}
  \caption{
    Spatially resolved spectroscopic maps made using the ellipse fitting method.
    These use the same signal to noise ratio as the contour binned maps in Fig.~\ref{fig:contbin_maps} but are zoomed to show a larger area and have slightly different colour scales.
    The ratio map in this case has been smoothed by a Gaussian with $\sigma=6$~arcsec.
    The ellipses selected have a similar area to the contour binning regions shown in Fig.~\ref{fig:contbin_maps} and have a similar aspect ratio, with many having the maximum 3:1 ratio.
  }
  \label{fig:ellipse_maps}
\end{figure*}

\subsection{Spatially resolved spectroscopic mapping}
\label{sect:mapping}
We made maps of the cluster spectral properties using two methods.
The first was to select regions using the contour binning algorithm \citep{SandersBin06}.
The second method is a new one based on fitting spectra from dynamically sized ellipses.
It has the advantage of producing a smoothly varying map, but the disadvantage of results not being statistically independent, except when points are separated by more than the size of these ellipses.

For the ellipse method, the input image is an adaptively smoothed image (from Section \ref{sect:bgsub}).
We take an initial set of grid points separated by 256 2-arcsec pixels.
For each of these grid points, we iterate over a set of ellipses rotated at intervals of 5 degrees and with a range of aspect ratios (here between 1 and 3 with 11 steps), where the ellipse size is increased until the required signal to noise threshold is reached (masking point sources) for each of these possibilities.
The ellipse chosen for a grid point is one where the standard deviation of the contained pixels in the adaptively smoothed map is minimised.

Given this set of initial ellipses on a grid, we construct a list of sets of four neighbouring grid points (quads).
For each axis in each quad we compute the distance between the grid points as a ratio of the size of an ellipse along that axis direction (i.e. where the ellipse intersects the axis).
There are four different ratios for each axis: one for each of the ellipses.
We take the largest ratio in each direction.
If the separation ratio for both directions is larger than some value $S$ (we use $S=0.5$ here), then we split the quad in both directions by two, by inserting five new grid points, making four new quads.
If the ratio for only one axis is larger than $S$ we split the grid along one axis by inserting two new points, making two new quads.
After doing this for all the quads, we take our new list of quads and repeat the process.
The effect of this procedure is to refine a grid so that the size of the grid is similar to the size of the ellipses, where the size of the ellipses is chosen to have the same signal to noise ratio and the shape and rotation of each ellipse is governed by the local morphology of the surface brightness.
For a perfectly radially declining surface brightness the ellipses increase in size with radius and rotate to be azimuthally aligned along their longest axes.

From each of these ellipses we extracted spectra and created responses, excluding point sources, using the \texttt{srctool} task.
These spectra were fitted using the method described below to produce a set of temperature, metallicity and normalisation values.
To generate output maps, we use the results for the ellipse whose centre lies closest to each pixel.

For the contour binning method, we binned the X-ray image following the contours on the same adaptively smoothed X-ray image as used in the ellipse fitting method.
We created regions applying a geometrical constraint factor of $1.7$, to prevent bins becoming too elongated.
Point sources were masked out during binning.
For our analysis we did our analysis using the maps binned using a signal to noise ratio of 75, for examining temperature and density, and again using a ratio of 100, for mapping metallicity.

As for the ellipse fitting analysis, from each region we extracted spectra and created responses using \texttt{srctool}.
Following Section \ref{sect:central} we create two sets of ARFs: one using the adaptively smoothed image as a source model to account for the cluster emission, and the second with a flat model for the XRB.
As previously, we simultaneously fitted the data for TMs 1, 2, 3, 4, and 6 with a model made up of a combination of cluster, XRB and NXB components.

In Fig.~\ref{fig:contbin_maps} we show the resulting temperature maps from the contour binning process compared to an X-ray residual map (Fig.~\ref{fig:img_resid}).
Also mapped is the pseudo-pressure, calculated by multiplying the temperature (in keV) by the square root of the \texttt{apec} norm per square-degree from \texttt{Xspec}.
The pseudo-pressure is related to the real pressure except it accounts neither for projection effects nor line-of-site depth through the cluster.
Another useful quantity we compute is the pseudo-entropy, calculated by multiplying the temperature by the \texttt{apec} normalisation per square degree to the power $-2/6$.
To aid comparison, we plot the same contours as shown in Fig.~\ref{fig:img_resid} on each map.

We similarly show the results from the ellipse fitting method in Fig.~\ref{fig:ellipse_maps}.
These maps have the same signal to noise ratio (and therefore statistical uncertainties) as the contour binned maps, but are zoomed out to show the $R_{500}$ radius.

Both the contour binned and ellipse method maps show very similar results.
Thanks to the design of the methods to trace the X-ray surface brightness, both show the same thermodynamic variations associated with the X-ray emitting structures seen in the cluster.
The maps show the central core is lower temperature and entropy, and connects to the lower temperature and entropy material which forms a fan shape (the NE structure).
The bright emission and filamentary structures to the west of the core (the W structure) also appear as lower temperature, lower entropy ICM.
Towards the NW we see cooler material associated with the NW structure.
Immediately to the east of the cool bright core we see a high temperature region, previously seen in \emph{XMM-Newton} data.
This region also has a high pressure in the pressure map.
We examine it in more detail in Section \ref{sect:core}.

\begin{figure}
  \centering
  \includegraphics[width=0.48\textwidth]{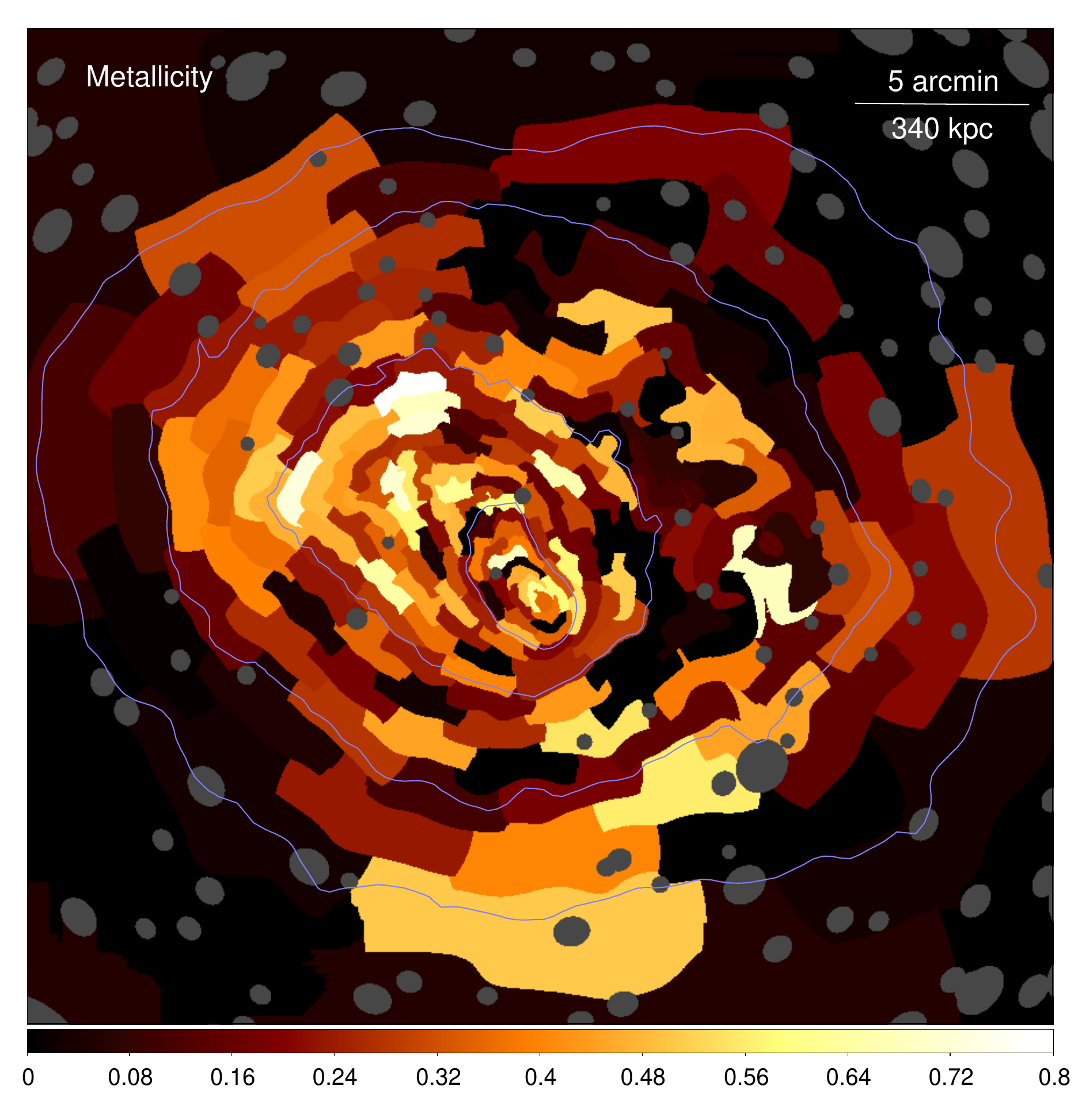}\\
  \includegraphics[width=0.48\textwidth]{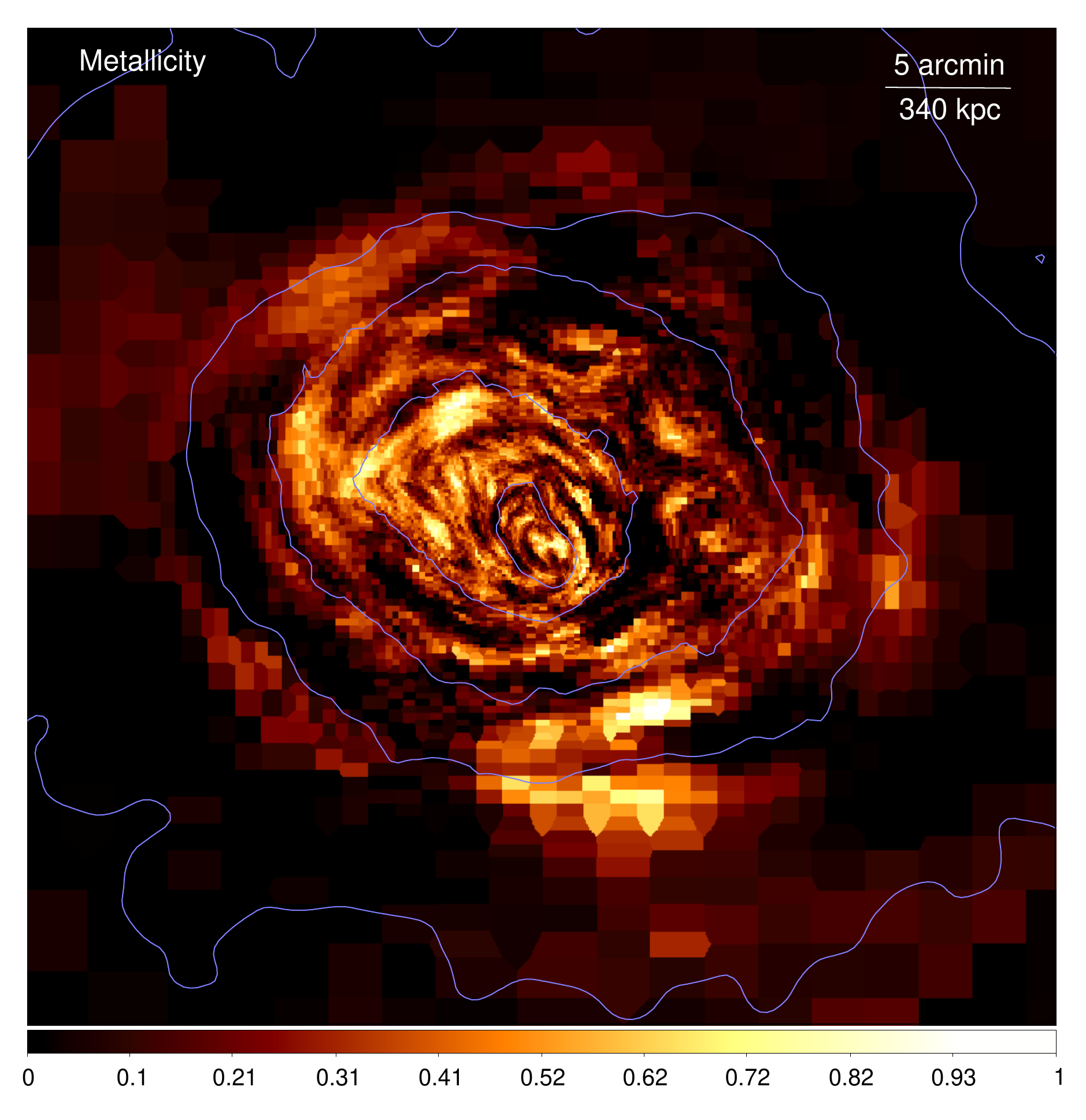}
  \caption{
    Metallicity maps of the cluster made using the contour binning (top) and ellipse (bottom) methods.
    The maps were created using spectra extracted from regions with a signal to noise ratio of 100.
    Typical statistical uncertainties are $0.17$~Z$_\odot$.
  }
  \label{fig:metallicity_maps}
\end{figure}

\subsection{Metallicity maps}
\label{sect:metals}
In Fig.~\ref{fig:metallicity_maps}, maps of the best fitting metallicity of the material, both using the contour binning and ellipse fitting methods, are shown.
Due to the relatively lower effective area of eROSITA at higher energies, these metallicity results are mostly sensitive to Fe-L emission, despite the relatively high temperatures of the ICM in this cluster.
We note that some caution might be necessary as the interpretation of the broad Fe-L complex is more model and calibration dependent than Fe-K.
The fits to the central region (Section \ref{sect:central}) show that there is some degeneracy between the absorbing column density and the metallicity.

The metallicity maps (Fig. \ref{fig:metallicity_maps}) show that the metallicity is not uniform within the cluster.
The maps show a higher region of metallicity in core and the fan shaped NE structure (marked High Z in Fig.~\ref{fig:schematic}).
\cite{Sauvageot05} and \cite{Finoguenov06} both found this region to have higher metallicities.
Our metallicity maps appears to match what they saw qualitatively.
Between the W structure and the central core is a region of low metallicity (marked Low Z in Fig.~\ref{fig:schematic}).
\cite{Sauvageot05} found a low metallicity path which partially coincides with this region.
The ellipse-smoothed map suggests the low metallicity region extends around from the W to the SE at the radius of the 2nd-innermost contour.
This is also seen in contour binned maps with larger bins.
\cite{Sauvageot05} also saw high metallicity regions to the south which are also found in our results.

\subsection{Cluster core in detail}
\label{sect:core}

\begin{figure}
  \centering
  \includegraphics[width=0.8\columnwidth]{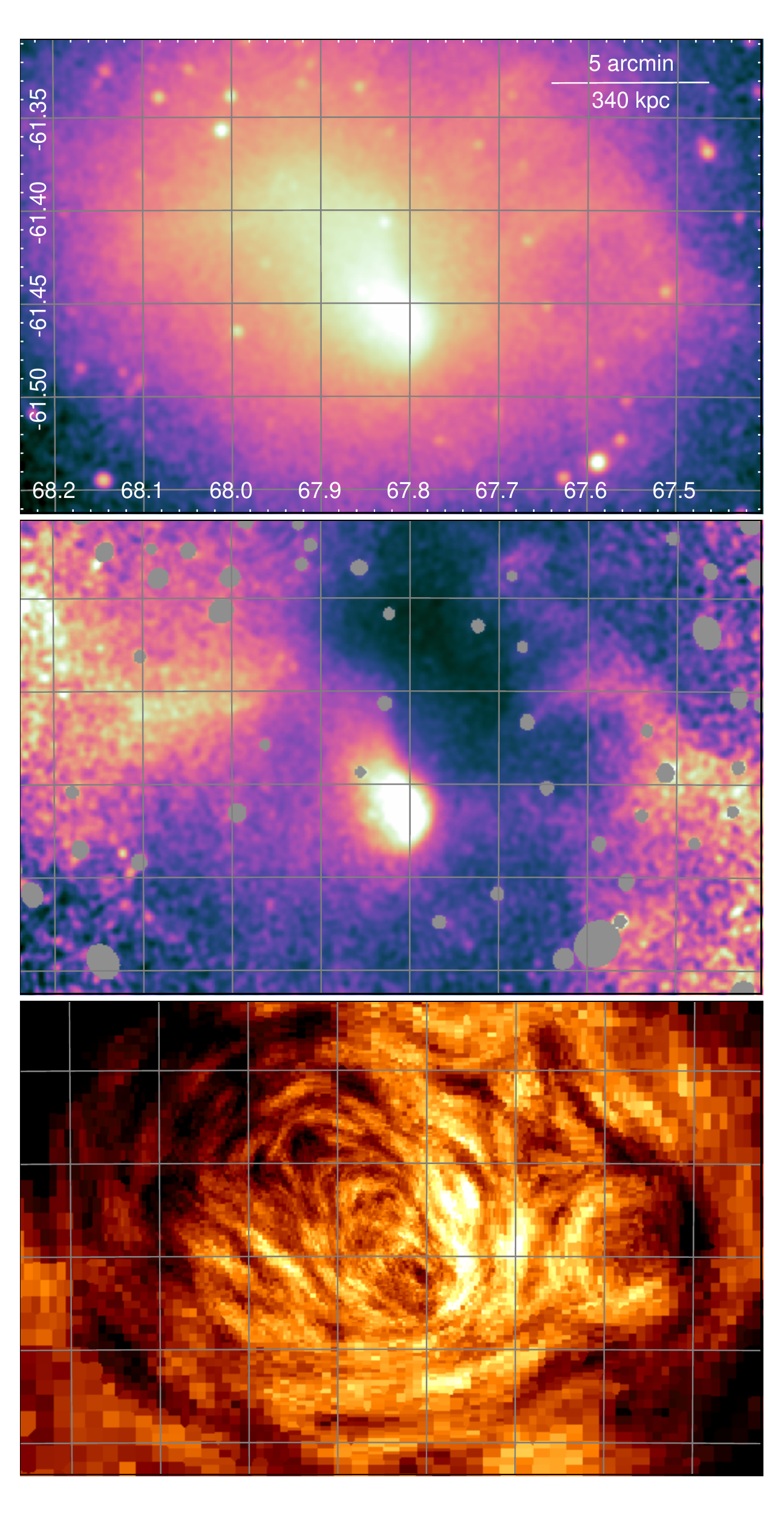}
  \caption{
    Close up of the central region, showing the X-ray image (top), residual map (centre), and temperature map (bottom).
    The eROSITA temperature and residual maps were taken from Fig.~\ref{fig:ellipse_maps}.
  }
  \label{fig:centre}
\end{figure}

\begin{figure}
  \centering
  \includegraphics[width=\columnwidth]{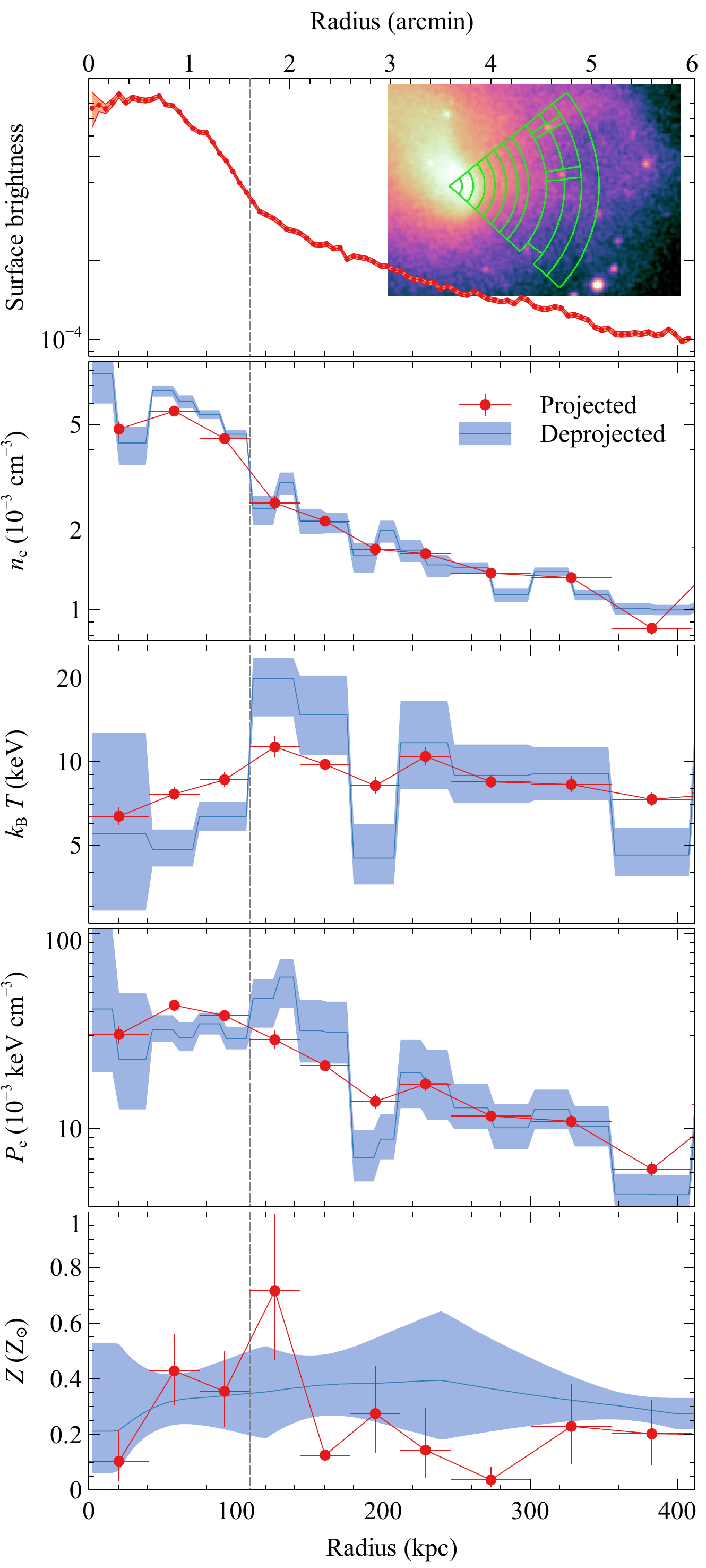}
  \caption{
    Profile across the dense inner core to the hotter region to the west, showing from top to bottom the surface brightness, electron density, temperature, electron pressure, and metallicity.
    Profiles created using projected spectral fits and deprojected using \texttt{MBProj2} are shown.
    The outer spectral bin is not shown due to edge effects.
    The surface brightness is background subtracted and exposure corrected in the 0.3--2.3~keV band using 2 arcsec pixels.
  }
  \label{fig:core_profile}
\end{figure}

Here the core and the hotter region to the west are examined in detail.
Figure~\ref{fig:centre} shows a zoomed views over the core region of the X-ray image, residual map, and temperature map.
As mentioned previously, there is a high-temperature region to the west of the core, although it seems to be broken into two with a cooler region between.
These higher temperature regions approximately follow the surface brightness contours, although there is no apparent surface brightness change associated with the cooler region between the two hotter ones.
The bright regions to the west in the residual map appear cooler, except for the bridge which appears to run directly westwards from the central core.

To help understand what is occurring to the west of the core, we extracted and fitted spectra from regions in a westward sector (Fig.~\ref{fig:core_profile}).
The same spectral modelling as in previously analysis of spectra was used, except here we chose a Jeffreys prior on the temperature parameter (equivalent to a transformation to using log of the temperature) and the metallicity was fixed to be $0.3$~Z$_{\odot}$.
A MCMC based on the best fitting model was run to produce a chain of parameters for each spatial region.
In the temperature profiles we show the median and $1\sigma$ percentiles for the temperature parameter.
To compute the density profile, we took random sets of model normalisations from the chains and deprojected these for each annulus.
This was done by multiplying an inverted matrix with the spherical projections volumes to compute the normalisation per unit area and then the electron density.
The densities plotted show the median and $1\sigma$ percentiles of these density values.
The pressure and entropy profiles were computed by taking each set of densities and projected temperatures and then looking at the resulting distributions.
We note that this analysis is not strictly correct, as projected rather than deprojected temperatures were used here to compute the pressure and entropy.
It is also not strictly correct to deproject model normalisations unless the temperatures are the same.

We also used \texttt{MBProj2} \citep{Sanders18} to make deprojected profiles for comparison.
\texttt{MBProj2} uses a forward modelling approach to create model surface brightness profiles in multiple energy bands, which are then fitted to the data.
For this analysis we extracted surface brightness profiles in different energy bands\footnote{The bands used were 0.30--0.60 (except for TM4), 0.60--0.70, 0.70--0.80, 0.80--0.90, 0.90--1.00, 1.00--1.10, 1.10--1.20, 1.20--1.30, 1.30--1.40, 1.40--1.50, 1.50--1.65, 1.65--1.80, 1.80--2.10, 2.10--2.30, 2.30--2.90, 2.90--3.50, 3.50--4.00, 4.00--4.50, 4.50--5.50, and 5.50--7.00 keV.}
for TMs 1, 2, 3, 4, and 6 with 4 arcsec binning out to 8 arcmin radius.
These profiles were fitted with a model which did not assume hydrostatic equilibrium, but parametrised the density and temperature profiles.
A temperature model with the same radial binning as the projected fits was chosen (although extended to 8 arcmin radius with the same binning).
The density model was binned twice as finely as the temperature model.
The metallicity was parametrised at radii of 20, 120, 240, and 400\,kpc using linear interpolation in log radius to calculate values between and using a flat prior between $0$ and $1$\,Z$_{\odot}$.
PSF corrections were included in this analysis by multiplying the uncorrected projected rates by a matrix in each energy band which was computed from PSF images (with linear interpolation in angle from the centre of the detector and in energy).

The bright central core has an electron density of around $6\times10^{-3}$~cm$^{-3}$ (the innermost drop is likely dependent on the exact centre chosen) and a temperature ranging between 3 and 6 keV.
These values give a mean radiative cooling time of $\sim 11$~Gyr and therefore would not be classified as a cool core.

The projected and deprojected profiles show that the temperature appears to increase by a factor of $\sim 2$ outside the bright central core (the edge of which seen as the vertical dotted line in the figure).
Although the deprojected pressure appears to rise outwards here this is not statistically significant.
The temperature drops down again in both the projected and deprojected profiles in the third bin after the edge, then rises again, as seen in the temperature maps.
There is weak $\sim 2 \sigma$ evidence for a pressure drop in this cooler region.

\subsection{Sector profiles}

\begin{figure}
  \centering
  \includegraphics[width=0.9\columnwidth]{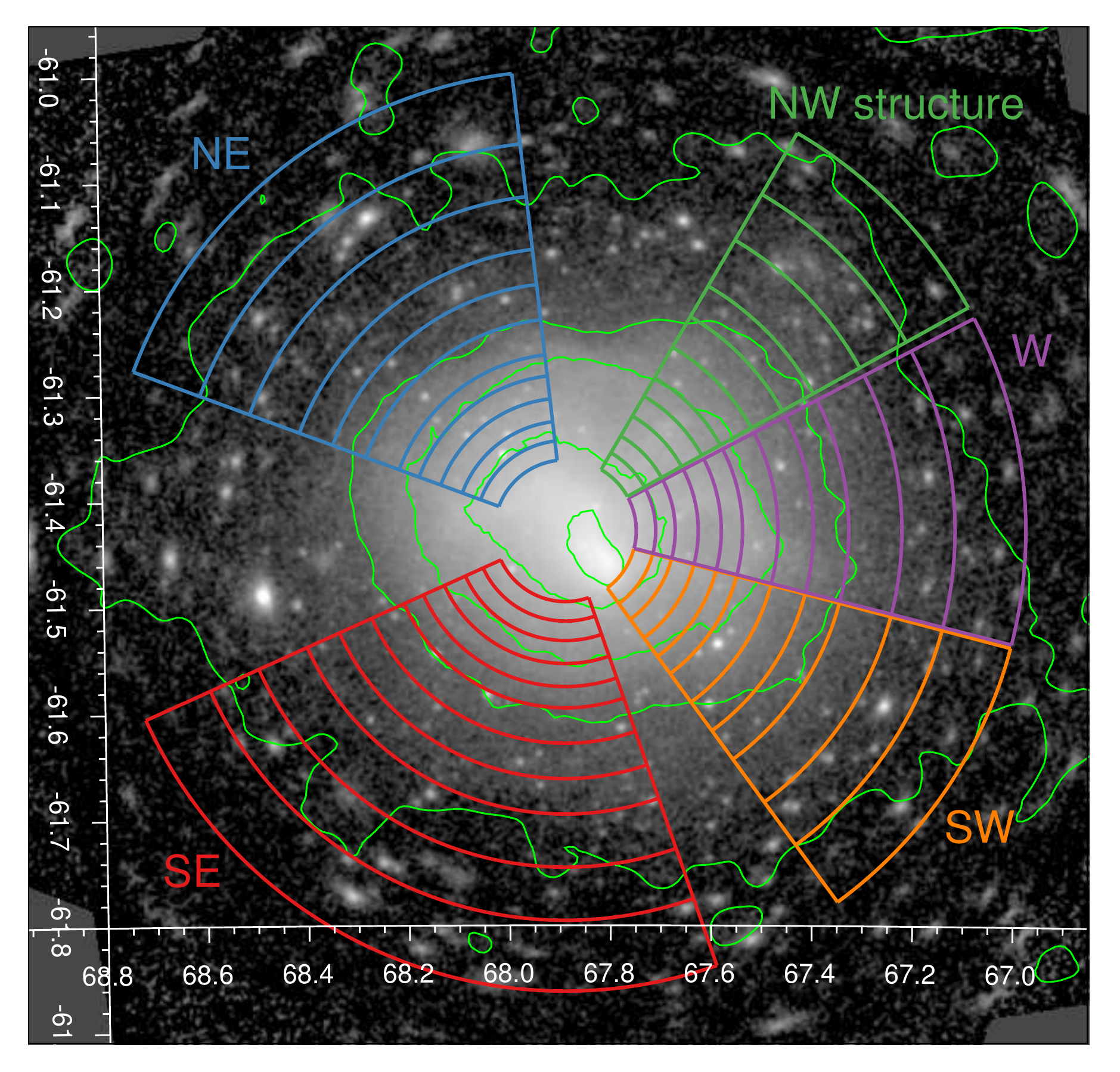}
  \caption{
    Sectors examined in radial profiles.
    The sectors have been chosen to lie in five interesting directions: over the relatively undisturbed SE region, the low entropy material in the NE structure, over the NW structure, and in the direction of the low entropy material to the west (W and SW).
    Surface brightness and spectral profiles are shown in Fig.~\ref{fig:sector}.
  }
  \label{fig:sector_img}
\end{figure}

\begin{figure*}
  \centering
  \includegraphics[width=\textwidth]{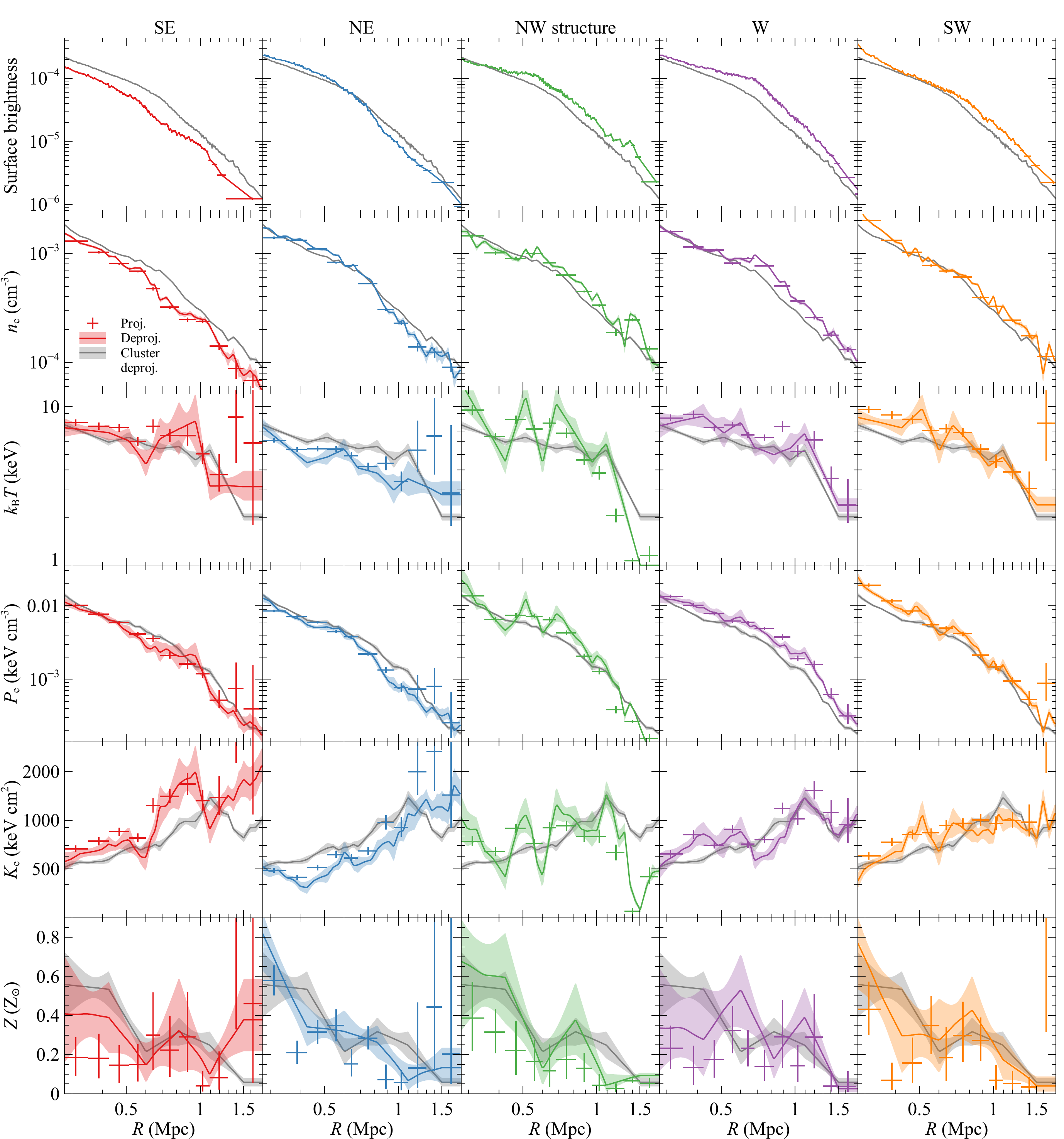}
  \caption{
    Profiles obtained in sectors.
    The columns contain the results for each sector (Fig.~\ref{fig:sector_img}).
    The rows show the surface brightness, electron density, temperature, electron pressure, electron entropy, and metallicity.
    The points were obtained by fitting projected spectra.
    The coloured shaded regions show the deprojected results.
    The grey shaded regions show the deprojected results for the whole cluster.
    Surface brightness profiles were obtained using the same technique as in Fig.~\ref{fig:sbprof}, using the 0.3 to 2.3 keV band and showing the count rate per 2 arcsec pixel.
  }
  \label{fig:sector}
\end{figure*}

To examine the thermodynamics associated with the features discussed previously, we computed spectral and surface brightness profiles in particular sectors.
Figure~\ref{fig:sector_img} shows sectors designed to cross a relatively undisturbed region of the cluster south-east (SE), a region towards the NE merging subcluster (NE), a sector over the NW merging subcluster (NW), and two sectors towards the northern and southern parts of the western structure (W and SW, respectively).
The results are shown in Fig.~\ref{fig:sector}.

Surface brightness profiles were extracted and rebinned as in Fig.~\ref{fig:sbprof}.
Spectral profiles are derived from fits to projected spectra extracted from the sectors as in Section \ref{sect:core}.
Similarly, we compute densities, pressures, and entropies from these projected fits.
We also used \texttt{MBProj2} to make deprojected profiles in each sector.
In these fits we parametrised the log density by subdividing each bin used for the projected density profile into three and parametrising it at the edges of these sub-bins.
The log temperature was parametrised at the projected bin edges, although each pair of bins was combined below 423 and above 1093 kpc to reduce statistical uncertainties.
We used a flat prior on the log temperature between 1 and 30 keV.
Linear interpolation in log radial space was to calculate the log density and log temperature values between these points.
The metallicity was parametrised at the edges of each pair of bins, using linear interpolation in log radius to calculate values between each pair of points and a flat prior on the parameters between 0 and $1$\,Z${_\odot}$.
PSF correction was not included in this analysis.

Examining these profiles we see that the cluster is systematically under-luminous and under-dense in the SE direction, but over-luminous and over-dense in the NW, W and SW directions.
There are also multiple breaks in the surface brightness profiles, similar to what is seen in the profiles with uniform sectors (Fig.~\ref{fig:sbprof}).
The inner edge is particularly visible.

In the SE direction moving outwards, there is a break in density at a radius of 560~kpc (the inner edge), where the density steepens until a radius of 700\,kpc, where it flattens out again.
At a radius of 1000\,kpc (the outer edge) the density profile steepens once more.
There is some evidence for a deprojected hotter region between radii of 600 and 1000~kpc.
The temperature and pressure appear to drop steeply at this outer edge.
The metallicity is enhanced in the core and in the hotter region.

Looking outwards towards the NE, around the inner edge (570\,kpc) there is a drop in density and the steepness of the profile increases.
This continues until a radius of around 900\,kpc, where the profile flattens once more.
Close to the inner edge there is also evidence for an increase in temperature and a flattening of the pressure profile.
Beyond this radius, the temperature and pressure profiles are fairly smooth.
The metallicity profile along this direction smoothly declines from the core of the cluster.

The profile towards the NW sub-structure is less certain due to this sub-structure.
There are peaks in deprojected temperature, pressure, and density at radii of 500 and 700 kpc.
They persist if the outer NW structure itself is masked out and the deprojection repeated.
However, if the temperature interpolation is done using two-times finer binning over this region, then the results show twice the number of temperature peaks.
Therefore this region of the cluster is likely prone to instabilities in the deprojection \citep{Russell08}, common in such `inverse problems' and likely because the model is not describing the data well enough (e.g. due to multiphase material).
When this happens, the temperature oscillates on the scale of the bin or interpolation scale.
At the location of the NW structure we see the temperature drops drastically.
Separate spectral fitting of this structure shows multiphase material (Section \ref{sect:ext}).
We therefore caution using the temperatures, pressures and entropies in this direction, although the densities are less affected by instabilities and temperature variation.

Westwards is seen a break in the density profile at around 670\,kpc radius, where the profile steepens, then continuing with a similar slope until the edge of the profile.
However, there is an edge in the pressure profile at a radius of 1100\,kpc, where is also an increase in the temperature and entropy profiles.

Towards the SW there is a more dense region towards the centre and central core.
This breaks to a flatter profile around a radius of 450\,kpc and then a steepens at 840\,kpc radius.
There are some peaks in the temperature profile, but these look similar to the instabilities seen along the direction of the NW structure.
The metallicity profile is peaked towards the centre, with a secondary peak at a radius of around 800\,kpc.

\subsection{Hydrostatic profiles}

\begin{figure}
  \centering
  \includegraphics[width=\columnwidth]{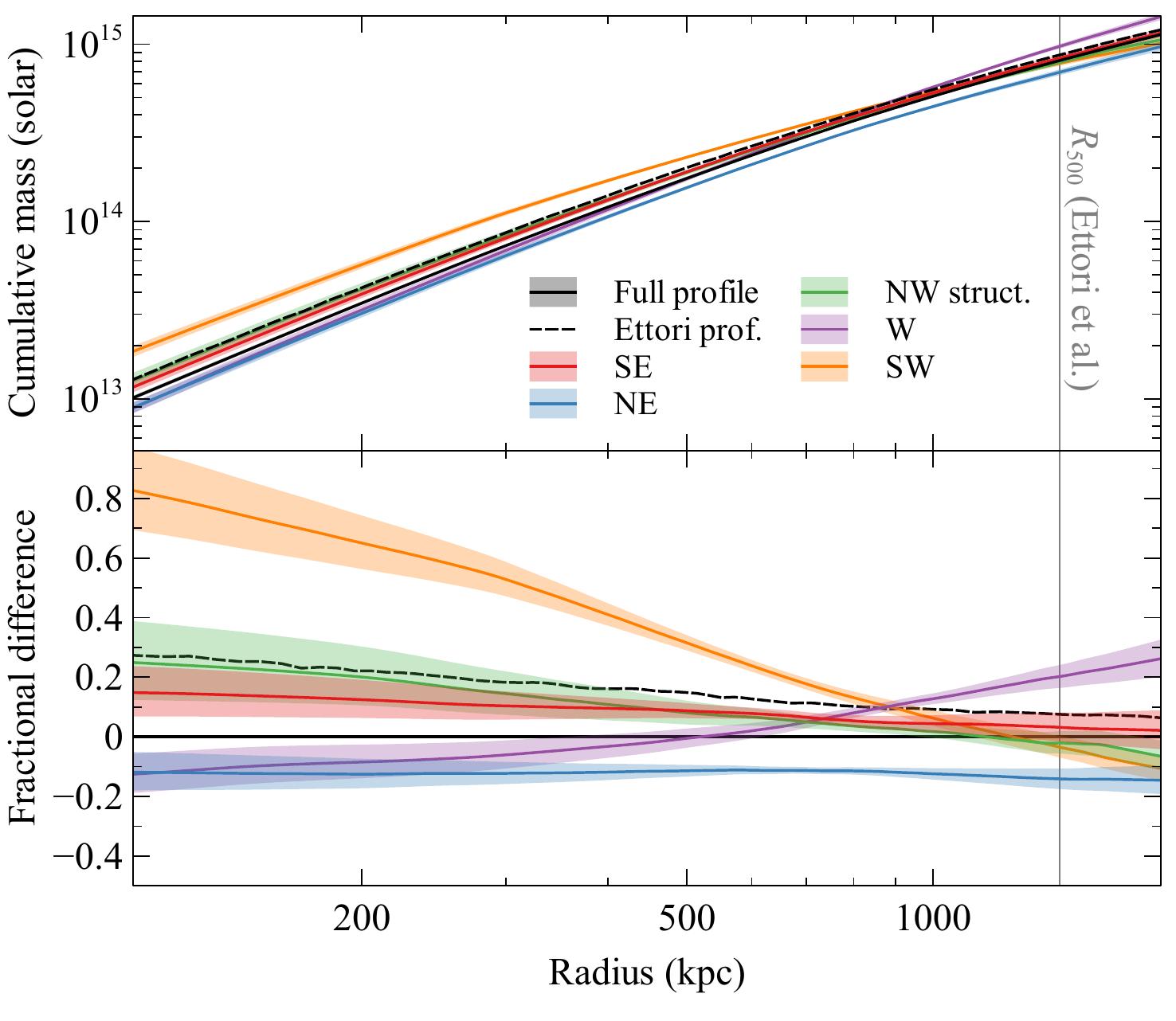}
  \includegraphics[width=\columnwidth]{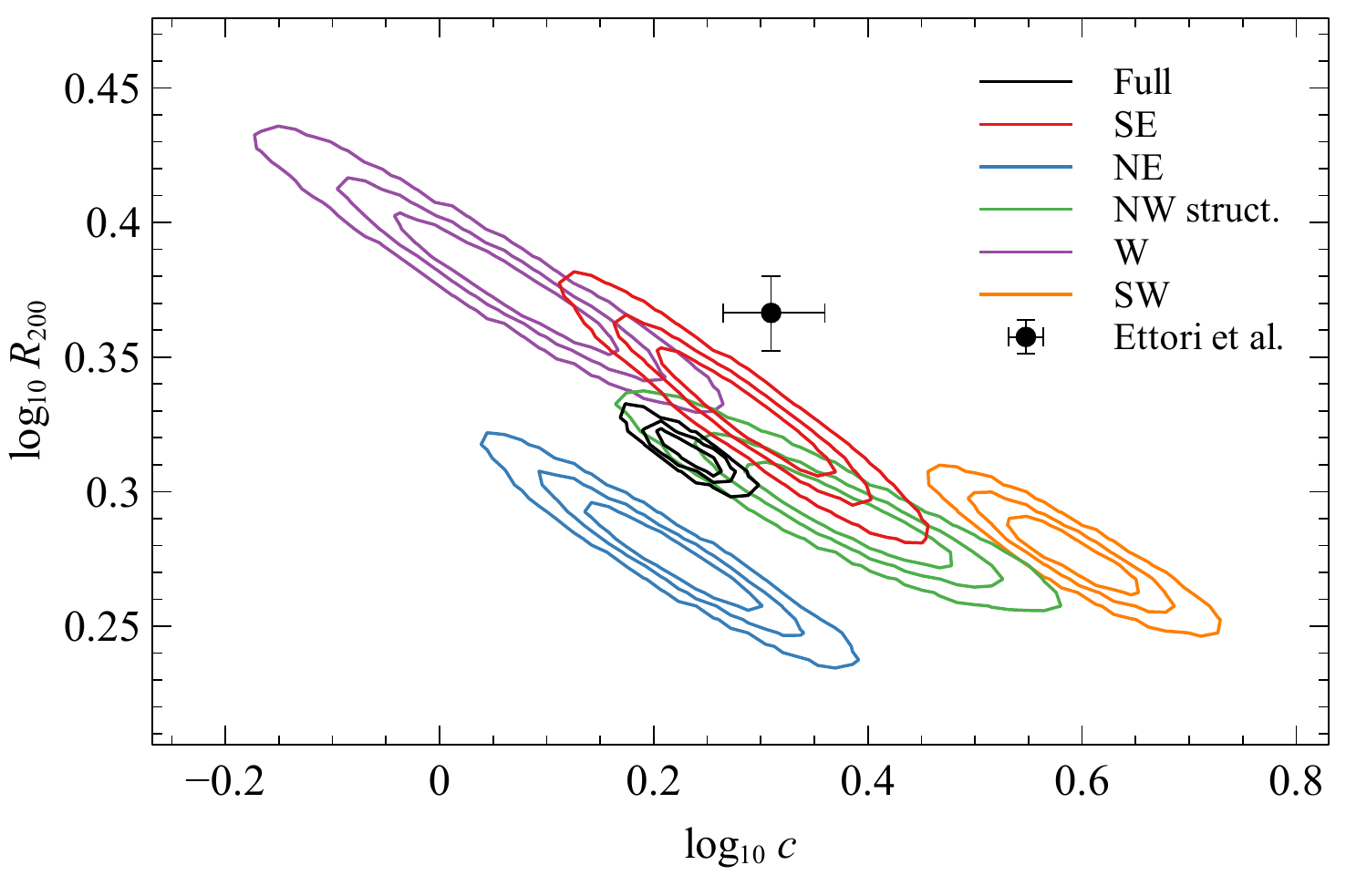}
  \caption{
  Obtained hydrostatic mass profiles and parameters.
  The top panel shows the mass profiles in different sectors and for the full cluster, compared to the results from \cite{Ettori19}.
  Below that is shown the fractional difference between each profile and the full profile.
  In the bottom panel, the posterior probability contours for the concentration $\log_{10} c$ and radius $\log_{10} R_{200}$, are shown, with levels containing 68, 90, and 99\% of the values.
  The \cite{Ettori19} point is plotted assuming radius and concentration are independent.
  }
  \label{fig:massprof}
\end{figure}

\texttt{MBProj2} was also used to obtain hydrostatic masses for the cluster.
A parametric mass model and gas density model were used to compute a temperature profile under the assumption of hydrostatic equilibrium, which is then used with the gas density model to predict the surface brightness profile in each energy band.
We extracted surface brightness, exposure, and background profiles in the same bands listed in Section~\ref{sect:core}.
These profiles were fitted assuming an NFW mass model, parametrised by $R_{200}$ and the concentration $c$, with flat priors in log space.
No additional mass was added to the NFW model from the ICM.
The outer pressure of the cluster was also parametrised by a log value with flat priors.
The ICM density model was sampled at radii three times more finely than the bins in Fig.~\ref{fig:sector_img}, assuming flat priors in log space, where spline interpolation in log space was used to compute the density at intermediate radii.
The metallicity was parametrised similarly to before by interpolating in log radius parametrised values at radii of 20, 200, 400, 650, 980, and 1500\,kpc using a flat prior between 0 and 1\,Z$_\odot$.
The model was fitted to the full cluster region (from the centre to 28 arcmin radius) and also to the angular ranges of the previous sectors (using the same radial range as for the full cluster).
This analysis did not include broadening due to the PSF.
MCMC was used to produce chains of model parameters given the data.

Figure \ref{fig:massprof} shows the resulting mass profiles in each sector and for the full cluster and the fractional difference between each profile and the median full profile.
The results of \cite{Ettori19} using \emph{XMM-Newton} data are also compared.
We also plot the contours in mass-concentration space of the NFW model from the MCMC.
The cumulative mass of our model (for the full profile) matches their cumulative mass to within 7\% at their value of $R_{500}$.
Taking our mass profile and using the critical density at the redshift of the cluster, we obtain $R_{500}=1364 \pm 10$~kpc and $M_{500}=(7.64\pm0.17)\times 10^{14}$~M$_\odot$, which is 16\% smaller (or $2\sigma$) smaller than the \cite{Ettori19} mass of $(8.80 \pm 0.57)\times 10^{14}$~M$_\odot$.
This discrepancy may be due to the lack of modelling of the PSF in this analysis, residual calibration differences or different relative sensitivity to different temperature components within the cluster.

Despite these total masses being in reasonable agreement, there are some differences between our different sectors, particularly in the concentration parameter.
The SW profile is most discrepant, seen in both the mass profile and parameters, with very high concentrations.
The cause of this may be that the SW sector contains most of the bright central region which has a high pressure, causing the concentration to increase.
At large radii the SW profile diverges to lower mass than the full profile.
The NW structure and W profiles are very similar and diverge to higher masses than the cluster whole.
Towards the NE the mass is low over most of the radial range compared to the other sectors and the concentration is also low, although it becomes in agreement with the main profile at $R_{500}$.
The discrepancies between the mass profiles in each sector imply that the dynamical structure in this cluster prevents reliable hydrostatic mass measurement.
However, it may be possible to mask out the sub-structures in the centre and the outskirts to improve the agreement between the different sectors.

\subsection{Optical images}
\label{sect:optical}
Our studies are complemented by high quality optical data in the five Sloan filters
$u$, $g$, $r$, $i$, and $z$. A\,3266 was observed with the Dark Energy Camera \citep[DECam; see][]{Flaugher_2015} mounted on the Blanco 4-metre Telescope at CTIO. The data were obtained under the Program-ID 2014B-0265 (P.I. Dell'Antonio). We retrieved all images in raw format from the NOAO Archive\footnote{\url{http://archive1.dm.noao.edu/}} and processed them with the THELI image-reduction pipeline \citep[see][]{Erben_05,Schirmer_13}. Our extensions and refinements of THELI to process DECam data are described in \citet{Reiprich21} and our processing was done in exactly the same way as outlined in that article.

The final co-added images consist of high quality, sub-arcsecond data (except for the $u$-band) obtained under photometric conditions.
In Table~\ref{tab:DECam_data} we list total exposure time, image seeing (Gaussian fit to point-like sources) and the limiting magnitude of our co-added data.
The limiting magnitude is defined as the 5$\sigma$ detection limit in a $2.0$~arcsec aperture via $m_{\rm lim}=ZP-2.5\log(5\sqrt{N_{\rm pix}}\sigma_{\rm sky})$, where $ZP$ is the magnitude zeropoint, $N_{\rm pix}$ is the number of pixels in a circle with radius 2.0~arcsec, and $\sigma_{\rm sky}$ is the sky
background noise variation.
\begin{table}
  \caption{Average quality parameters of our final DECam science tiles.}           
  \label{tab:DECam_data}      
  \centering          
  \begin{tabular}{ccccc}    
    \hline\hline 
    Filter  & exp. time (s)  &
      seeing (arcsec) & $m_{\rm lim}$ (AB mag) \\ 
    \hline
    $u$ & 5000 & 1.16 & 24.48 & \\ 
    $g$ & 1800 & 0.88 & 24.70 & \\ 
    $r$ & 4500 & 0.85 & 24.74 & \\ 
    $i$ & 2100 & 0.84 & 23.80 & \\ 
    $z$ & 2880 & 0.77 & 23.21 & \\ 
    \hline             
  \end{tabular}
  \tablefoot{
  Please see the text for a description of the columns.
  }
\end{table}
\begin{figure*}
  \centering
  \includegraphics[width=\textwidth]{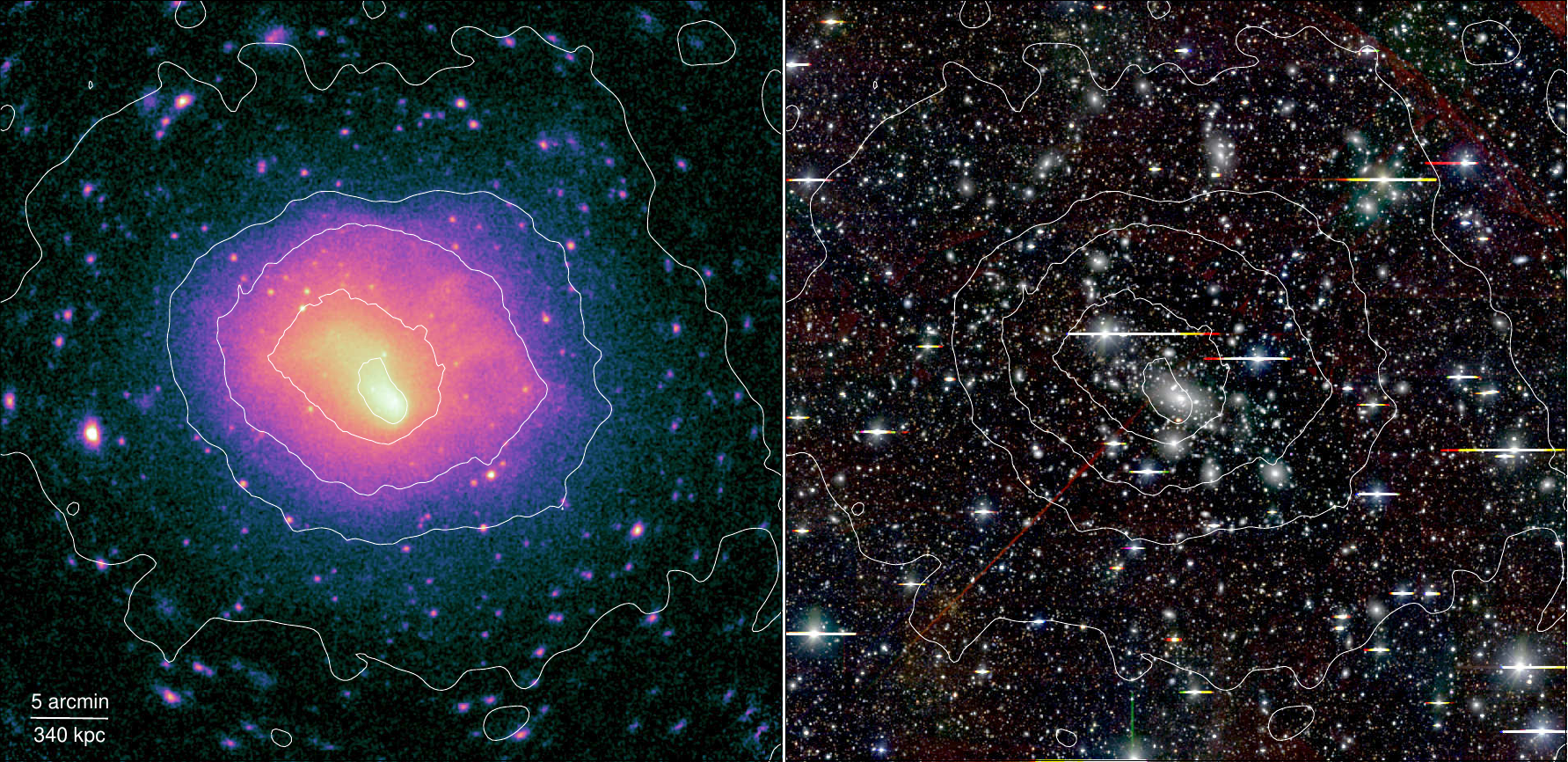}
  \caption{
    Comparison of X-ray and optical image of cluster centre.
    X-ray image is between 0.3 and 2.3 keV and has 2 arcsec pixels, smoothed by a Gaussian with $\sigma=2$ pixels.
    Optical is an RGB composite using $i$, $r$ and $g$ bands, as red, green and blue, respectively.
    The X-ray contours are those in Fig.~\ref{fig:img_resid}.
  }
  \label{fig:optical_compar}
\end{figure*}

Figure~\ref{fig:optical_compar} compares the X-ray emission in the centre of the cluster with an RGB image of the DECam data.
The main central cD cluster galaxy appears located within the bright core region.
Several of the other brighter galaxies appear to be offset towards the SW.
A group of galaxies associated with the extended source X8 can be seen to the north-west of this image.

\subsection{Extended sources and associations}
\label{sect:ext}
As stated in Section \ref{sect:src_detn}, a number of extended objects were detected within the observed field.
The multi-component matched filter \citep[MCMF;][]{Klein18,Klein19} cluster confirmation tool was used to confirm the extended X-ray sources in the field as clusters and to determine their photometric redshift. MCMF utilises a cluster red sequence technique \citep{gladders00} to identify overdensities of red galaxies in redshift space. The found overdensities around cluster candidates are then compared to MCMF runs on random line of sights.
MCMF derives the estimator $f_\mathrm{cont}$, which is correlated with the probability of being a chance association. In context of eROSITA selected clusters, candidates with $f_\mathrm{cont}>0.3$ are typically considered as unconfirmed, $0.2<f_\mathrm{cont}<0.3$ as weakly confirmed, and $f_\mathrm{cont}<0.2$ as confirmed.
MCMF was run using the photometric catalogues from the legacy survey DR8 \citep[LS;][]{legacysurveys}, which itself contains data from Dark Energy Survey \citep[DES;][]{Abbott21} at the location of this field. For more details on MCMF using legacy survey and eROSITA sources, we refer the interested reader to \cite{Klein21}.
For the correct calculation of $f_\mathrm{cont}$ as function of cluster redshift and richness a larger number of cluster candidates are needed than those eleven sources found in this field. We therefore used the results in \cite{Klein21} to assign $f_\mathrm{cont}$ to the candidates.

\begin{table*}
  \caption{
    Extended sources in field, MCMF associations, and derived quantities.
  }
  \scriptsize
  \centering
\begin{tabular}{ccccccccccccc}
\hline\hline
ID & Name & $\alpha$ & $\delta$ & $z_\mathrm{phot}$ & $f_\mathrm{cont}$ & $k_\mathrm{B}T$ & $f_\mathrm{X}$ & $L_\mathrm{X}$ & $M_{500,LX}$ & $M_{500,kT}$ \\ \hline
X1 & SRGE\,J043217.5-610359 & $68.0729$ & $-61.0666$ & $0.364$ & $0.00$ & $2.7^{+0.3}_{-0.2}$ & $-13.17 \pm 0.01$ & $43.49 \pm 0.01$ & $14.2 \pm 0.1$ & $14.3 \pm 0.2$ \\
X2 & SRGE\,J043324.0-610801 & $68.3498$ & $-61.1336$ & $0.745$ & $0.00$ & $4.1^{+1.0}_{-0.7}$ & $-13.41 \pm 0.02$ & $43.95^{+0.02}_{-0.03}$ & $14.4 \pm 0.1$ & $14.5 \pm 0.2$ \\
X3 & SRGE\,J043208.6-615106 & $68.0358$ & $-61.8518$ & $1.05$ & $0.07$ & $3.9^{+1.1}_{-0.7}$ & $-13.44 \pm 0.02$ & $44.28^{+0.03}_{-0.04}$ & $14.5 \pm 0.1$ & $14.5 \pm 0.3$ \\
X4 & SRGE\,J043317.2-614035 & $68.3217$ & $-61.6765$ & $0.532$ & $0.00$ & $3.2^{+1.8}_{-0.7}$ & $-13.58 \pm 0.03$ & $43.46^{+0.03}_{-0.05}$ & $14.2 \pm 0.1$ & $14.4 \pm 0.3$ \\
X5 & SRGE\,J043327.9-612722 & $68.3664$ & $-61.4563$ & $0.717$ & $0.00$ & $3.7^{+1.5}_{-0.8}$ & $-13.73 \pm 0.03$ & $43.61^{+0.03}_{-0.05}$ & $14.2 \pm 0.1$ & $14.5 \pm 0.3$ \\
X6 & SRGE\,J043224.4-614519 & $68.1016$ & $-61.7553$ & $0.907$ & $0.06$ & $3.6^{+1.1}_{-0.7}$ & $-13.75 \pm 0.03$ & $43.84^{+0.03}_{-0.04}$ & $14.3 \pm 0.1$ & $14.4 \pm 0.3$ \\
X7 & SRGE\,J042900.3-610243 & $67.2512$ & $-61.0455$ & $0.0458$ & $0.35$ & $22^{+22}_{-12}$ & $-13.69^{+0.03}_{-0.04}$ & $41.03^{+0.03}_{-0.04}$ & $13.0 \pm 0.1$ & - \\
X8 & SRGE\,J042943.9-611038 & $67.4330$ & $-61.1773$ & $0.0487$ & $0.14$ & $0.78^{+0.03}_{-0.04}$ & $-13.20^{+0.05}_{-0.03}$ & $41.60^{+0.05}_{-0.03}$ & $13.3 \pm 0.1$ & $13.7 \pm 0.2$ \\
X9 & SRGE\,J042746.8-611858 & $66.9451$ & $-61.3162$ & $0.457$ & $0.28$ & $2.5^{+1.3}_{-0.6}$ & $-13.88^{+0.03}_{-0.04}$ & $43.04^{+0.03}_{-0.06}$ & $14.0 \pm 0.1$ & $14.3 \pm 0.3$ \\
X10 & SRGE\,J042734.6-611613 & $66.8941$ & $-61.2705$ & $0.338$ & $0.53$ & $5^{+14}_{-3}$ & $-14.03^{+0.05}_{-0.07}$ & $42.55^{+0.03}_{-0.09}$ & $13.7 \pm 0.1$ & - \\
X11 & SRGE\,J043432.5-612153 & $68.6353$ & $-61.3649$ & $0.672$ & $0.03$ & $10^{+24}_{-7}$ & $-14.31^{+0.08}_{-0.10}$ & $42.93^{+0.07}_{-0.14}$ & $13.9 \pm 0.2$ & - \\
\hline
\end{tabular}
\tablefoot{
    Shown is the short ID used in this paper, the \emph{SRG} source name, the J2000 position of the source in decimal degrees $(\alpha, \delta)$, the photometric redshift of the most likely MCMF identification ($z_\mathrm{phot}$), the MCMF probability of contamination of the most likely identification ($f_\mathrm{cont}$), the best fitting X-ray temperature ($k_\mathrm{B}T$; keV), the 0.5--2.0 keV X-ray flux ($f_\mathrm{X}$; $\log_{10}$~erg~cm$^{-2}$~s$^{-1}$), the 0.5--2.0 keV rest X-ray luminosity ($L_\mathrm{X}$; $\log_{10}$~erg~s$^{-1}$), and $M_{500}$ cluster masses assuming luminosity-mass ($M_{500,LX}$; $\log_{10} M_\odot$) and temperature-mass ($M_{500,kT}$; $\log_{10} M_\odot$) scaling relations. X8 was fitted by a two \texttt{apec} components, of which we only show the properties of the coolest one here.
}
\label{tab:identifications}
\end{table*}

\begin{figure*}
  \centering
  \includegraphics[width=0.325\textwidth]{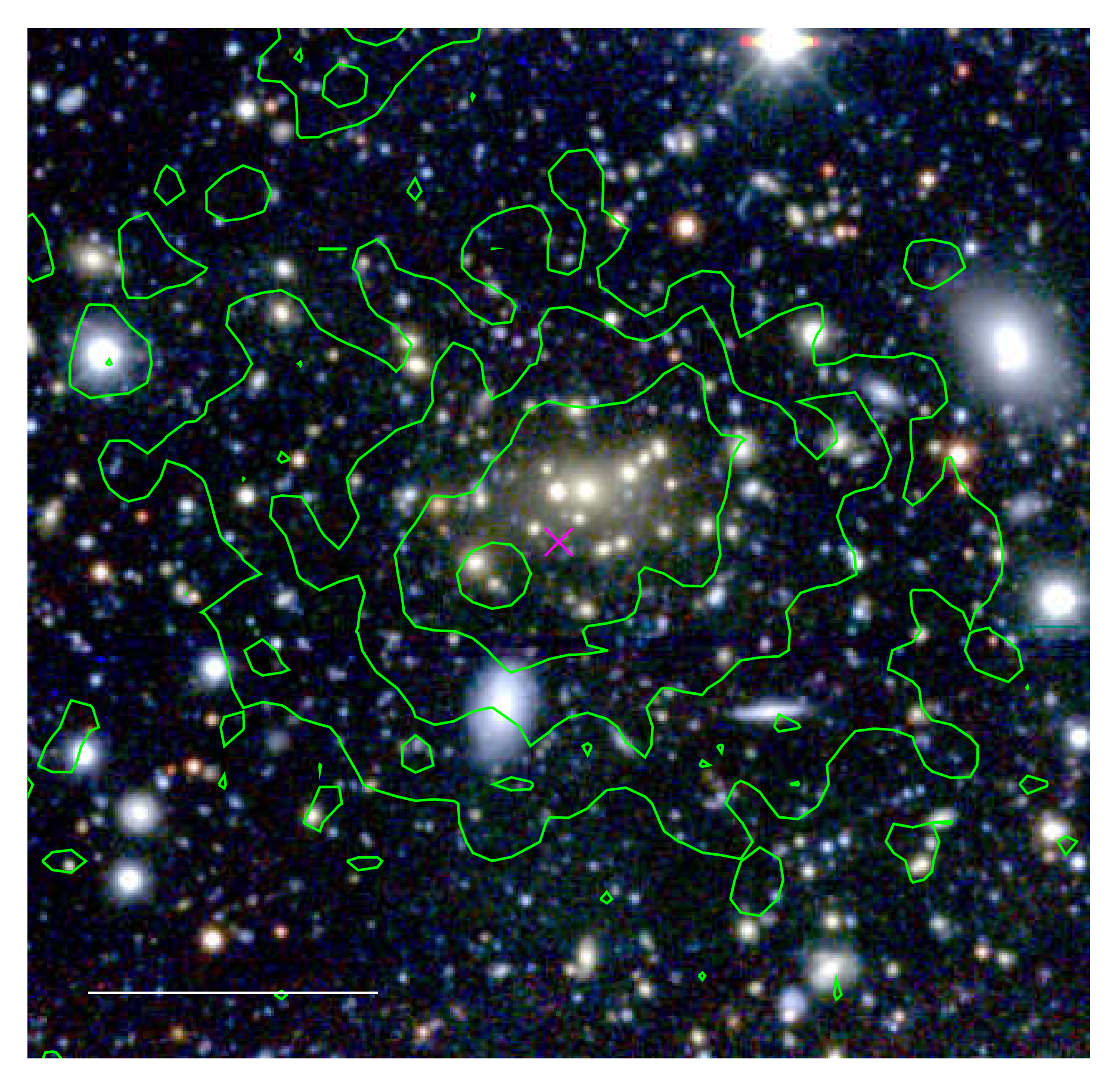}
  \includegraphics[width=0.325\textwidth]{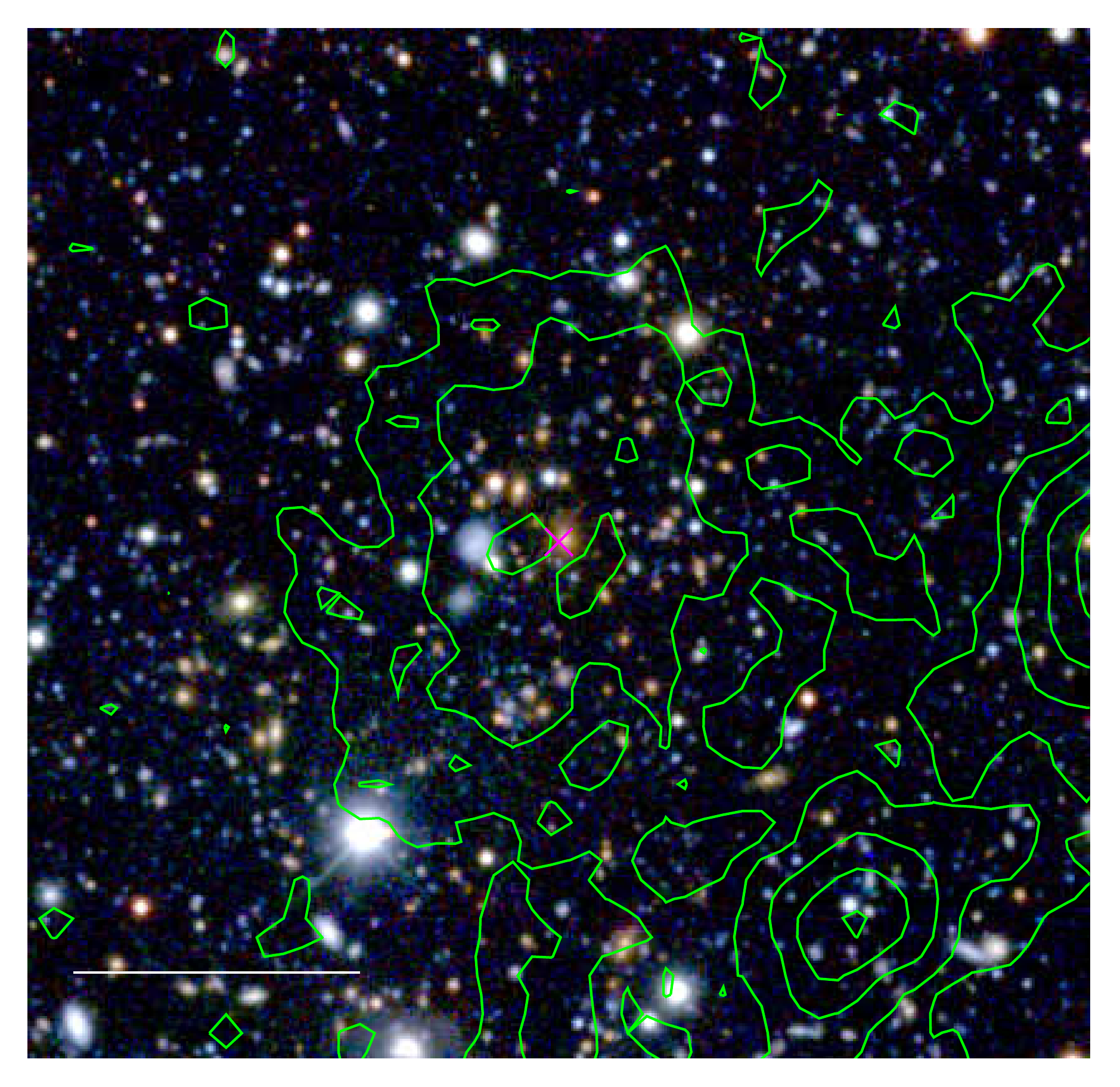}
  \includegraphics[width=0.325\textwidth]{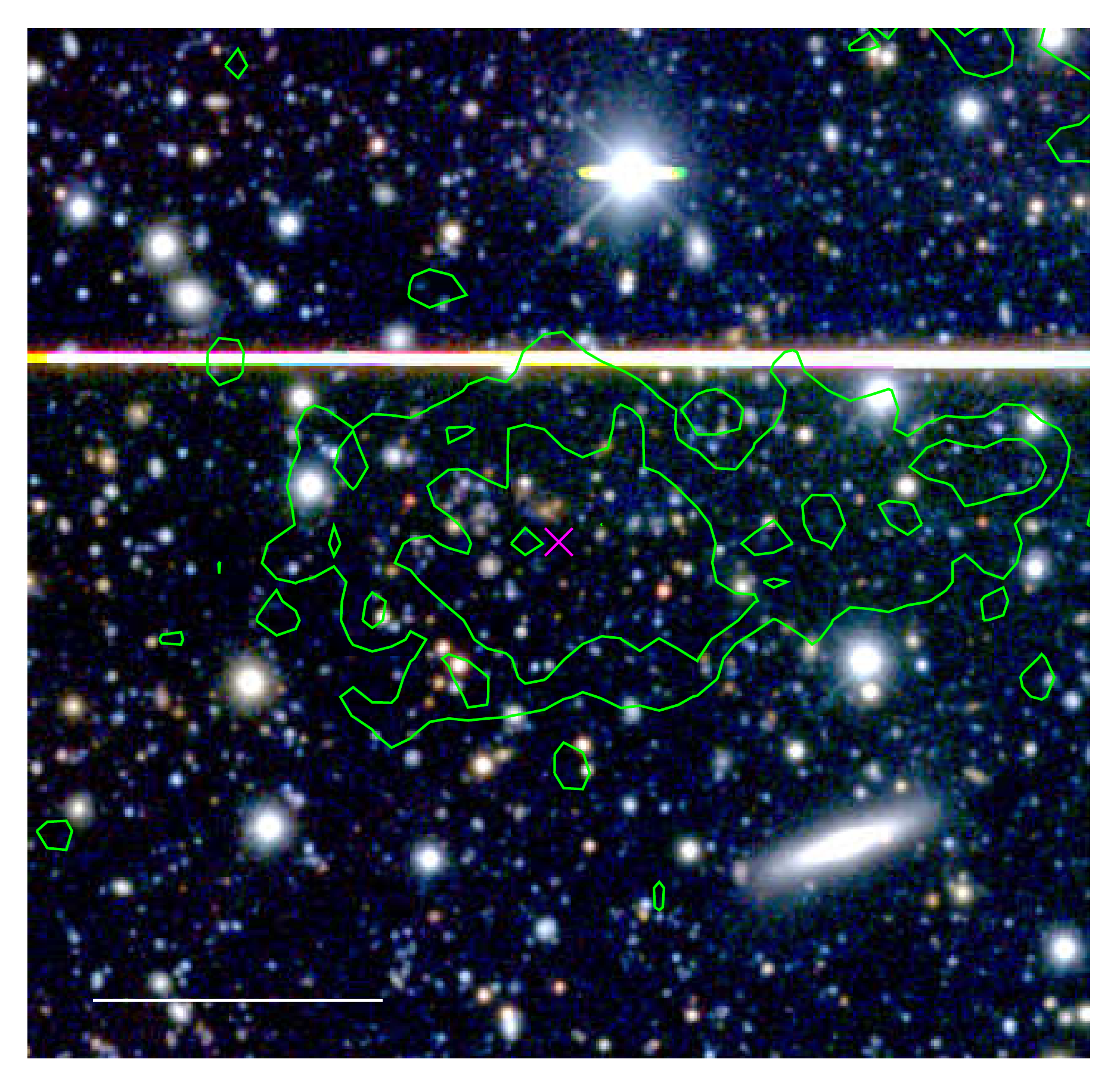}\\
  \includegraphics[width=0.325\textwidth]{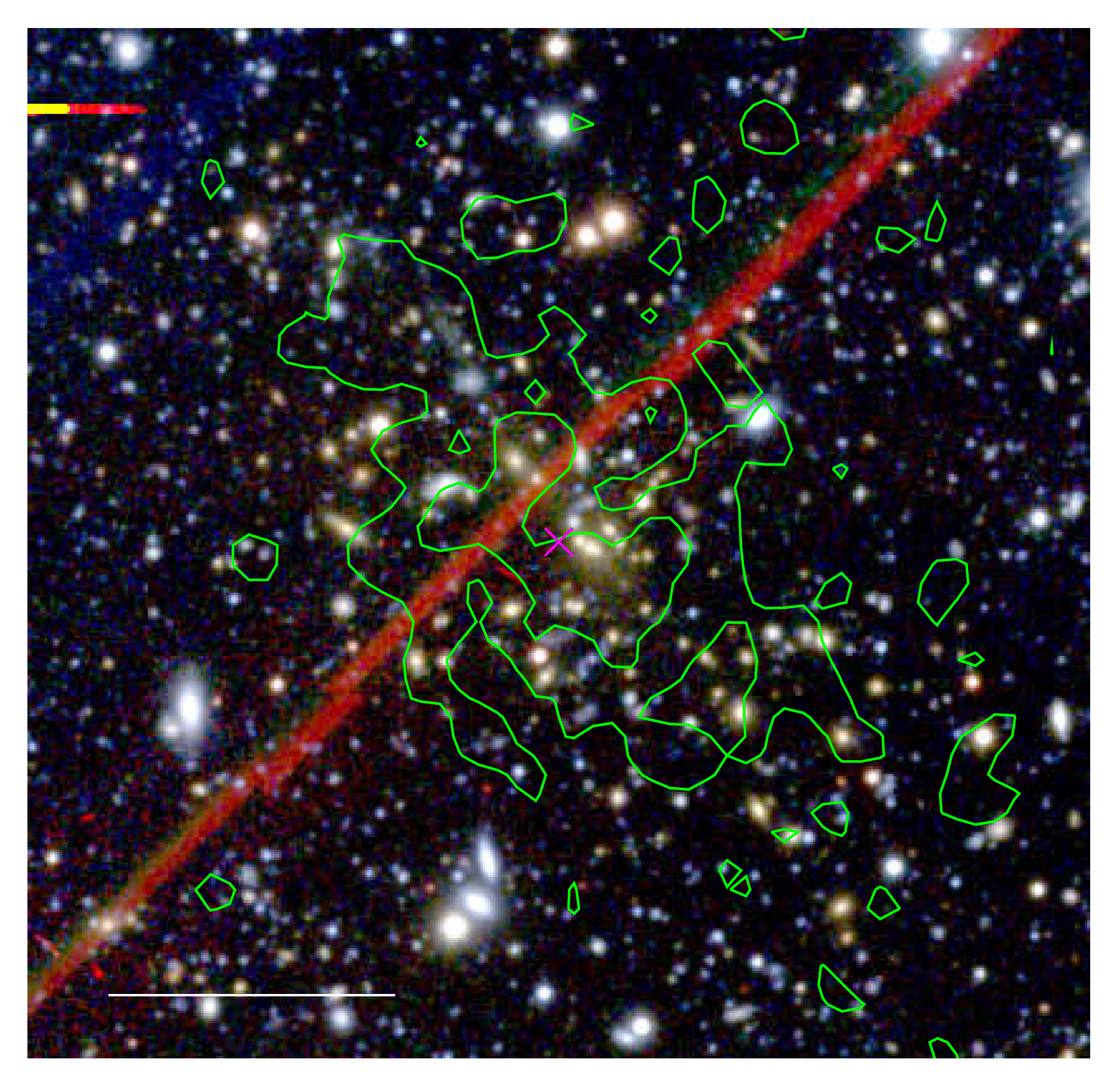}
  \includegraphics[width=0.325\textwidth]{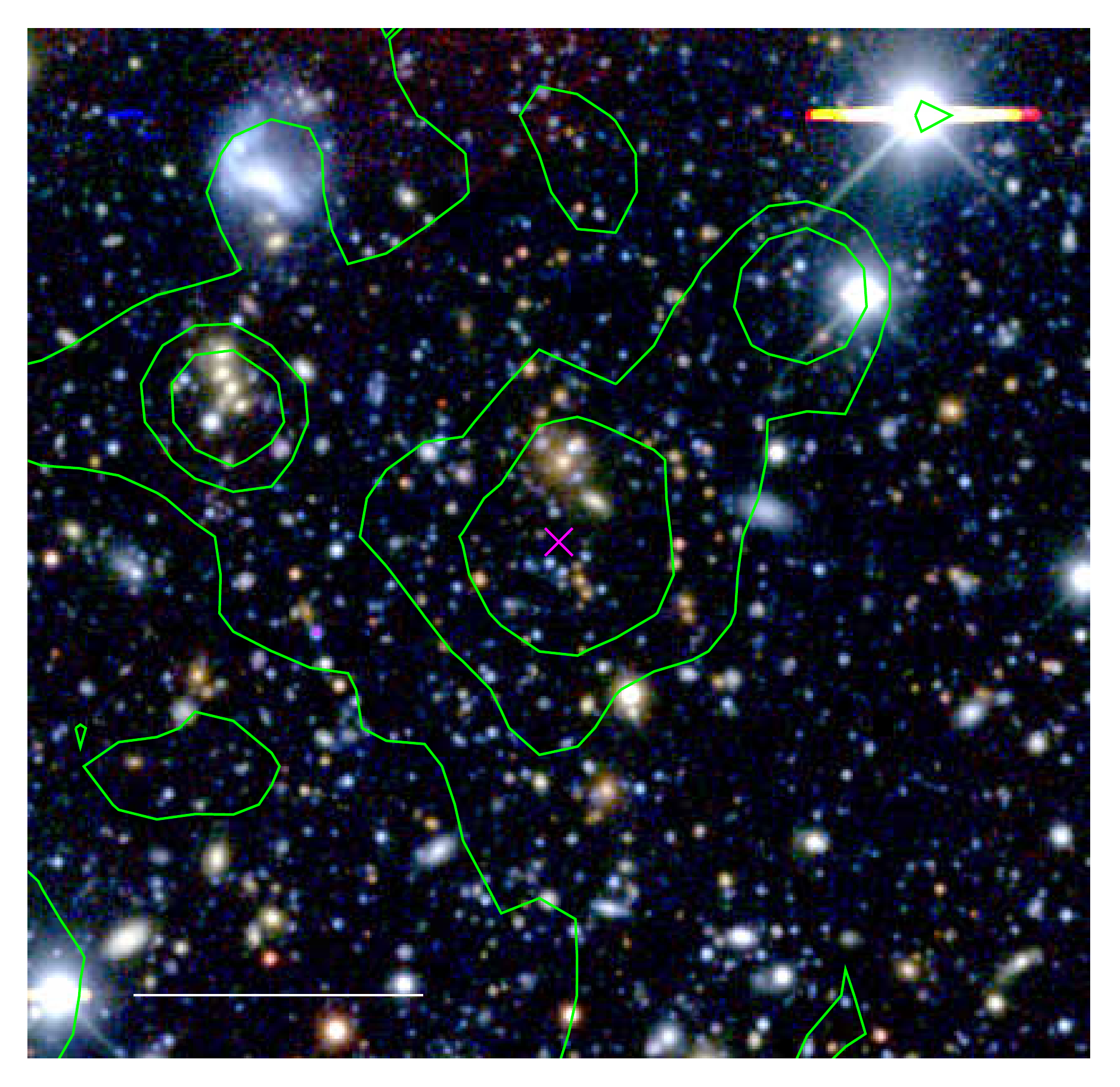}
  \includegraphics[width=0.325\textwidth]{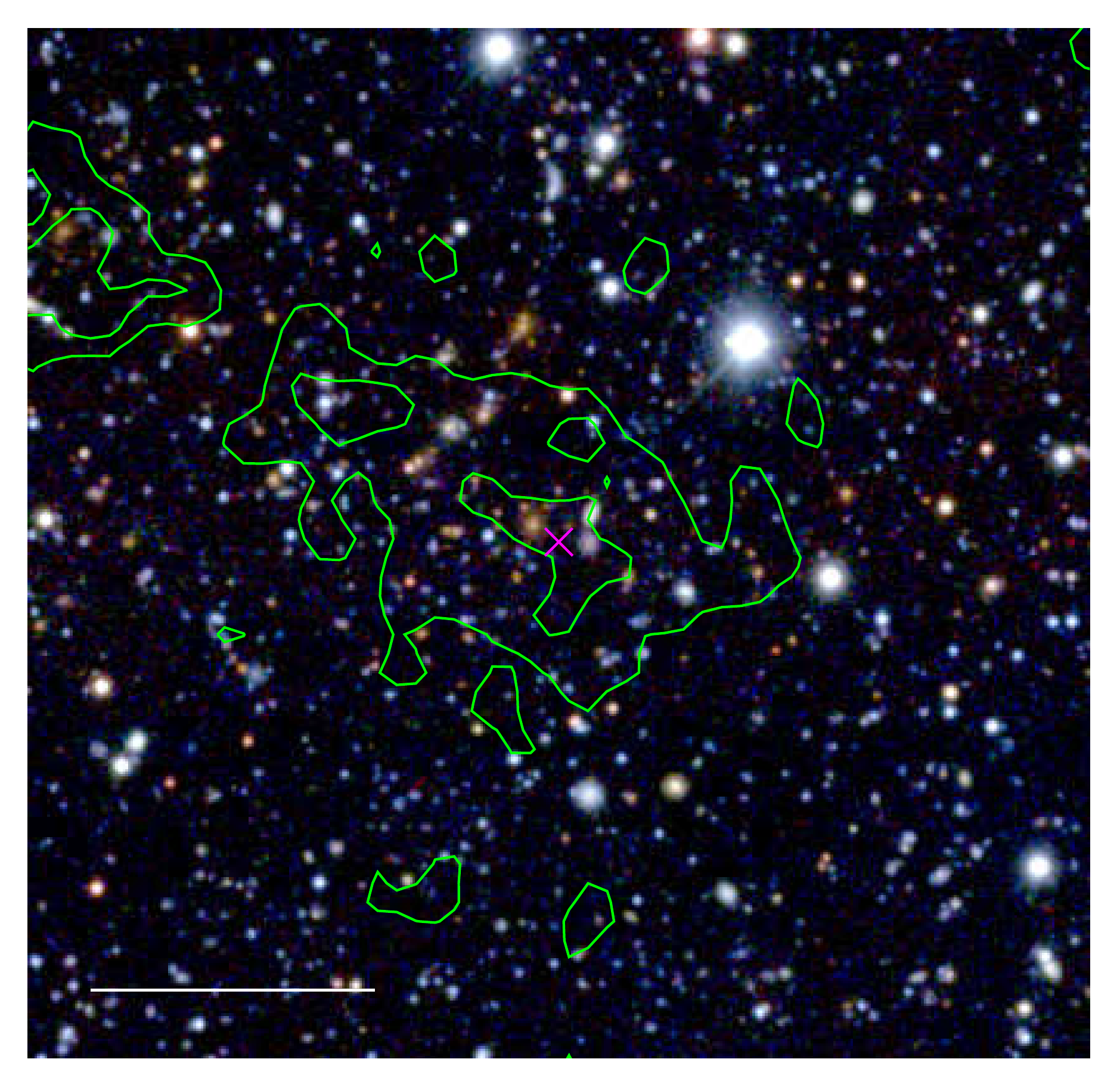}\\
  \includegraphics[width=0.325\textwidth]{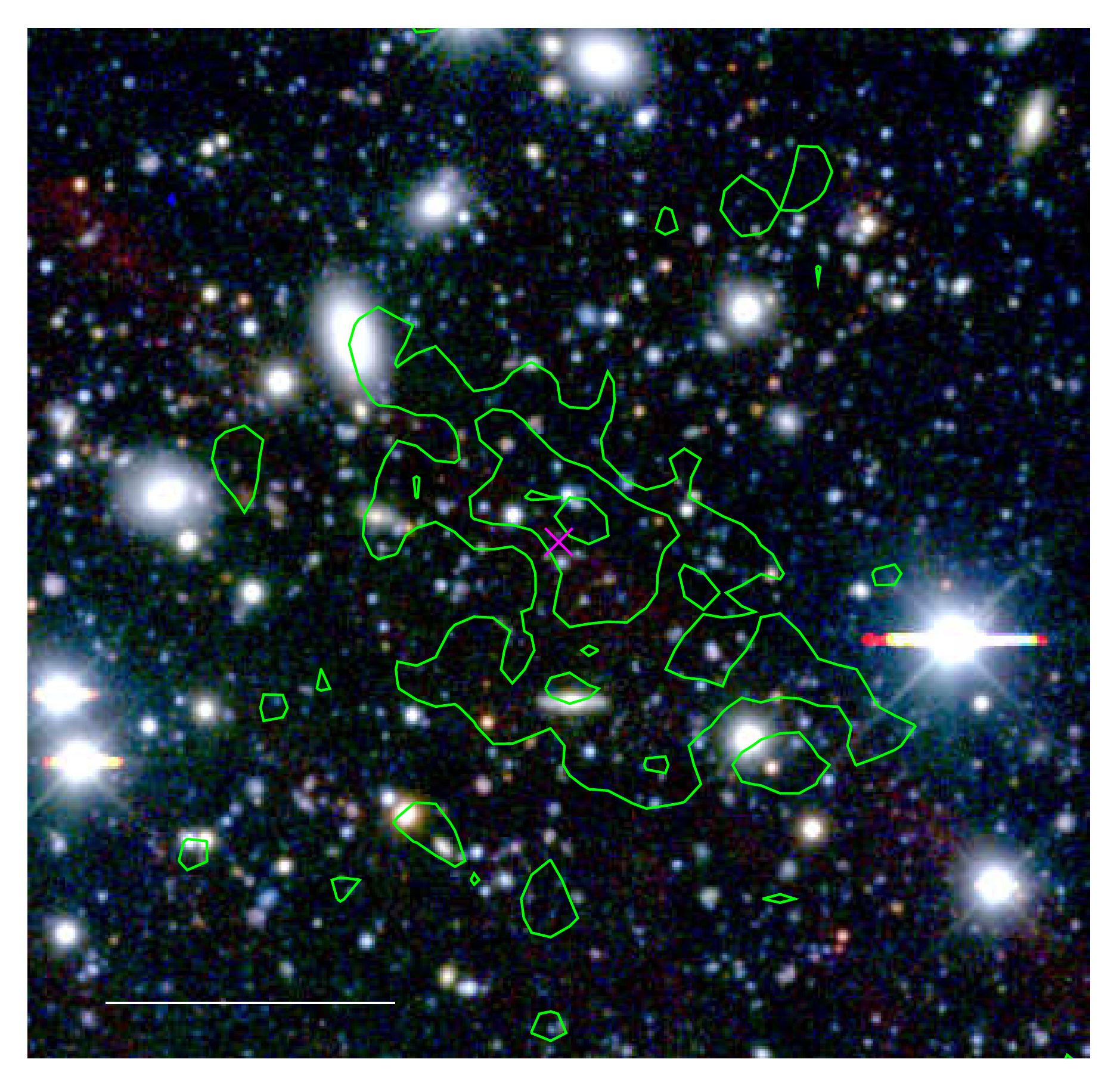}
  \includegraphics[width=0.325\textwidth]{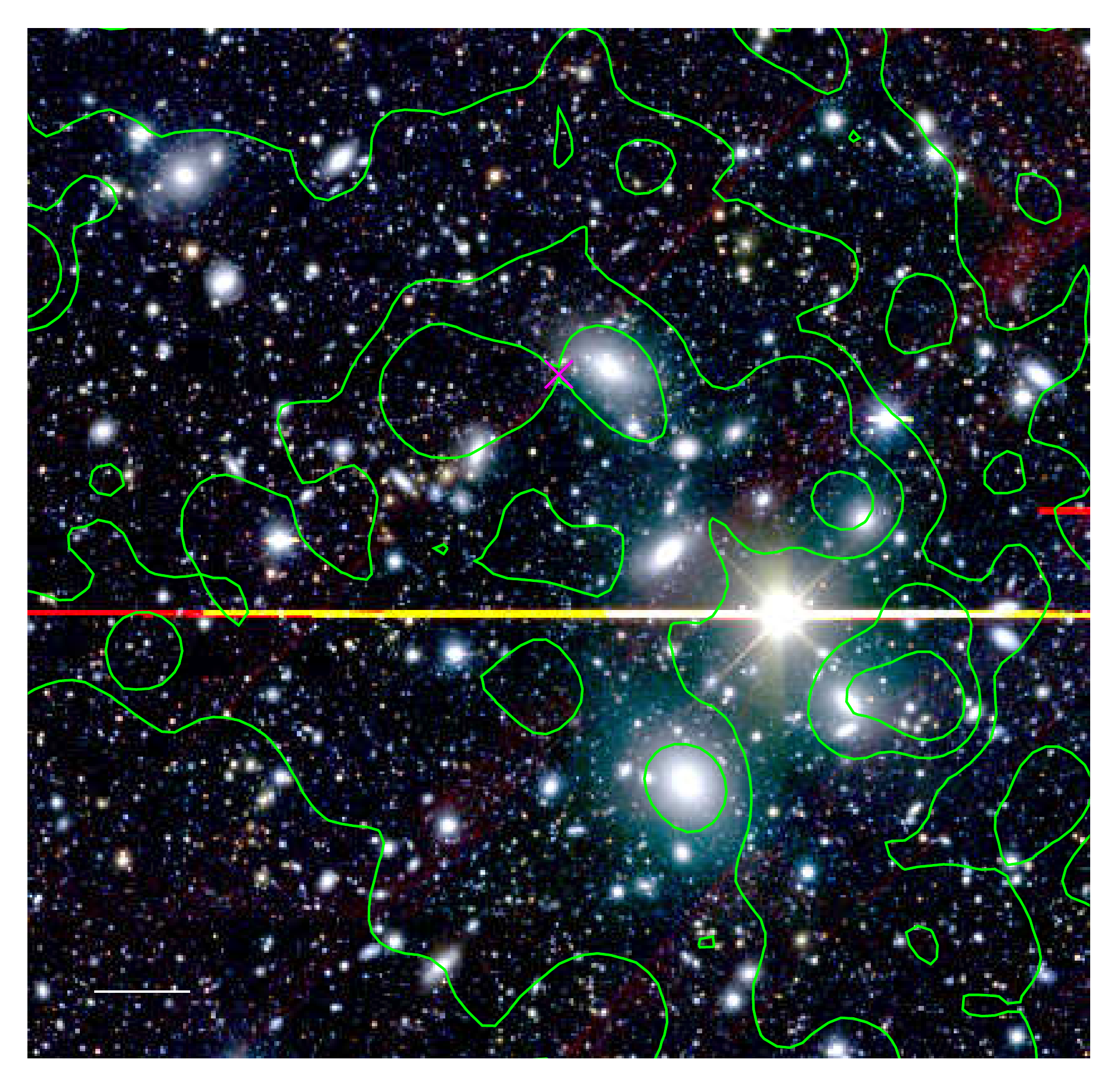}
  \includegraphics[width=0.325\textwidth]{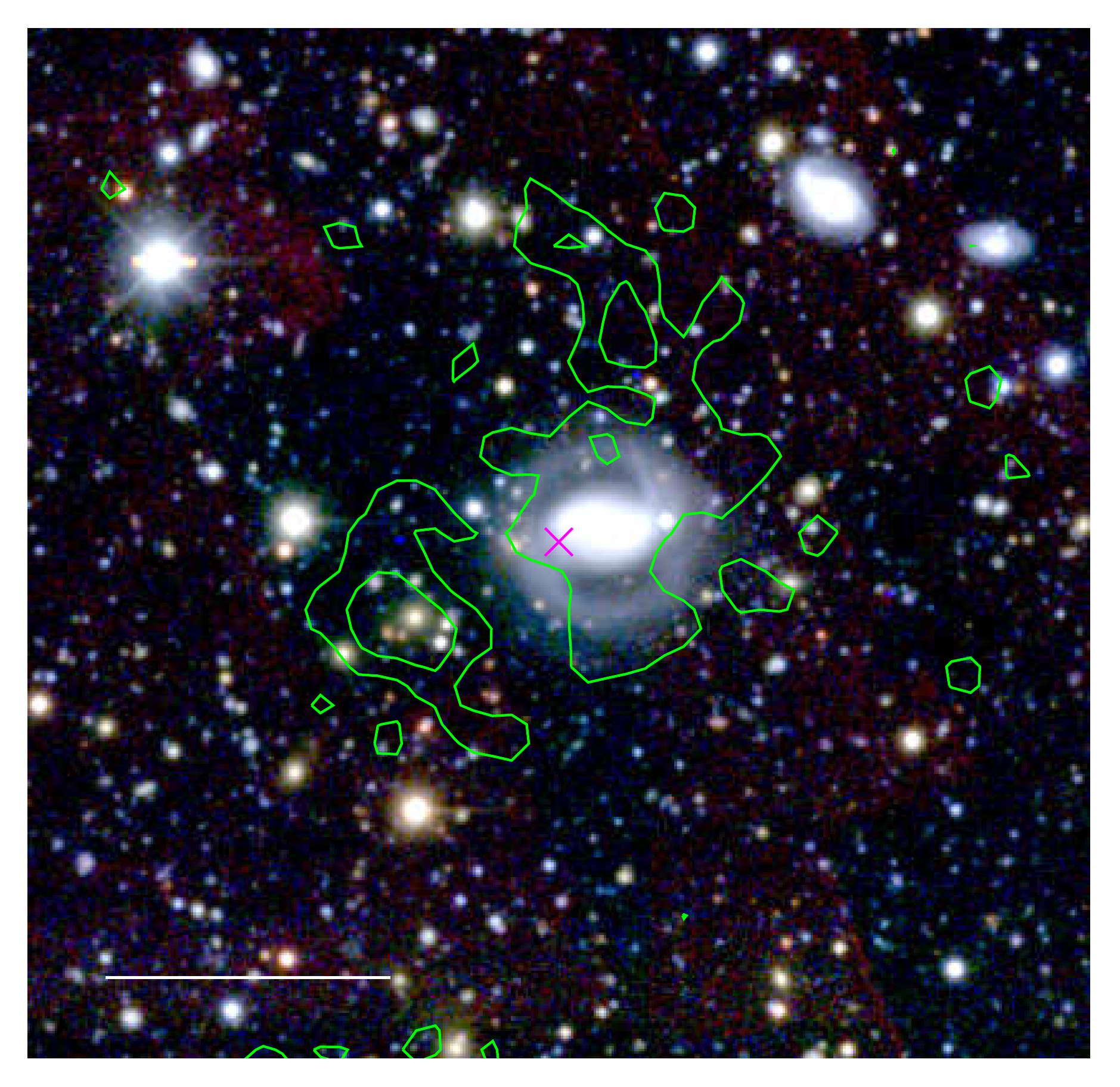}\\
  \includegraphics[width=0.325\textwidth]{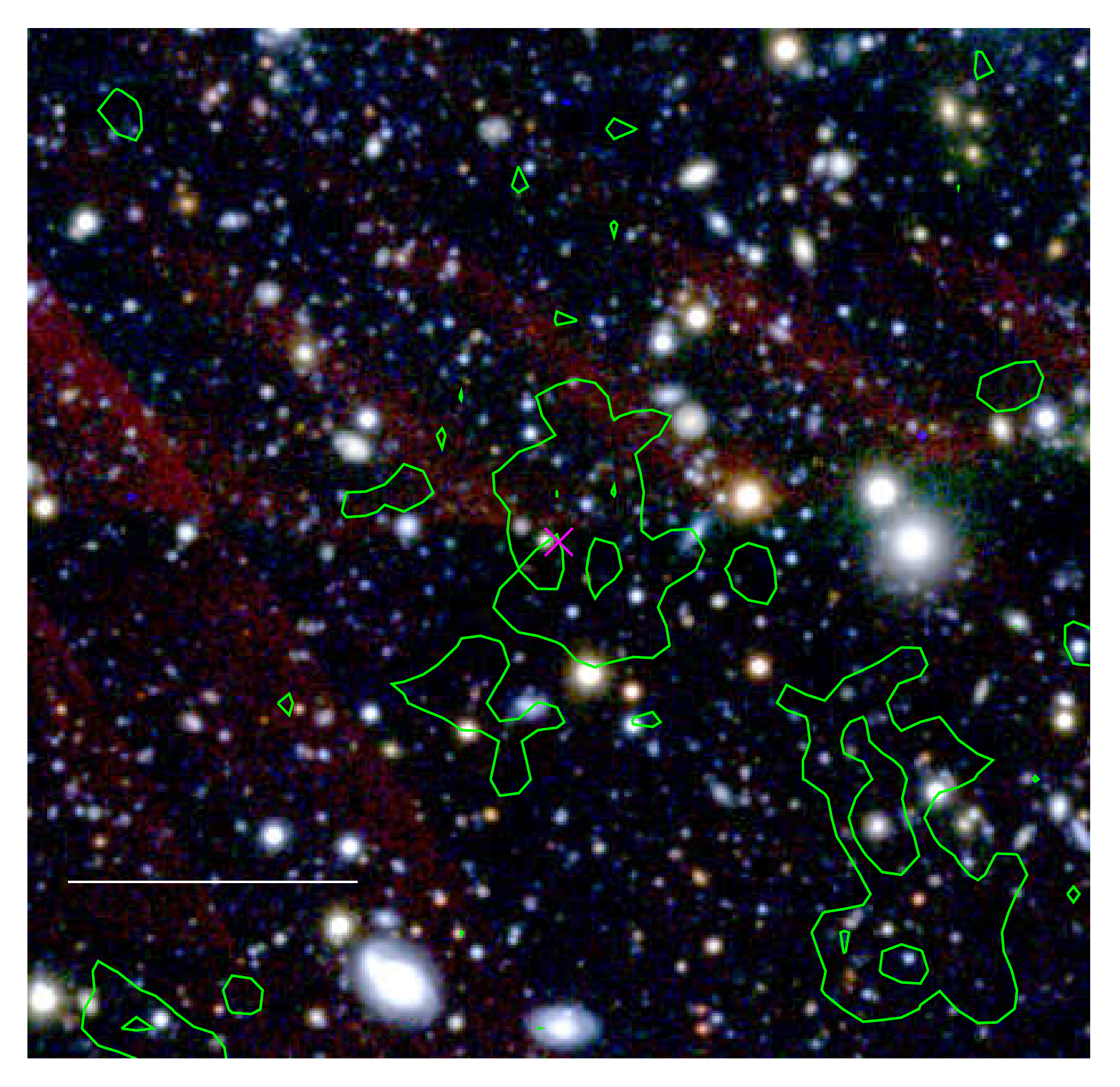}
  \includegraphics[width=0.325\textwidth]{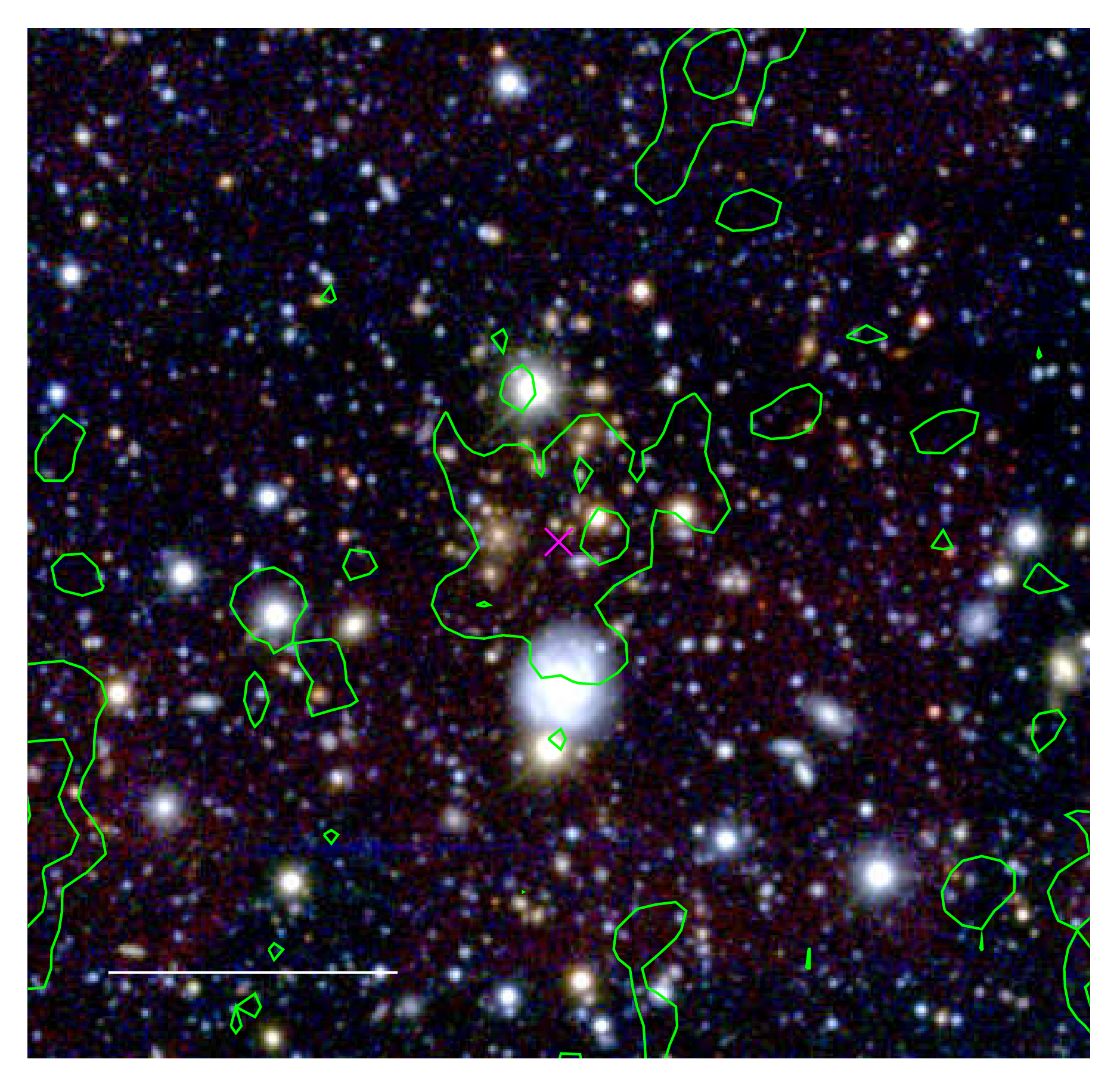}\\
    \caption{
    Zoomed DECam images of the extended sources in the cluster.
    The panels show extended sources X1 to X11, ordered from left to right, then top to bottom.
    The white bars have a length of 1 arcmin.
    The contours are derived from the X-ray image.
    The magenta crosses show the X-ray position.
  }
  \label{fig:ext_src_opt}
\end{figure*}

Table \ref{tab:identifications} lists the identified sources in the system, giving the position, the photometric redshift of the most likely identification, and the MCMF probability of the identification being contamination.
Only one of these objects (X8) appears to be at a similar distance to the cluster itself (ignoring X7 as it has a large value of $f_\mathrm{cont}$).
The systematic uncertainties on the photometric redshift means that it cannot be used to compute a relative velocity of X8 to the cluster.
The source locations are shown in Fig.~\ref{fig:regions}, while in Fig.~\ref{fig:ext_src_opt}, the region around each source in the DECam data, are shown.
Contours from the 0.3--2.3 keV X-ray image are also plotted for comparison.

In the table we also give the temperature, flux, and luminosity of each source, based on spectral fitting with the listed redshift.
To do the spectral fitting, we extracted spectra from the source and a neighbouring background region (see Appendix \ref{appen:regions} for the regions).
The spectra for TMs 1, 2, 3, 4, and 6 were fitted for each source.
We used background spectra in \texttt{Xspec}, rather than modelling a background, due to the difficult of properly modelling projected emission from A\,3266.
We assumed a metallicity of $0.3$~Z$_\odot$ and our previous best fitting column density value.
For source X8, the spectra could not be well fitted by a single temperature component.
For this source we fitted two temperature components, obtaining best-fitting temperatures of $0.79^{+0.03}_{-0.04}$ and $5.4_{-0.6}^{+2.4}$ keV (the cooler component accounts for around 30\% of the flux between 0.5 and 2.0 keV).
The higher temperature could be due to projected cluster emission (although we chose background regions at a similar radius to the structure) or it could be material within the sub-structure which is at a similar temperature to the main cluster (for example, heated or mixed by the merging process).

Based on the obtained temperatures and luminosities, we estimated the source masses using the scaling relations of \cite{Bulbul19}, specifically their equations 18 and 26, also including the intrinsic scatters in the relations and our measurement uncertainties.
The obtained $M_{500}$ values are given in Table \ref{tab:identifications}.
Some of the sources have large temperature uncertainties, for which we do not compute temperature derived masses.
For many of the systems there is reasonable agreement between the results of the two scaling relations.

The low values of $f_\mathrm{cont}$ for most of the sources implies their associations are secure.
However, sources X7 and X10 have $f_\mathrm{cont}>0.3$ and are likely invalid, while X9 is $0.2<f_\mathrm{cont}<0.3$.
X9 appears contaminated, as it lies very close to galaxy at low redshift which also appears to have X-ray emission.
X11 is a rather faint source and has no constraint on its temperature, although the optical image shows a number of red galaxies at its location, justifying its low $f_\mathrm{cont}$.

Source X8 is also the NW structure.
Its temperature and luminosity are more uncertain, as it is unclear where the boundaries of the object are.
The object also has several masked point sources, but without a better PSF it is not obvious that these are not part of its extended emission.
As noted above, a two component fit is required to fit the spectrum and therefore it is unclear how to apply the scaling relations.
For X8 the luminosity derived mass is low compared to the temperature derived one.
However, if the luminosity of both temperature components is used instead of just the lower temperature value, then a mass of $10^{13.6 \pm 0.1}$\,M$_\odot$ is obtained, which is consistent with the one found using the lower temperature value and temperature-mass scaling relation.

The mass range of the other secure objects has a relatively small mass range, varying between $1.6$ and $3.2 \times 10^{14}$ M$_\odot$, although the fainter source X11 may have a mass of $8\times 10^{13}$ M$_\odot$.
The redshift range spans 0.36 to 1.05.
The highest redshift source, X2, is SPT-CLJ0432-6150 \citep{Bleem15}, for which the SPT mass is M$_{500}$ is $2.9^{+0.4}_{-0.7} \times 10^{14}$ M$_\odot$, which is in reasonable agreement with our luminosity-based mass of $2.5 \times 10^{14}$ M$_\odot$ and the temperature-based mass of $3.2 \times 10^{14}$ M$_\odot$.

\begin{figure}
  \centering
  \includegraphics[width=\columnwidth]{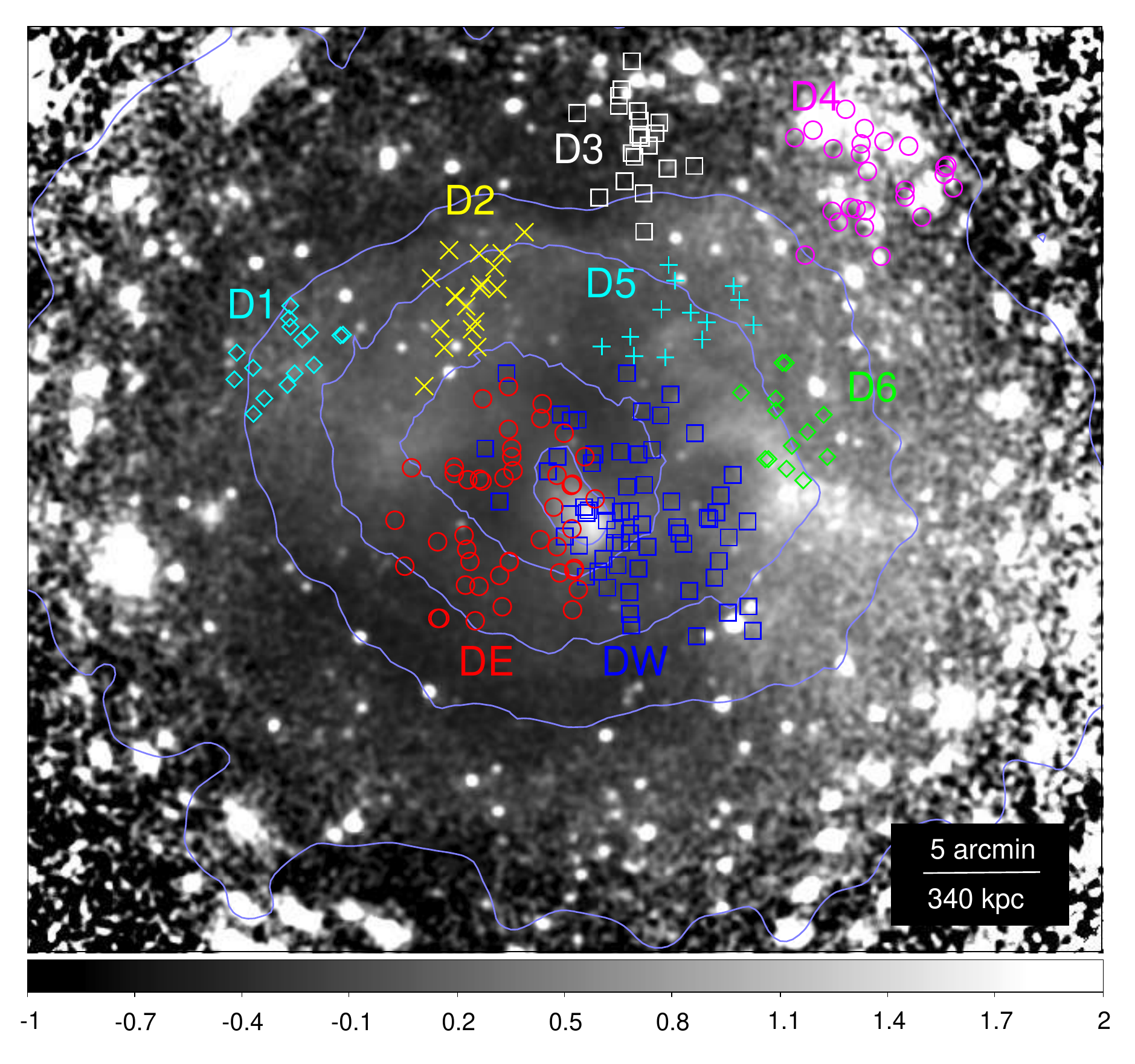}
  \caption{
    Residual map showing location of the galaxy sub-structure subcomponents from \cite{Dehghan17} plotted on the X-ray residual image from Fig.~\ref{fig:ellipse_maps}.
    Different subcomponents are plotted with a different colour or marker style and labelled D1--D6, DE (core east component), and DW (core west component).
    In the core of the cluster there are two different subcomponents.
    The X-ray contours were taken from Fig.~\ref{fig:img_resid}.
  }
  \label{fig:substructure}
\end{figure}

\section{Discussion}
The wide field X-ray view by eROSITA of A\,3266 shows it to be a complex merging system, rather than a simple merger of a cluster with a smaller object.
The data suggest there are at least three different systems currently merging with the main system, which include the NE, W, and NW structures.

\subsection{NE structure}
Extending NE from the cluster core is the low entropy material previously seen in \emph{Chandra} and \emph{XMM-Newton} observations \citep{Henriksen02,Sauvageot05,Finoguenov06}, which we see as tail connecting to a fan shaped structure in our entropy maps (Figures \ref{fig:contbin_maps} and \ref{fig:ellipse_maps}).
This tail is likely part of the baryons associated with a subcluster merging with the main cluster.
The end of fan extends at least to a radius of $\sim 1.1$~Mpc.
In addition to being low entropy, it has high metallicity compared to the main cluster (Fig.~\ref{fig:metallicity_maps}).
\cite{Sauvageot05} put forward two possible scenarios for this merger: either the NE structure is on its first passage through the core or it is returning on its second after entering from the SW.
Comparing our temperature and entropy maps with their simulations, the first passage scenario looks a closer match, particularly given the fan shaped low entropy region.
However, this cluster is not a simple merger and therefore further simulations would be helpful for interpretation.

The galaxy dynamical sub-structures \cite{Dehghan17} identifies, plotted in Fig.~\ref{fig:substructure}, show two different core components.
The western (DW) and eastern (DE) components have a velocities of $18261\pm164$ and $17091\pm228$~km~s$^{-1}$, and velocity dispersions of $1390\pm99$ and $1449\pm115$~km~s$^{-1}$, respectively.
The two core components are offset in velocity by $1170\pm150$~km~s$^{-1}$, which is large if the subcomponents are associated with the main cluster and NE merging subcluster, as the geometry indicates a merger in the plane of the sky.
The optical map shows the central cD galaxy is located within the bright central cluster region.

\subsection{Possible inner shock}
Looking west of the bright central region (Fig.~\ref{fig:core_profile}) the temperature increases by roughly a factor of two outside the core and the density decreases by 28\%, although there is no significant change in pressure.
At a radius of 180 kpc the deprojected temperature and pressure then drop by around 50\%, although the significance is only around $2\sigma$.
If this hot region is a bow shock generated by the merging subcluster and the cooler region is the unshocked material, the Rankine-Hugoniot jump conditions give a Mach number of approximately 1.4 given the pressure jump value and a ratio of specific heats, $\gamma=5/3$.
However, we see no clear associated jump in density or X-ray surface brightness in our profiles, although there are hints that there is an edge in surface brightness located at the second-innermost contour in Fig.~\ref{fig:centre} and an edge in the GGM-filtered image (Fig.~\ref{fig:ggm} top-left panel).
The temperature maps of this region appear complex (for example the hotter-cooler-hotter variation discussed in Section \ref{sect:core}) and there is filamentary X-ray emission located here, too.
It is possible that if a shock were propagating into a complex environment it could be difficult to detect, particularly as eROSITA is less sensitive to hotter material. 

\subsection{Possible outer shock}
The thermodynamic profiles in different sectors (Fig.~\ref{fig:sector}) can differ from the average profile by up to 60\% in density, 40\% in temperature, and factors of two in pressure and entropy.
As seen in the metallicity maps, there are also significant variations in metallicity between the sectors.
However, it is challenging to interpret the profiles physically in a completely unambiguous fashion.
The cleanest jump in the pressure profiles is by a factor of $\sim 3.4$ at 1100\,kpc towards the SE, which is the sector which shows the least disturbance.
The high deprojected temperature region behind this edge is suggestive of a shock.
In the W direction there is also a jump in the pressure profile around 1100\,kpc, which also indicates hotter material in deprojection behind the edge.
Pressure jumps may also occur at a cold front due to ram pressure, although this is rarely seen.
We see the pressure jump at the same radius in two different directions, also arguing against this possibility.
Taking the pressure jump in the SE sector, this would imply a Mach number of $\sim 1.71$, assuming the ratio of specific heats of $5/3$.
Taking the $\sim 76$\% increase in density inwards of this radius would imply a Mach number of $\sim 1.54$.
This is consistent with the pressure jump within the measurement uncertainties.
% e.g. Sarazin (2001), section 3.1. https://arxiv.org/abs/astro-ph/0105418
% g=5/3.; P2_P1=3.4
% math.sqrt((P2_P1*(g+1)+g-1)/(2*g)) = 1.71
% ne2_ne1=1.76
% ((1/ne2_ne1 - (g-1)/(g+1))*(g+1)/2)**-0.5 = 1.54

\subsection{Radio emission}
GLEAM survey radio images of the cluster \citep{Wayth15,HurleyWalker17} appear to show extended halo-like radio emission, which would be consistent with the cluster being a merger.
However, the extended emission is not real, but is made up of resolved sources and instrumental sidelobes \citep[M.~Johnston-Hollitt, private communication;][]{DehghanThesis}.

\subsection{NW structure}
As previously stated, the low entropy, fan shaped structure towards the NE is not the only merger within the cluster.
Seen around 20 arcmin (1350 kpc) is the NW structure, identified as extended object X8, which is also associated with dynamical subcluster D4.
This object appears to be a group in the optical images and appears cool in the X-ray maps and profiles.
From the luminosity scaling relation it has a mass of around $3 \times 10^{14} M_\odot$, although this could be an underestimate if some gas has been stripped.
The galaxies (Fig. \ref{fig:ext_src_opt}) are coincident with the X-ray emission from the group and appear to have retained some of their hot group gas.
\cite{Dehghan17} obtained a relative line-of-sight velocity for this object of $\sim 220$~km~s$^{-1}$ redshifted with respect to the main cluster.
The X-ray contours of the main cluster bend outwards towards the group, suggesting that it has passed at least once through the cluster.
In addition, one part of the filamentary structure which makes up the low entropy western material points directly towards the group, although this is associated with a different dynamic structure D6.
As it retains hot baryons the NW structure was not completely stripped, so likely did not pass through the cluster core, unless it was initially denser than the core itself, which appears unlikely based on the images.

\subsection{W structure}
There are low entropy filaments, the W structure, to the west of the cluster core (Fig.~\ref{fig:centre}).
They appear to connect to an X-ray bright region 9.5 arcmin (650 kpc) from the core in a similar location to dynamical structure D6 and several galaxies in the optical image.
The X-ray emission of D6 appears connected with a brighter region at the location of dynamical structure D5.
Part of the low entropy material to the W has a remarkable similarity with the cavities seen in X-ray images caused by AGN feedback (see e.g. Fig.~\ref{fig:ggm}).
Alternatively, given their association with the dynamical sub-structures and their large physical sizes, they could be stripped intragroup medium from one or two groups which have passed through the cluster centre, associated with D5 and D6.
\cite{Dehghan17} find consistent average velocities for D5 and D6, so it is unclear whether these were two separate objects originally.
If these western filamentary structures are associated with a merger, it suggests they are material which was stripped from the group as it passed near the core of the cluster.
The western structure also appears higher in metallicity in the region immediately adjacent to the central core.

% numbers:
%  radius bubble 125 kpc (assume spherical)
%  distance from centre: 6.5' or 440 kpc
%  g = 9.5e-9
% terminal velocity is around 1140 km/s : (2*g*V/S/C)**0.5

\subsection{Possible AGN cavities}
Bubbles, generated by the jets of AGN, are filled with relativistic particles and are balanced in pressure with the surrounding thermal gas.
They provide a source of mechanical heating within the cluster.
If we consider an alternative explanation for the W filamentary structures that they are the rims of bubbles, we can estimate the bubble enthalpy ($4PV$) from the total thermal pressure of the surrounding material ($P$) and their volume ($V$).
Taking a radius of 110~arcsec (125~kpc) for the westward structure (the possible cavity north of this has a less distinct structure, but perhaps a similar size), and an electron pressure of $7.7 \times 10^{-3}$~keV~cm$^{-3}$ (Fig.~\ref{fig:sector}), we estimate that its enthalpy would be $\sim 2 \times 10^{61}$~erg.
Following \cite{Churazov01} and using our hydrostatic model to obtain the gravitational acceleration ($\sim 9.5\times 10^{-9}$~cm~s$^{-2}$), we estimate the buoyancy rise time to be $400$~Myr.
The mechanical heating power would therefore be $2 \times 10^{45}$~erg~s$^{-1}$.

Such a cavity would be one of the most powerful known, in terms of mechanical heating \citep[for comparison see][]{Birzan04,HlavacekLarrondo12}, although it would be weaker than the extreme case of MS\,0735.6+7421 \citep{Vantyghem14}.
These AGN-driven cavities are almost always found in systems with cool cores, where the centre has a short mean radiative cooling time.
A\,3266 has a long central cooling time ($\sim 11$~Gyr) and is not a cool core system.
However, the \emph{Chandra} data show a weak source at the nucleus of the central galaxy.
Fitting its spectrum, we find the 0.5--2.0 keV flux is $1.0 \times 10^{-14}$~erg~cm$^{-2}$~s$^{-1}$, assuming a power-law photon index of 1.7 and our corrected-HI4PI absorption, which implies a rest-frame 0.5--2.0 keV luminosity of $9\times 10^{40}$~erg~s$^{-1}$.
\cite{MerloniHeinz07} found that the estimated kinetic power and the nuclear luminosity of a cavity-inflating AGN followed a relation expected for radiatively inefficient flows, after normalising by black hole mass.
Although we do not know the mass of the black hole in A\,3266, if we assume a mass of $10^9$ or $10^{10}$~M$_\odot$, then the X-ray nucleus is too faint by $4$ or $2.5$ orders of magnitude, respectively, to account for the estimated power if the structure is a cavity. 
Given the long cooling time and only weak X-ray nucleus, the hypothesis that the cavity structure is stripped material from a group appears the most likely one, unless radio emission is later found at the location of the apparent cavity.

\subsection{Metallicity sub-structure}
The metallicity maps (Fig. \ref{fig:metallicity_maps}) show that the metallicity is not uniform within the cluster.
Some of these are associated with the merging system (in particular the NE structure).
There are also even higher metallicity regions inside the NE structure, and a previously reported low metallicity region to the W of the central core.
The maps also show a high metallicity region to the south which was also previously seen by \emph{XMM-Newton} \citep{Sauvageot05}.
Other clusters also show embedded high and low metallicity regions, including A2204, Perseus \citep{SandersPer07}, M87 \citep{Simionescu08,Million10}, Centaurus \citep{SandersCent16}, and NGC 4636 \citep{OSullivan05}.
The presence of these metallicity variations may imply that mixing or diffusion processes in the ICM in clusters are relatively slow compared to the merging processes in the cluster.
The metallicity variation we see is likely due to the merging and stripping processes.
AGN activity is unlikely to be able to uplift metals over the scale of a whole cluster.
Sloshing could displace metals from the centre over some scales, but will also be seen as surface brightness and temperature variations.

\section{Conclusions}
The data presented here show the power of eROSITA to image in X-rays a large region of the sky and map a galaxy cluster in almost its entirety.
These X-ray data confirm that A\,3266 is not a simple merging cluster.
There are at least three different systems merging with the main body of the cluster: the NE structure, the NW structure, and the W structure.
These three systems are seen as low entropy material in our spectroscopic maps and appear associated with higher-metallicity gas (in the case of the NE and W structure).
These merging systems have also been seen previously in a dynamical analysis of the cluster.
The NW structure appears as if it has already passed through the centre of the cluster, as seen by the extension in the main cluster in the direction of the group.
The W structure appears to be partially made up of stripped material as it passed through the cluster.
However, an alternative possibility for the filamentary material to the W could be the rims of an AGN bubble.
The NE structure looks as if it were on its first passage into the cluster, although other scenarios cannot yet be ruled out.
The central core, which seems connected to the NE structure, has a high temperature region on its W side.
Unfortunately our data are unable to conclusively say whether this is a merger shock or not.
Towards the outskirts of the cluster we see a pressure jump in the SE and W directions at a radius of 1100\,kpc.
Taking the values towards the SE where the edge is cleanest, this is consistent with a shock with a Mach number of $1.5$--$1.7$.
%
%For the first time we report the detection of a radio halo in the galaxy cluster.
%The GLEAM low-frequency data show signal up to $R_{500}$ with extension towards $R_{200}$.
%The presence of this halo confirms that this is a merging galaxy cluster.

Comparing the eROSITA data to that of other instruments, we find that eROSITA favours a higher absorbing column density than is expected from HI surveys.
Using a best fitting column density brings the eROSITA measured temperatures in the cluster into reasonable agreement with previous \emph{XMM-Newton} values.
The difference in preferred column density could be due to residual calibration uncertainties.
There are also differences in temperature obtained from the eRASS1 survey and this pointed observation, although this may be due to multiphase material.
Both of these issues require a further, more detailed investigation.
If we use the best fitting column density, we find hydrostatic masses in the cluster that agree well with what was found by \emph{XMM-Newton}.
However, due to the disturbed nature of the cluster, the mass profiles derived from different sectors are not in agreement.

We detect seven background galaxy clusters within our field with secure associations identified using MCMF.
These span a redshift range of 0.364 to 1.05 and a mass range of $8 \times 10^{13}$ to $3 \times 10^{14}$~M$_{\odot}$.
Red sequence galaxies can be seen in the DECam observations of this cluster.
Only one of these systems was previously known (SPT-CLJ0432-6150).

\begin{acknowledgements}
MB acknowledges support from the Deutsche Forschungsgemeinschaft under Germany's Excellence Strategy - EXC 2121 "Quantum Universe" - 390833306.
K.~Dolag acknowledges support by the Deutsche Forschungsgemeinschaft (DFG, German Research Foundation) under Germany’s Excellence Strategy -- EXC-2094 -- 3907833 and by the funding for the COMPLEX project from the European Research Council (ERC) under the European Union's Horizon 2020 research and innovation program grant agreement ERC-2019-AdG860744.
Part of this work has been funded by the Deutsche Forschungsgemeinschaft (DFG, German Research Foundation) - 415510302. DNH acknowledges support from the ERC through the grant ERC-Stg DRANOEL n. 714245.
JAZ acknowledges support from the Chandra X-ray Center, which is operated by the Smithsonian Astrophysical Observatory for and on behalf of NASA under contract NAS8-03060.

This work is based on data from eROSITA, the soft X-ray instrument aboard \emph{SRG}, a joint Russian-German science mission supported by the Russian Space Agency (Roskosmos), in the interests of the Russian Academy of Sciences represented by its Space Research Institute (IKI), and the Deutsches Zentrum für Luft- und Raumfahrt (DLR). The \emph{SRG} spacecraft was built by Lavochkin Association (NPOL) and its subcontractors, and is operated by NPOL with support from the Max Planck Institute for Extraterrestrial Physics (MPE).

The development and construction of the eROSITA X-ray instrument was led by MPE, with contributions from the Dr. Karl Remeis Observatory Bamberg \& ECAP (FAU Erlangen-Nuernberg), the University of Hamburg Observatory, the Leibniz Institute for Astrophysics Potsdam (AIP), and the Institute for Astronomy and Astrophysics of the University of Tübingen, with the support of DLR and the Max Planck Society. The Argelander Institute for Astronomy of the University of Bonn and the Ludwig Maximilians Universität Munich also participated in the science preparation for eROSITA.
The eROSITA data shown here were processed using the eSASS/NRTA software system developed by the German eROSITA consortium.
\end{acknowledgements}

\bibliographystyle{aa}
\bibliography{refs}

\begin{thebibliography}{67}
\expandafter\ifx\csname natexlab\endcsname\relax\def\natexlab#1{#1}\fi

\bibitem[{{Abbott} {et~al.}(2021){Abbott}, {Adamow}, {Aguena}, {Allam}, {Amon},
  {Avila}, {Bacon}, {Banerji}, {Bechtol}, {Becker}, {Bernstein}, {Bertin},
  {Bhargava}, {Bridle}, {Brooks}, {Burke}, {Carnero Rosell}, {Carrasco Kind},
  {Carretero}, {Castander}, {Cawthon}, {Chang}, {Choi}, {Conselice},
  {Costanzi}, {Crocce}, {da Costa}, {Davis}, {De Vicente}, {DeRose}, {Desai},
  {Diehl}, {Dietrich}, {Drlica-Wagner}, {Eckert}, {Elvin-Poole}, {Everett},
  {Evrard}, {Ferrero}, {Fert{\'e}}, {Flaugher}, {Fosalba}, {Friedel},
  {Frieman}, {Garc{\'\i}a-Bellido}, {Gelman}, {Gerdes}, {Giannantonio}, {Gill},
  {Gruen}, {Gruendl}, {Gschwend}, {Gutierrez}, {Hartley}, {Hinton},
  {Hollowood}, {Huterer}, {James}, {Jeltema}, {Johnson}, {Kent}, {Kron},
  {Kuehn}, {Kuropatkin}, {Lahav}, {Li}, {Lidman}, {Lin}, {MacCrann}, {Maia},
  {Manning}, {March}, {Marshall}, {Martini}, {Melchior}, {Menanteau}, {Miquel},
  {Morgan}, {Myles}, {Neilsen}, {Ogando}, {Palmese}, {Paz-Chinch{\'o}n},
  {Petravick}, {Pieres}, {Plazas}, {Pond}, {Rodriguez-Monroy}, {Romer},
  {Roodman}, {Rykoff}, {Sako}, {Sanchez}, {Santiago}, {Serrano},
  {Sevilla-Noarbe}, {Allyn. Smith}, {Smith}, {Soares-Santos}, {Suchyta},
  {Swanson}, {Tarle}, {Thomas}, {To}, {Tremblay}, {Troxel}, {Tucker}, {Turner},
  {Varga}, {Walker}, {Wechsler}, {Weller}, {Wester}, {Wilkinson}, {Yanny},
  {Zhang}, {Nikutta}, {Fitzpatrick}, {Jacques}, {Scott}, {Olsen}, {Huang},
  {Herrera}, {Juneau}, {Nidever}, {Weaver}, {Adean}, {Correia}, {de Freitas},
  {Freitas}, {Singulani}, \& {Vila-Verde}}]{Abbott21}
{Abbott}, T.~M.~C., {Adamow}, M., {Aguena}, M., {et~al.} 2021, \apjs, 255, 20

\bibitem[{{Anders} \& {Grevesse}(1989)}]{AndersGrevesse89}
{Anders}, E. \& {Grevesse}, N. 1989, \gca, 53, 197

\bibitem[{{Arnaud}(1996)}]{ArnaudXspec}
{Arnaud}, K.~A. 1996, in ASP Conf. Ser. 101: Astronomical Data Analysis
  Software and Systems V, ed. G.~H. {Jacoby} \& J.~{Barnes}, 17

\bibitem[{{Asplund} {et~al.}(2009){Asplund}, {Grevesse}, {Sauval}, \&
  {Scott}}]{Asplund09}
{Asplund}, M., {Grevesse}, N., {Sauval}, A.~J., \& {Scott}, P. 2009, \araa, 47,
  481

\bibitem[{{Biffi} {et~al.}(2016){Biffi}, {Borgani}, {Murante}, {Rasia},
  {Planelles}, {Granato}, {Ragone-Figueroa}, {Beck}, {Gaspari}, \&
  {Dolag}}]{Biffi16}
{Biffi}, V., {Borgani}, S., {Murante}, G., {et~al.} 2016, \apj, 827, 112

\bibitem[{{B{\^i}rzan} {et~al.}(2004){B{\^i}rzan}, {Rafferty}, {McNamara},
  {Wise}, \& {Nulsen}}]{Birzan04}
{B{\^i}rzan}, L., {Rafferty}, D.~A., {McNamara}, B.~R., {Wise}, M.~W., \&
  {Nulsen}, P.~E.~J. 2004, \apj, 607, 800

\bibitem[{{Bleem} {et~al.}(2015){Bleem}, {Stalder}, {de Haan}, {Aird}, {Allen},
  {Applegate}, {Ashby}, {Bautz}, {Bayliss}, {Benson}, {Bocquet}, {Brodwin},
  {Carlstrom}, {Chang}, {Chiu}, {Cho}, {Clocchiatti}, {Crawford}, {Crites},
  {Desai}, {Dietrich}, {Dobbs}, {Foley}, {Forman}, {George}, {Gladders},
  {Gonzalez}, {Halverson}, {Hennig}, {Hoekstra}, {Holder}, {Holzapfel},
  {Hrubes}, {Jones}, {Keisler}, {Knox}, {Lee}, {Leitch}, {Liu}, {Lueker},
  {Luong-Van}, {Mantz}, {Marrone}, {McDonald}, {McMahon}, {Meyer}, {Mocanu},
  {Mohr}, {Murray}, {Padin}, {Pryke}, {Reichardt}, {Rest}, {Ruel}, {Ruhl},
  {Saliwanchik}, {Saro}, {Sayre}, {Schaffer}, {Schrabback}, {Shirokoff},
  {Song}, {Spieler}, {Stanford}, {Staniszewski}, {Stark}, {Story}, {Stubbs},
  {Vanderlinde}, {Vieira}, {Vikhlinin}, {Williamson}, {Zahn}, \&
  {Zenteno}}]{Bleem15}
{Bleem}, L.~E., {Stalder}, B., {de Haan}, T., {et~al.} 2015, \apjs, 216, 27

\bibitem[{{B{\"o}hringer} \& {Werner}(2010)}]{Bohringer10}
{B{\"o}hringer}, H. \& {Werner}, N. 2010, \aapr, 18, 127

\bibitem[{{Brunner} {et~al.}(2021){Brunner}, {Liu}, {Lamer}, {Georgakakis},
  {Merloni}, {Brusa}, {Bulbul}, {Dennerl}, {Friedrich}, {Liu}, {Maitra},
  {Nandra}, {Ramos-Ceja}, {Sanders}, {Stewart}, {Boller}, {Buchner}, {Clerc},
  {Comparat}, {Dwelly}, {Eckert}, {Finoguenov}, {Freyberg}, {Ghirardini},
  {Gueguen}, {Haberl}, {Kreykenbohm}, {Krumpe}, {Osterhage}, {Pacaud},
  {Predehl}, {Reiprich}, {Robrade}, {Salvato}, {Santangelo}, {Schrabback},
  {Schwope}, \& {Wilms}}]{Brunner21}
{Brunner}, H., {Liu}, T., {Lamer}, G., {et~al.} 2021, \aap, submitted

\bibitem[{{Bulbul} {et~al.}(2019){Bulbul}, {Chiu}, {Mohr}, {McDonald},
  {Benson}, {Bautz}, {Bayliss}, {Bleem}, {Brodwin}, {Bocquet}, {Capasso},
  {Dietrich}, {Forman}, {Hlavacek-Larrondo}, {Holzapfel}, {Khullar}, {Klein},
  {Kraft}, {Miller}, {Reichardt}, {Saro}, {Sharon}, {Stalder}, {Schrabback}, \&
  {Stanford}}]{Bulbul19}
{Bulbul}, E., {Chiu}, I.~N., {Mohr}, J.~J., {et~al.} 2019, \apj, 871, 50

\bibitem[{{Cappelluti} {et~al.}(2017){Cappelluti}, {Li}, {Ricarte}, {Agarwal},
  {Allevato}, {Tasnim Ananna}, {Ajello}, {Civano}, {Comastri}, {Elvis},
  {Finoguenov}, {Gilli}, {Hasinger}, {Marchesi}, {Natarajan}, {Pacucci},
  {Treister}, \& {Urry}}]{Cappelluti17}
{Cappelluti}, N., {Li}, Y., {Ricarte}, A., {et~al.} 2017, \apj, 837, 19

\bibitem[{{Churazov} {et~al.}(2001){Churazov}, {Br{\"u}ggen}, {Kaiser},
  {B{\"o}hringer}, \& {Forman}}]{Churazov01}
{Churazov}, E., {Br{\"u}ggen}, M., {Kaiser}, C.~R., {B{\"o}hringer}, H., \&
  {Forman}, W. 2001, \apj, 554, 261

\bibitem[{{Dehghan}(2014)}]{DehghanThesis}
{Dehghan}, S. 2014, PhD thesis, Victoria University of Wellington

\bibitem[{{Dehghan} {et~al.}(2017){Dehghan}, {Johnston-Hollitt}, {Colless}, \&
  {Miller}}]{Dehghan17}
{Dehghan}, S., {Johnston-Hollitt}, M., {Colless}, M., \& {Miller}, R. 2017,
  \mnras, 468, 2645

\bibitem[{{Dennerl} {et~al.}(2020){Dennerl}, {Andritschke}, {Br{\"a}uninger},
  {Burkert}, {Burwitz}, {Emberger}, {Freyberg}, {Friedrich}, {Gaida},
  {Granato}, {Hartner}, {von Kienlin}, {Meidinger}, {Menz}, \&
  {Predehl}}]{Dennerl20}
{Dennerl}, K., {Andritschke}, R., {Br{\"a}uninger}, H., {et~al.} 2020, in
  Society of Photo-Optical Instrumentation Engineers (SPIE) Conference Series,
  Vol. 11444, Society of Photo-Optical Instrumentation Engineers (SPIE)
  Conference Series, 114444Q

\bibitem[{{Dey} {et~al.}(2019){Dey}, {Schlegel}, {Lang}, {Blum}, {Burleigh},
  {Fan}, {Findlay}, {Finkbeiner}, {Herrera}, {Juneau}, {Landriau}, {Levi},
  {McGreer}, {Meisner}, {Myers}, {Moustakas}, {Nugent}, {Patej}, {Schlafly},
  {Walker}, {Valdes}, {Weaver}, {Y{\`e}che}, {Zou}, {Zhou}, {Abareshi},
  {Abbott}, {Abolfathi}, {Aguilera}, {Alam}, {Allen}, {Alvarez}, {Annis},
  {Ansarinejad}, {Aubert}, {Beechert}, {Bell}, {BenZvi}, {Beutler}, {Bielby},
  {Bolton}, {Brice{\~n}o}, {Buckley-Geer}, {Butler}, {Calamida}, {Carlberg},
  {Carter}, {Casas}, {Castander}, {Choi}, {Comparat}, {Cukanovaite}, {Delubac},
  {DeVries}, {Dey}, {Dhungana}, {Dickinson}, {Ding}, {Donaldson}, {Duan},
  {Duckworth}, {Eftekharzadeh}, {Eisenstein}, {Etourneau}, {Fagrelius},
  {Farihi}, {Fitzpatrick}, {Font-Ribera}, {Fulmer}, {G{\"a}nsicke},
  {Gaztanaga}, {George}, {Gerdes}, {Gontcho}, {Gorgoni}, {Green}, {Guy},
  {Harmer}, {Hernandez}, {Honscheid}, {Huang}, {James}, {Jannuzi}, {Jiang},
  {Joyce}, {Karcher}, {Karkar}, {Kehoe}, {Kneib}, {Kueter-Young}, {Lan},
  {Lauer}, {Le Guillou}, {Le Van Suu}, {Lee}, {Lesser}, {Perreault Levasseur},
  {Li}, {Mann}, {Marshall}, {Mart{\'\i}nez-V{\'a}zquez}, {Martini}, {du Mas des
  Bourboux}, {McManus}, {Meier}, {M{\'e}nard}, {Metcalfe},
  {Mu{\~n}oz-Guti{\'e}rrez}, {Najita}, {Napier}, {Narayan}, {Newman}, {Nie},
  {Nord}, {Norman}, {Olsen}, {Paat}, {Palanque-Delabrouille}, {Peng},
  {Poppett}, {Poremba}, {Prakash}, {Rabinowitz}, {Raichoor}, {Rezaie},
  {Robertson}, {Roe}, {Ross}, {Ross}, {Rudnick}, {Safonova}, {Saha},
  {S{\'a}nchez}, {Savary}, {Schweiker}, {Scott}, {Seo}, {Shan}, {Silva},
  {Slepian}, {Soto}, {Sprayberry}, {Staten}, {Stillman}, {Stupak}, {Summers},
  {Sien Tie}, {Tirado}, {Vargas-Maga{\~n}a}, {Vivas}, {Wechsler}, {Williams},
  {Yang}, {Yang}, {Yapici}, {Zaritsky}, {Zenteno}, {Zhang}, {Zhang}, {Zhou}, \&
  {Zhou}}]{legacysurveys}
{Dey}, A., {Schlegel}, D.~J., {Lang}, D., {et~al.} 2019, \aj, 157, 168

\bibitem[{{Edge} {et~al.}(1990){Edge}, {Stewart}, {Fabian}, \&
  {Arnaud}}]{Edge90}
{Edge}, A.~C., {Stewart}, G.~C., {Fabian}, A.~C., \& {Arnaud}, K.~A. 1990,
  \mnras, 245, 559

\bibitem[{{Erben} {et~al.}(2005){Erben}, {Schirmer}, {Dietrich}, {Cordes},
  {Haberzettl}, {Hetterscheidt}, {Hildebrandt}, {Schmithuesen}, {Schneider},
  {Simon}, {Deul}, {Hook}, {Kaiser}, {Radovich}, {Benoist}, {Nonino}, {Olsen},
  {Prandoni}, {Wichmann}, {Zaggia}, {Bomans}, {Dettmar}, \&
  {Miralles}}]{Erben_05}
{Erben}, T., {Schirmer}, M., {Dietrich}, J.~P., {et~al.} 2005, Astronomische
  Nachrichten, 326, 432

\bibitem[{{Ettori} {et~al.}(2019){Ettori}, {Ghirardini}, {Eckert},
  {Pointecouteau}, {Gastaldello}, {Sereno}, {Gaspari}, {Ghizzardi},
  {Roncarelli}, \& {Rossetti}}]{Ettori19}
{Ettori}, S., {Ghirardini}, V., {Eckert}, D., {et~al.} 2019, \aap, 621, A39

\bibitem[{{Fabian}(2012)}]{Fabian12}
{Fabian}, A.~C. 2012, \araa, 50, 455

\bibitem[{{Finoguenov} {et~al.}(2006){Finoguenov}, {Henriksen}, {Miniati},
  {Briel}, \& {Jones}}]{Finoguenov06}
{Finoguenov}, A., {Henriksen}, M.~J., {Miniati}, F., {Briel}, U.~G., \&
  {Jones}, C. 2006, \apj, 643, 790

\bibitem[{{Flaugher} {et~al.}(2015){Flaugher}, {Diehl}, {Honscheid}, {Abbott},
  {Alvarez}, {Angstadt}, {Annis}, {Antonik}, {Ballester}, {Beaufore},
  {Bernstein}, {Bernstein}, {Bigelow}, {Bonati}, {Boprie}, {Brooks},
  {Buckley-Geer}, {Campa}, {Cardiel-Sas}, {Castander}, {Castilla}, {Cease},
  {Cela-Ruiz}, {Chappa}, {Chi}, {Cooper}, {da Costa}, {Dede}, {Derylo},
  {DePoy}, {de Vicente}, {Doel}, {Drlica-Wagner}, {Eiting}, {Elliott}, {Emes},
  {Estrada}, {Fausti Neto}, {Finley}, {Flores}, {Frieman}, {Gerdes},
  {Gladders}, {Gregory}, {Gutierrez}, {Hao}, {Holland}, {Holm}, {Huffman},
  {Jackson}, {James}, {Jonas}, {Karcher}, {Karliner}, {Kent}, {Kessler},
  {Kozlovsky}, {Kron}, {Kubik}, {Kuehn}, {Kuhlmann}, {Kuk}, {Lahav}, {Lathrop},
  {Lee}, {Levi}, {Lewis}, {Li}, {Mandrichenko}, {Marshall}, {Martinez},
  {Merritt}, {Miquel}, {Mu{\~n}oz}, {Neilsen}, {Nichol}, {Nord}, {Ogando},
  {Olsen}, {Palaio}, {Patton}, {Peoples}, {Plazas}, {Rauch}, {Reil}, {Rheault},
  {Roe}, {Rogers}, {Roodman}, {Sanchez}, {Scarpine}, {Schindler}, {Schmidt},
  {Schmitt}, {Schubnell}, {Schultz}, {Schurter}, {Scott}, {Serrano}, {Shaw},
  {Smith}, {Soares-Santos}, {Stefanik}, {Stuermer}, {Suchyta}, {Sypniewski},
  {Tarle}, {Thaler}, {Tighe}, {Tran}, {Tucker}, {Walker}, {Wang}, {Watson},
  {Weaverdyck}, {Wester}, {Woods}, {Yanny}, \& {DES
  Collaboration}}]{Flaugher_2015}
{Flaugher}, B., {Diehl}, H.~T., {Honscheid}, K., {et~al.} 2015, \aj, 150, 150

\bibitem[{{Freyberg} {et~al.}(2020){Freyberg}, {Perinati}, {Pacaud}, {Eraerds},
  {Churazov}, {Dennerl}, {Predehl}, {Merloni}, {Meidinger}, {Bulbul},
  {Friedrich}, {Gilfanov}, {Tenzer}, {Pommranz}, {Eckert}, {Schmitt}, {Brusa},
  \& {Santangelo}}]{Freyberg20}
{Freyberg}, M., {Perinati}, E., {Pacaud}, F., {et~al.} 2020, in Society of
  Photo-Optical Instrumentation Engineers (SPIE) Conference Series, Vol. 11444,
  Society of Photo-Optical Instrumentation Engineers (SPIE) Conference Series,
  114441O

\bibitem[{{Fruscione} {et~al.}(2006){Fruscione}, {McDowell}, {Allen},
  {Brickhouse}, {Burke}, {Davis}, {Durham}, {Elvis}, {Galle}, {Harris},
  {Huenemoerder}, {Houck}, {Ishibashi}, {Karovska}, {Nicastro}, {Noble},
  {Nowak}, {Primini}, {Siemiginowska}, {Smith}, \& {Wise}}]{Fruscione06}
{Fruscione}, A., {McDowell}, J.~C., {Allen}, G.~E., {et~al.} 2006, in Society
  of Photo-Optical Instrumentation Engineers (SPIE) Conference Series, Vol.
  6270, Society of Photo-Optical Instrumentation Engineers (SPIE) Conference
  Series, 1

\bibitem[{{Ghirardini} {et~al.}(2019){Ghirardini}, {Eckert}, {Ettori},
  {Pointecouteau}, {Molendi}, {Gaspari}, {Rossetti}, {De Grandi}, {Roncarelli},
  {Bourdin}, {Mazzotta}, {Rasia}, \& {Vazza}}]{Ghirardini19}
{Ghirardini}, V., {Eckert}, D., {Ettori}, S., {et~al.} 2019, \aap, 621, A41

\bibitem[{{Gladders} \& {Yee}(2000)}]{gladders00}
{Gladders}, M.~D. \& {Yee}, H. K.~C. 2000, \aj, 120, 2148

\bibitem[{{Green}(2011)}]{GreenCubehelix11}
{Green}, D.~A. 2011, Bulletin of the Astronomical Society of India, 39, 289

\bibitem[{{Henriksen} {et~al.}(2000){Henriksen}, {Donnelly}, \&
  {Davis}}]{Henriksen00}
{Henriksen}, M., {Donnelly}, R.~H., \& {Davis}, D.~S. 2000, \apj, 529, 692

\bibitem[{{Henriksen} \& {Tittley}(2002)}]{Henriksen02}
{Henriksen}, M.~J. \& {Tittley}, E.~R. 2002, \apj, 577, 701

\bibitem[{{HI4PI Collaboration} {et~al.}(2016){HI4PI Collaboration}, {Ben
  Bekhti}, {Fl{\"o}er}, {Keller}, {Kerp}, {Lenz}, {Winkel}, {Bailin},
  {Calabretta}, {Dedes}, {Ford}, {Gibson}, {Haud}, {Janowiecki}, {Kalberla},
  {Lockman}, {McClure-Griffiths}, {Murphy}, {Nakanishi}, {Pisano}, \&
  {Staveley-Smith}}]{HI4PI16}
{HI4PI Collaboration}, {Ben Bekhti}, N., {Fl{\"o}er}, L., {et~al.} 2016, \aap,
  594, A116

\bibitem[{{Hitomi Collaboration}(2016)}]{Hitomi16}
{Hitomi Collaboration}. 2016, \nat, 535, 117

\bibitem[{{Hlavacek-Larrondo} {et~al.}(2012){Hlavacek-Larrondo}, {Fabian},
  {Edge}, {Ebeling}, {Sanders}, {Hogan}, \& {Taylor}}]{HlavacekLarrondo12}
{Hlavacek-Larrondo}, J., {Fabian}, A.~C., {Edge}, A.~C., {et~al.} 2012, \mnras,
  421, 1360

\bibitem[{{Hurley-Walker} {et~al.}(2017){Hurley-Walker}, {Callingham},
  {Hancock}, {Franzen}, {Hindson}, {Kapi{\'n}ska}, {Morgan}, {Offringa},
  {Wayth}, {Wu}, {Zheng}, {Murphy}, {Bell}, {Dwarakanath}, {For}, {Gaensler},
  {Johnston-Hollitt}, {Lenc}, {Procopio}, {Staveley-Smith}, {Ekers}, {Bowman},
  {Briggs}, {Cappallo}, {Deshpande}, {Greenhill}, {Hazelton}, {Kaplan},
  {Lonsdale}, {McWhirter}, {Mitchell}, {Morales}, {Morgan}, {Oberoi}, {Ord},
  {Prabu}, {Shankar}, {Srivani}, {Subrahmanyan}, {Tingay}, {Webster},
  {Williams}, \& {Williams}}]{HurleyWalker17}
{Hurley-Walker}, N., {Callingham}, J.~R., {Hancock}, P.~J., {et~al.} 2017,
  \mnras, 464, 1146

\bibitem[{{Klein} {et~al.}(2019){Klein}, {Grandis}, {Mohr}, {Paulus}, {Abbott},
  {Annis}, {Avila}, {Bertin}, {Brooks}, {Buckley-Geer}, {Rosell}, {Kind},
  {Carretero}, {Castander}, {Cunha}, {D'Andrea}, {da Costa}, {De Vicente},
  {Desai}, {Diehl}, {Dietrich}, {Doel}, {Evrard}, {Flaugher}, {Fosalba},
  {Frieman}, {Garc{\'\i}a-Bellido}, {Gaztanaga}, {Giles}, {Gruen}, {Gruendl},
  {Gschwend}, {Gutierrez}, {Hartley}, {Hollowood}, {Honscheid}, {Hoyle},
  {James}, {Jeltema}, {Kuehn}, {Kuropatkin}, {Lima}, {Maia}, {March},
  {Marshall}, {Menanteau}, {Miquel}, {Ogando}, {Plazas}, {Romer}, {Roodman},
  {Sanchez}, {Scarpine}, {Schindler}, {Serrano}, {Sevilla-Noarbe}, {Smith},
  {Smith}, {Soares-Santos}, {Sobreira}, {Suchyta}, {Swanson}, {Tarle},
  {Thomas}, {Vikram}, \& {DES Collaboration}}]{Klein19}
{Klein}, M., {Grandis}, S., {Mohr}, J.~J., {et~al.} 2019, \mnras, 488, 739

\bibitem[{{Klein} {et~al.}(2018){Klein}, {Mohr}, {Desai}, {Israel}, {Allam},
  {Benoit-L{\'e}vy}, {Brooks}, {Buckley-Geer}, {Carnero Rosell}, {Carrasco
  Kind}, {Cunha}, {da Costa}, {Dietrich}, {Eifler}, {Evrard}, {Frieman},
  {Gruen}, {Gruendl}, {Gutierrez}, {Honscheid}, {James}, {Kuehn}, {Lima},
  {Maia}, {March}, {Melchior}, {Menanteau}, {Miquel}, {Plazas}, {Reil},
  {Romer}, {Sanchez}, {Santiago}, {Scarpine}, {Schubnell}, {Sevilla-Noarbe},
  {Smith}, {Soares-Santos}, {Sobreira}, {Suchyta}, {Swanson}, {Tarle}, \& {DES
  Collaboration}}]{Klein18}
{Klein}, M., {Mohr}, J.~J., {Desai}, S., {et~al.} 2018, \mnras, 474, 3324

\bibitem[{{Klein} {et~al.}(2021){Klein}, {Oguri}, {Mohr}, {Grandis},
  {Ghirardini}, {Liu}, {Liu}, {Bulbul}, {Wolf}, {Comparat}, {Ramos-Ceja},
  {Buchner}, {Chiu}, {Clerc}, {Merloni}, {Miyatake}, {Miyazaki}, {Okabe},
  {Ota}, {Pacaud}, {Salvato}, \& {Driver}}]{Klein21}
{Klein}, M., {Oguri}, M., {Mohr}, J.~J., {et~al.} 2021, \aap, accepted

\bibitem[{{Lau} {et~al.}(2009){Lau}, {Kravtsov}, \& {Nagai}}]{Lau09}
{Lau}, E.~T., {Kravtsov}, A.~V., \& {Nagai}, D. 2009, \apj, 705, 1129

\bibitem[{{Markevitch} \& {Vikhlinin}(2007)}]{MarkevitchCFShock07}
{Markevitch}, M. \& {Vikhlinin}, A. 2007, \physrep, 443, 1

\bibitem[{{Meidinger} {et~al.}(2020){Meidinger}, {Andritschke}, {Dennerl},
  {Emberger}, {Eraerds}, {Haelker}, {Hartner}, {Pietschner}, \&
  {Reiffers}}]{Meidinger20}
{Meidinger}, N., {Andritschke}, R., {Dennerl}, K., {et~al.} 2020, in Society of
  Photo-Optical Instrumentation Engineers (SPIE) Conference Series, Vol. 11444,
  Society of Photo-Optical Instrumentation Engineers (SPIE) Conference Series,
  114444O

\bibitem[{{Merloni} \& {Heinz}(2007)}]{MerloniHeinz07}
{Merloni}, A. \& {Heinz}, S. 2007, \mnras, 381, 589

\bibitem[{{Million} {et~al.}(2010){Million}, {Werner}, {Simionescu}, {Allen},
  {Nulsen}, {Fabian}, {B{\"o}hringer}, \& {Sanders}}]{Million10}
{Million}, E.~T., {Werner}, N., {Simionescu}, A., {et~al.} 2010, \mnras, 407,
  2046

\bibitem[{{Navarro} {et~al.}(1996){Navarro}, {Frenk}, \& {White}}]{NFW96}
{Navarro}, J.~F., {Frenk}, C.~S., \& {White}, S.~D.~M. 1996, \apj, 462, 563

\bibitem[{{O'Sullivan} {et~al.}(2005){O'Sullivan}, {Vrtilek}, \&
  {Kempner}}]{OSullivan05}
{O'Sullivan}, E., {Vrtilek}, J.~M., \& {Kempner}, J.~C. 2005, \apjl, 624, L77

\bibitem[{{Predehl} {et~al.}(2021){Predehl}, {Andritschke}, {Arefiev},
  {Babyshkin}, {Batanov}, {Becker}, {B{\"o}hringer}, {Bogomolov}, {Boller},
  {Borm}, {Bornemann}, {Br{\"a}uninger}, {Br{\"u}ggen}, {Brunner}, {Brusa},
  {Bulbul}, {Buntov}, {Burwitz}, {Burkert}, {Clerc}, {Churazov}, {Coutinho},
  {Dauser}, {Dennerl}, {Doroshenko}, {Eder}, {Emberger}, {Eraerds},
  {Finoguenov}, {Freyberg}, {Friedrich}, {Friedrich}, {F{\"u}rmetz},
  {Georgakakis}, {Gilfanov}, {Granato}, {Grossberger}, {Gueguen}, {Gureev},
  {Haberl}, {H{\"a}lker}, {Hartner}, {Hasinger}, {Huber}, {Ji}, {Kienlin},
  {Kink}, {Korotkov}, {Kreykenbohm}, {Lamer}, {Lomakin}, {Lapshov}, {Liu},
  {Maitra}, {Meidinger}, {Menz}, {Merloni}, {Mernik}, {Mican}, {Mohr},
  {M{\"u}ller}, {Nandra}, {Nazarov}, {Pacaud}, {Pavlinsky}, {Perinati},
  {Pfeffermann}, {Pietschner}, {Ramos-Ceja}, {Rau}, {Reiffers}, {Reiprich},
  {Robrade}, {Salvato}, {Sanders}, {Santangelo}, {Sasaki}, {Scheuerle},
  {Schmid}, {Schmitt}, {Schwope}, {Shirshakov}, {Steinmetz}, {Stewart},
  {Str{\"u}der}, {Sunyaev}, {Tenzer}, {Tiedemann}, {Tr{\"u}mper}, {Voron},
  {Weber}, {Wilms}, \& {Yaroshenko}}]{Predehl21}
{Predehl}, P., {Andritschke}, R., {Arefiev}, V., {et~al.} 2021, \aap, 647, A1

\bibitem[{{Rebusco} {et~al.}(2005){Rebusco}, {Churazov}, {B{\"o}hringer}, \&
  {Forman}}]{RebuscoDiff05}
{Rebusco}, P., {Churazov}, E., {B{\"o}hringer}, H., \& {Forman}, W. 2005,
  \mnras, 359, 1041

\bibitem[{{Reiprich} {et~al.}(2013){Reiprich}, {Basu}, {Ettori}, {Israel},
  {Lovisari}, {Molendi}, {Pointecouteau}, \& {Roncarelli}}]{Reiprich13}
{Reiprich}, T.~H., {Basu}, K., {Ettori}, S., {et~al.} 2013, \ssr, 177, 195

\bibitem[{{Reiprich} \& {B{\"o}hringer}(2002)}]{Reiprich02}
{Reiprich}, T.~H. \& {B{\"o}hringer}, H. 2002, \apj, 567, 716

\bibitem[{{Reiprich} {et~al.}(2021){Reiprich}, {Veronica}, {Pacaud},
  {Ramos-Ceja}, {Ota}, {Sanders}, {Kara}, {Erben}, {Klein}, {Erler}, {Kerp},
  {Hoang}, {Br{\"u}ggen}, {Marvil}, {Rudnick}, {Biffi}, {Dolag},
  {Aschersleben}, {Basu}, {Brunner}, {Bulbul}, {Dennerl}, {Eckert}, {Freyberg},
  {Gatuzz}, {Ghirardini}, {K{\"a}fer}, {Merloni}, {Migkas}, {Nandra},
  {Predehl}, {Robrade}, {Salvato}, {Whelan}, {Diaz-Ocampo}, {Hernandez-Lang},
  {Zenteno}, {Brown}, {Collier}, {Diego}, {Hopkins}, {Kapinska}, {Koribalski},
  {Mroczkowski}, {Norris}, {O'Brien}, \& {Vardoulaki}}]{Reiprich21}
{Reiprich}, T.~H., {Veronica}, A., {Pacaud}, F., {et~al.} 2021, \aap, 647, A2

\bibitem[{{Russell} {et~al.}(2008){Russell}, {Sanders}, \&
  {Fabian}}]{Russell08}
{Russell}, H.~R., {Sanders}, J.~S., \& {Fabian}, A.~C. 2008, \mnras, 390, 1207

\bibitem[{{Sanders}(2006)}]{SandersBin06}
{Sanders}, J.~S. 2006, \mnras, 371, 829

\bibitem[{{Sanders} \& {Fabian}(2007)}]{SandersPer07}
{Sanders}, J.~S. \& {Fabian}, A.~C. 2007, \mnras, 381, 1381

\bibitem[{{Sanders} {et~al.}(2018){Sanders}, {Fabian}, {Russell}, \&
  {Walker}}]{Sanders18}
{Sanders}, J.~S., {Fabian}, A.~C., {Russell}, H.~R., \& {Walker}, S.~A. 2018,
  \mnras, 474, 1065

\bibitem[{{Sanders} {et~al.}(2016{\natexlab{a}}){Sanders}, {Fabian}, {Russell},
  {Walker}, \& {Blundell}}]{SandersGGM16}
{Sanders}, J.~S., {Fabian}, A.~C., {Russell}, H.~R., {Walker}, S.~A., \&
  {Blundell}, K.~M. 2016{\natexlab{a}}, \mnras, 460, 1898

\bibitem[{{Sanders} {et~al.}(2010){Sanders}, {Fabian}, {Smith}, \&
  {Peterson}}]{Sanders10_A1835}
{Sanders}, J.~S., {Fabian}, A.~C., {Smith}, R.~K., \& {Peterson}, J.~R. 2010,
  \mnras, 402, L11

\bibitem[{{Sanders} {et~al.}(2016{\natexlab{b}}){Sanders}, {Fabian}, {Taylor},
  {Russell}, {Blundell}, {Canning}, {Hlavacek-Larrondo}, {Walker}, \&
  {Grimes}}]{SandersCent16}
{Sanders}, J.~S., {Fabian}, A.~C., {Taylor}, G.~B., {et~al.}
  2016{\natexlab{b}}, \mnras, 457, 82

\bibitem[{{Sauvageot} {et~al.}(2005){Sauvageot}, {Belsole}, \&
  {Pratt}}]{Sauvageot05}
{Sauvageot}, J.~L., {Belsole}, E., \& {Pratt}, G.~W. 2005, \aap, 444, 673

\bibitem[{{Schellenberger} {et~al.}(2015){Schellenberger}, {Reiprich},
  {Lovisari}, {Nevalainen}, \& {David}}]{Schellenberger15}
{Schellenberger}, G., {Reiprich}, T.~H., {Lovisari}, L., {Nevalainen}, J., \&
  {David}, L. 2015, \aap, 575, A30

\bibitem[{{Schirmer}(2013)}]{Schirmer_13}
{Schirmer}, M. 2013, \apjs, 209, 21

\bibitem[{{Simionescu} {et~al.}(2008){Simionescu}, {Werner}, {Finoguenov},
  {B{\"o}hringer}, \& {Br{\"u}ggen}}]{Simionescu08}
{Simionescu}, A., {Werner}, N., {Finoguenov}, A., {B{\"o}hringer}, H., \&
  {Br{\"u}ggen}, M. 2008, \aap, 482, 97

\bibitem[{{Smith} {et~al.}(2001){Smith}, {Brickhouse}, {Liedahl}, \&
  {Raymond}}]{SmithApec01}
{Smith}, R.~K., {Brickhouse}, N.~S., {Liedahl}, D.~A., \& {Raymond}, J.~C.
  2001, \apjl, 556, L91

\bibitem[{{Struble} \& {Rood}(1999)}]{StrubleRood99}
{Struble}, M.~F. \& {Rood}, H.~J. 1999, \apjs, 125, 35

\bibitem[{{Vantyghem} {et~al.}(2014){Vantyghem}, {McNamara}, {Russell}, {Main},
  {Nulsen}, {Wise}, {Hoekstra}, \& {Gitti}}]{Vantyghem14}
{Vantyghem}, A.~N., {McNamara}, B.~R., {Russell}, H.~R., {et~al.} 2014, \mnras,
  442, 3192

\bibitem[{{Wayth} {et~al.}(2015){Wayth}, {Lenc}, {Bell}, {Callingham},
  {Dwarakanath}, {Franzen}, {For}, {Gaensler}, {Hancock}, {Hindson},
  {Hurley-Walker}, {Jackson}, {Johnston-Hollitt}, {Kapi{\'n}ska}, {McKinley},
  {Morgan}, {Offringa}, {Procopio}, {Staveley-Smith}, {Wu}, {Zheng}, {Trott},
  {Bernardi}, {Bowman}, {Briggs}, {Cappallo}, {Corey}, {Deshpande}, {Emrich},
  {Goeke}, {Greenhill}, {Hazelton}, {Kaplan}, {Kasper}, {Kratzenberg},
  {Lonsdale}, {Lynch}, {McWhirter}, {Mitchell}, {Morales}, {Morgan}, {Oberoi},
  {Ord}, {Prabu}, {Rogers}, {Roshi}, {Shankar}, {Srivani}, {Subrahmanyan},
  {Tingay}, {Waterson}, {Webster}, {Whitney}, {Williams}, \&
  {Williams}}]{Wayth15}
{Wayth}, R.~B., {Lenc}, E., {Bell}, M.~E., {et~al.} 2015, \pasa, 32, e025

\bibitem[{{Willingale} {et~al.}(2013){Willingale}, {Starling}, {Beardmore},
  {Tanvir}, \& {O'Brien}}]{Willingale13}
{Willingale}, R., {Starling}, R.~L.~C., {Beardmore}, A.~P., {Tanvir}, N.~R., \&
  {O'Brien}, P.~T. 2013, \mnras, 431, 394

\bibitem[{{Wilms} {et~al.}(2000){Wilms}, {Allen}, \& {McCray}}]{Wilms00}
{Wilms}, J., {Allen}, A., \& {McCray}, R. 2000, \apj, 542, 914

\bibitem[{{Zhuravleva} {et~al.}(2014){Zhuravleva}, {Churazov}, {Schekochihin},
  {Lau}, {Nagai}, {Gaspari}, {Allen}, {Nelson}, \&
  {Parrish}}]{ZhuravlevaFluct14}
{Zhuravleva}, I., {Churazov}, E.~M., {Schekochihin}, A.~A., {et~al.} 2014,
  \apjl, 788, L13

\bibitem[{{ZuHone} \& {Roediger}(2016)}]{ZuHoneRoediger16}
{ZuHone}, J.~A. \& {Roediger}, E. 2016, Journal of Plasma Physics, 82,
  535820301

\end{thebibliography}

\appendix

\section{Adaptive GGM filtering}
\label{appen:ggm}
The accuracy to which a gradient can be measured in an X-ray image is constrained by the number of counts within the size scale probed.
Filtering the data to calculate the gradient on one particular spatial scale has the disadvantage that the noise in the measurement becomes stronger in low brightness image regions (for example, in the outskirts of a galaxy cluster).
\cite{SandersCent16} described a method where images filtered on different size scales could be added together with radial weighting to prevent the noise becoming strong in lower surface brightness regions of the image.
The disadvantage of this method is that it is slow and subjective, because the weighting profiles of each input image with radius must be manually chosen.

We describe here a modification of the GGM filtering process to dynamically choose an appropriate size scale which makes the noise in the gradient measurement much more uniform across the field. The technique is to measure the gradient magnitude on a logarithmic (or non-logarithmic) adaptively smoothed image of the object.
An implementation of this technique is available within the GGM software package\footnote{\url{https://github.com/jeremysanders/ggm}}.

\begin{figure*}
    \includegraphics[width=0.74\textwidth]{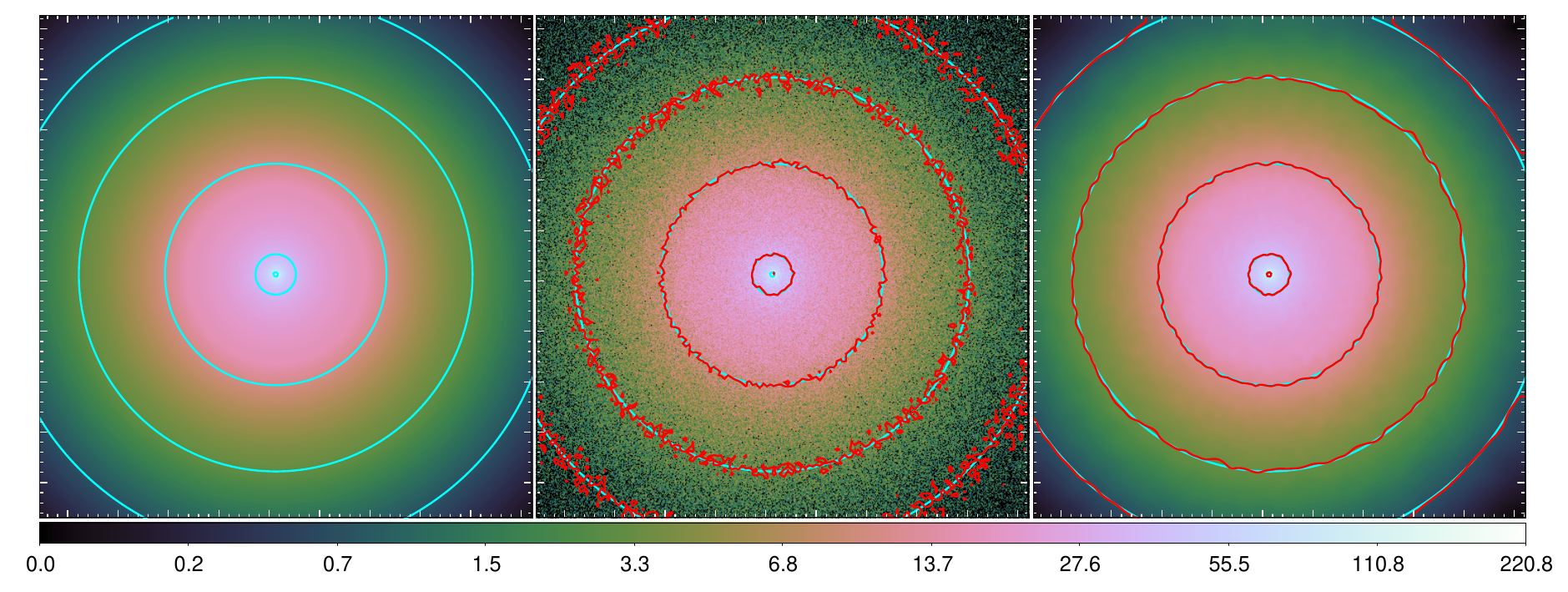}
    \includegraphics[width=0.24\textwidth]{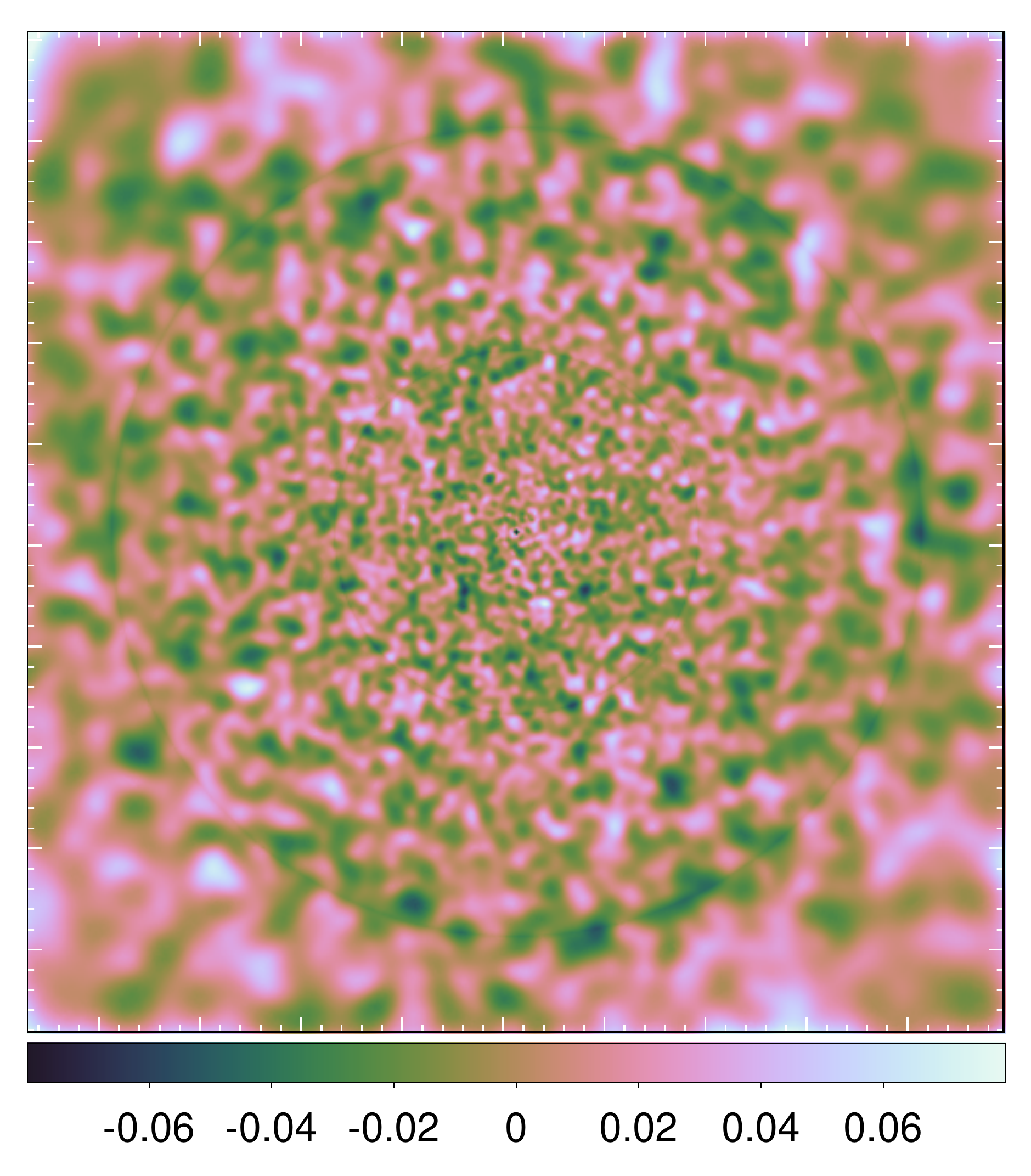}
    \caption{
    Surface brightness images and residuals.
    (Left panel) Input surface brightness model, with five logarithmic cyan contours between 1 and 100 counts per pixel.
    (Second panel) Poisson realisation of input model (signal to noise ratio of 30).
    Shown are the model contours with red contours at the same levels.
    (Third panel) Reconstruction of input model using adaptive smoothing.
    Shown are the model contours with red contours at the same levels.
    (Right panel) Fractional difference between reconstructed image and input model.
    }
    \label{fig:ggm_img}
\end{figure*}

To help describe the method we constructed an idealised surface brightness model to use as input.
We note that this model does not represent a realistic 3D object and does not include background components.
Figure \ref{fig:ggm_img} (left panel) shows this input surface brightness model, constructed with dimensions of $1024\times 1024$ pixels.
Inside a radius $r=180$ pixels from the centre of the image, the model surface brightness is given by $200 \, r^{-0.5}$ counts per pixel.
The slope changes to $r^{-2}$ between $r=180$ and $400$ and to $r^{-3.5}$ beyond this radius (choosing normalisations as to be continuous).
The second panel shows a Poisson realisation of this surface brightness model.

To construct an adaptively smoothed reconstruction of the input model we use the \texttt{accumulate\_counts} program from the contour binning package \citep{SandersBin06}.
This program has two modes of operation which are applied in turn.
The first is to compute a scale map, which is the radius-squared around each pixel which contains a minimum signal to noise ratio.
This currently does not include background in the computation and therefore the minimum number of counts is given by the signal to noise ratio squared.
The signal to noise ratio is the only parameter in the process and controls the level of noise in the output gradient map.
We note that the radius-squared used here can grow by less than one pixel as it expanded to include further neighbouring pixels.
The second step is to apply this scale map to an input image to smooth it.
Each pixel is smoothed by a Gaussian kernel, where $\sigma$ (or the radius) is given by the square root of the input scale map.
We note that the program can also apply top-hat smoothing, but here we use Gaussian smoothing to reduce the discontinuities between different smoothing scales and reduce noise.
The output adaptively smoothed example image using a signal to noise ratio of 30 is shown in Fig.~\ref{fig:ggm_img} (third panel).
The fractional difference between this output image and input model is shown in Fig.~\ref{fig:ggm_img} (right panel).

The log value of this smoothed surface brightness map is taken so that the measured gradient is the gradient in the $\log_{10}$ quantity.
This step is optional if the user wishes to instead measure the linear gradient.
The log gradient is often more useful as the dynamic range of input images can be large.
To compute the gradient from the log or non-log adaptively smoothed X-ray images the per-pixel gradients are computed along the two different image axes and added in quadrature.

\begin{figure*}
    \includegraphics[width=0.66\textwidth]{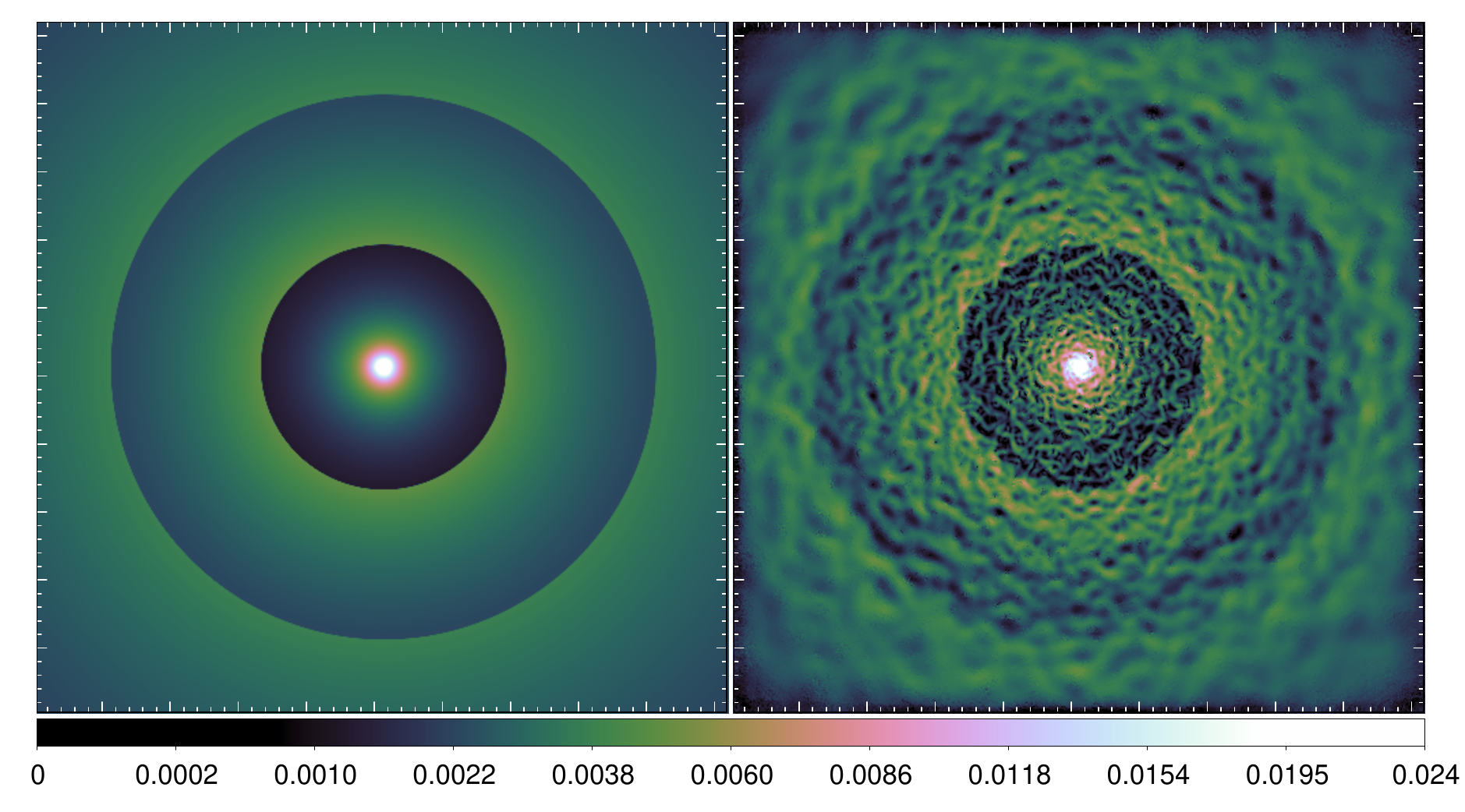}
    \includegraphics[width=0.33\textwidth]{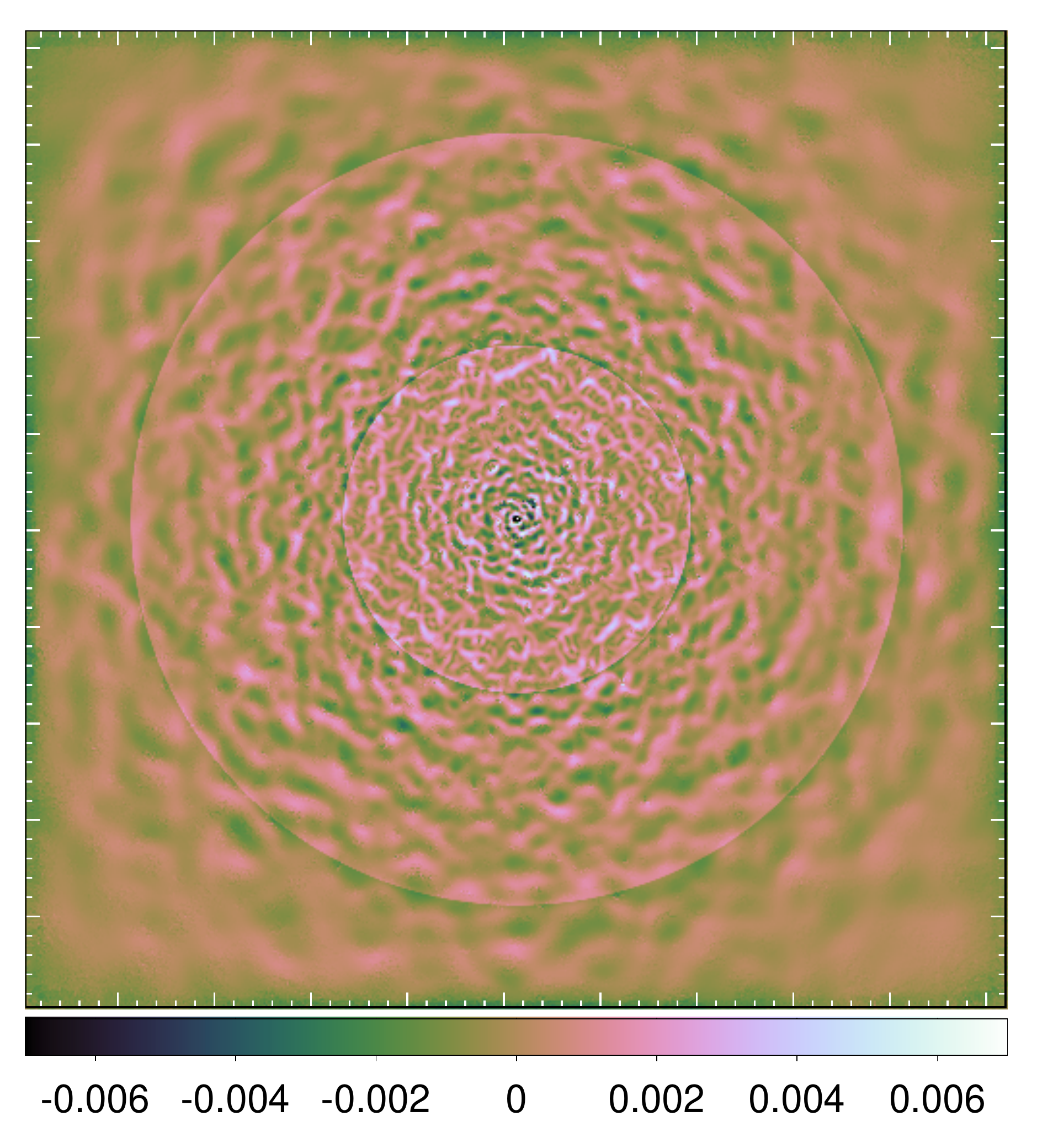}
    \caption{
    Gradient magnitude images and residuals.
    (Left panel) Gradient magnitude of the input model from Fig.~\ref{fig:ggm_img}, computing the absolute magnitude of the gradient of the $\log_{10}$ quantity per pixel.
    (Centre panel) Gradient of the reconstructed image.
    (Right panel) Difference between the reconstructed and input gradients.
    }
    \label{fig:ggm_grad}
\end{figure*}

\begin{figure}
    \centering
    \includegraphics[width=\columnwidth]{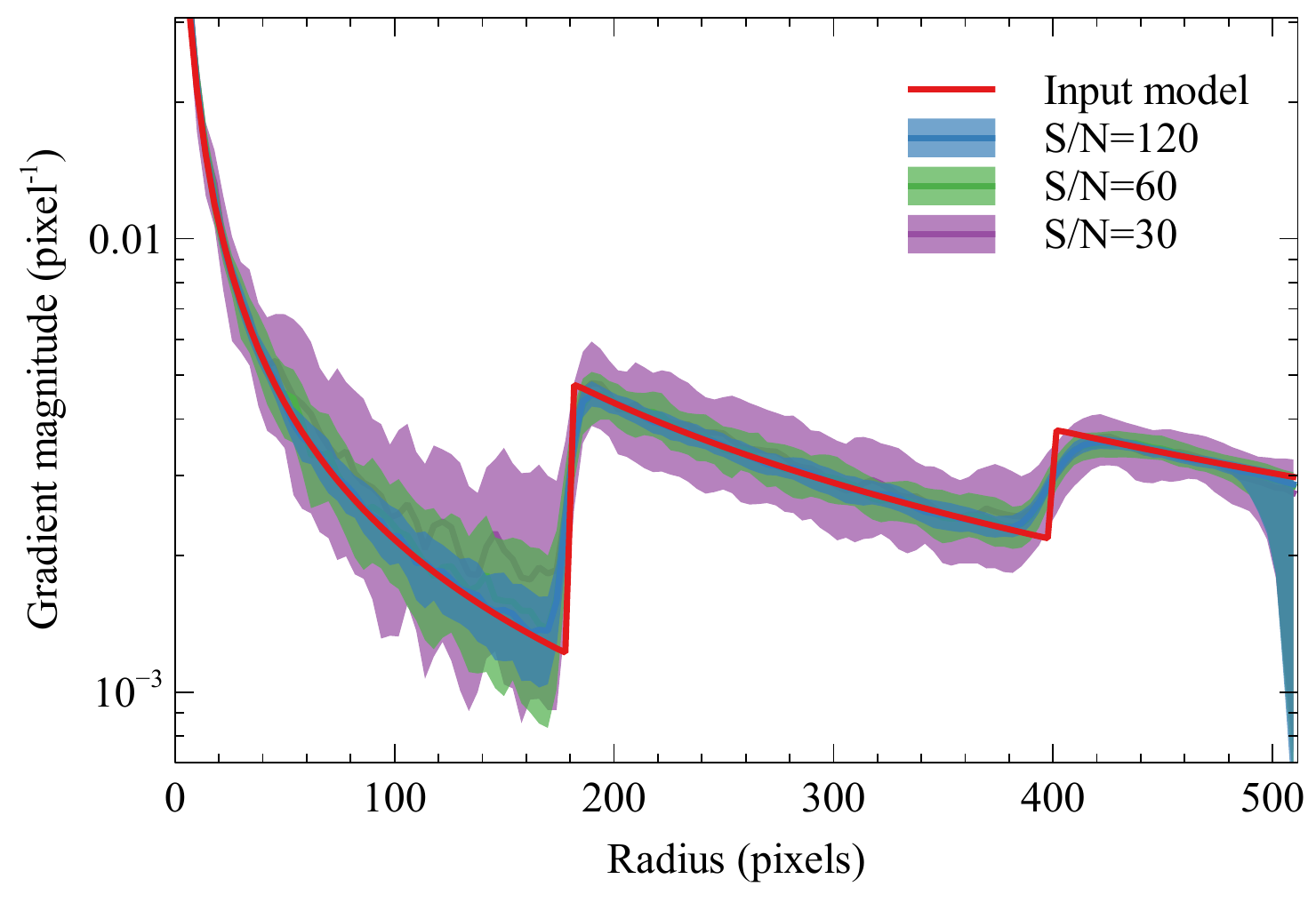}
    \caption{
    Comparison of the input and recovered gradient magnitudes.
    The gradient of the input model is plotted in 128 radial bins.
    The shaded regions show the median and $1\sigma$ percentiles of the pixels in each radial bin from recovered gradient maps.
    The S/N=30 results are taken from \ref{fig:ggm_grad}.
    Also shown is the effect of increasing the signal to noise ratio of the input image and smoothing by factors of 2 and 4.
    }
    \label{fig:prof_grad}
\end{figure}

Figure \ref{fig:ggm_grad} shows the gradient magnitude of the log input surface brightness model (left panel), the reconstructed gradient magnitude from the adaptively smoothed image (centre panel) and the difference between the two (right panel).
It can be seen that the residual image is relatively uniform over the image, showing that the adaptive method works well.
However, in the central region the residuals are biased higher than zero.
A radial profile of the gradient in the input model and reconstructed gradient shows this more clearly (Fig.~\ref{fig:prof_grad}).
Where the gradient is low, the median reconstructed gradient magnitude is higher than the input because a gradient magnitude can only be positive.
This effect is reduced if the data are smoothed with a larger signal to noise ratio, shown in the results where the input data and smoothing scale signal to noise are increased by factors of 2 or 4.

We note that the range of residuals in the gradient map or profile are not completely uniform across this test field (e.g. Fig.~\ref{fig:prof_grad}).
Producing a uniform scatter in the smoothed X-ray image does not give completely uniform residuals in the gradient. 
The level of the scatter, however, is much more uniform than if a single smoothing scale is used.
As with any kind of gradient filtering, features in the filtered maps can be due to noise and some effort may be required to assess the significance of structures.
However, due to the more uniform noise level across the filtered image and the gradient of the log quantity, it should be much easier to assess the significance of features directly by comparison with neighbouring regions.
The features described in the adaptive GGM map for A\,3266 (Fig.~\ref{fig:ggm}) are much more significant than the noise level.

\section{Extended source spectral extraction}
\label{appen:regions}
\begin{figure*}
    \centering
    \includegraphics[width=0.32\textwidth]{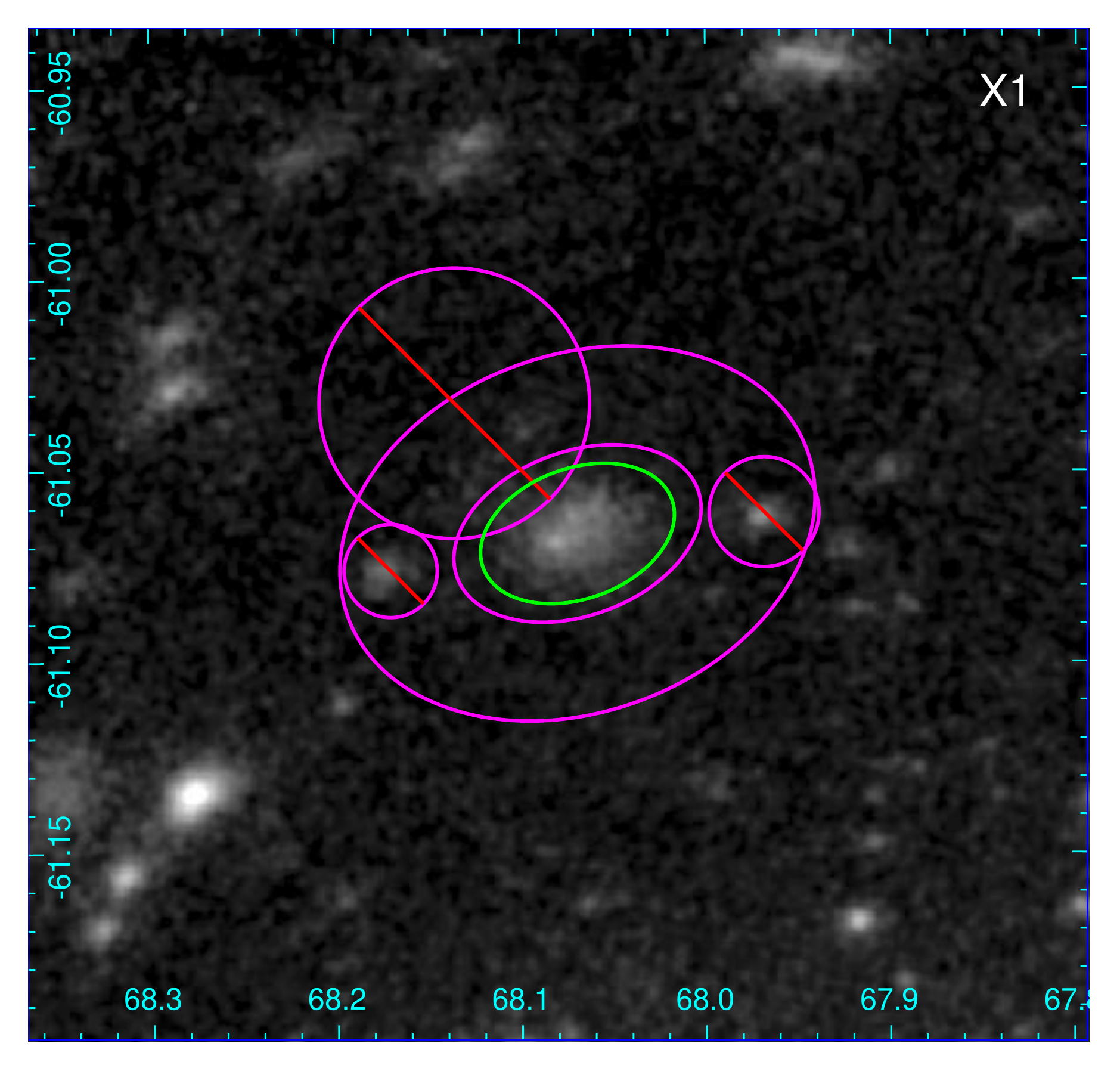}
    \includegraphics[width=0.32\textwidth]{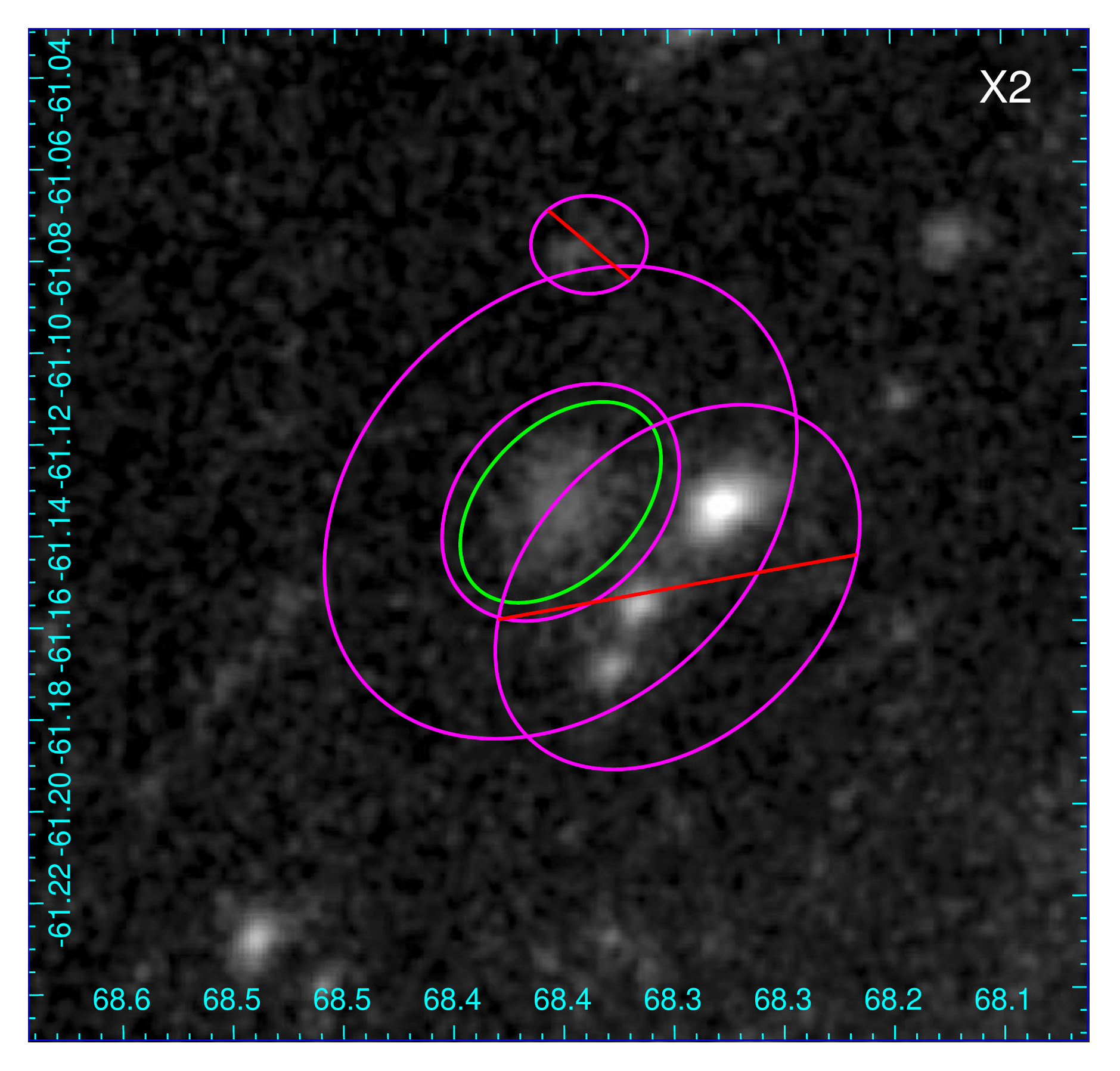}
    \includegraphics[width=0.32\textwidth]{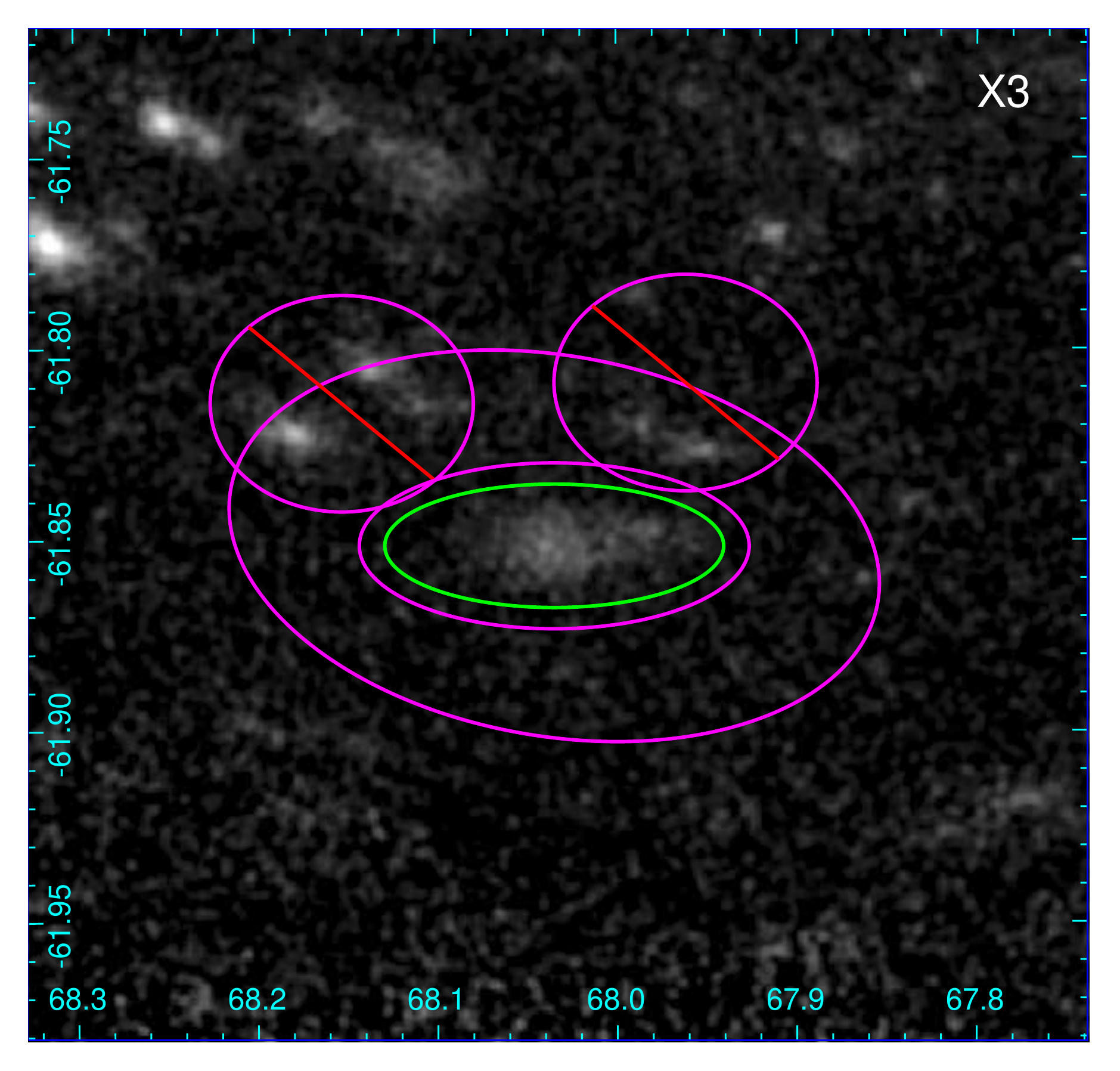}\\
    \includegraphics[width=0.32\textwidth]{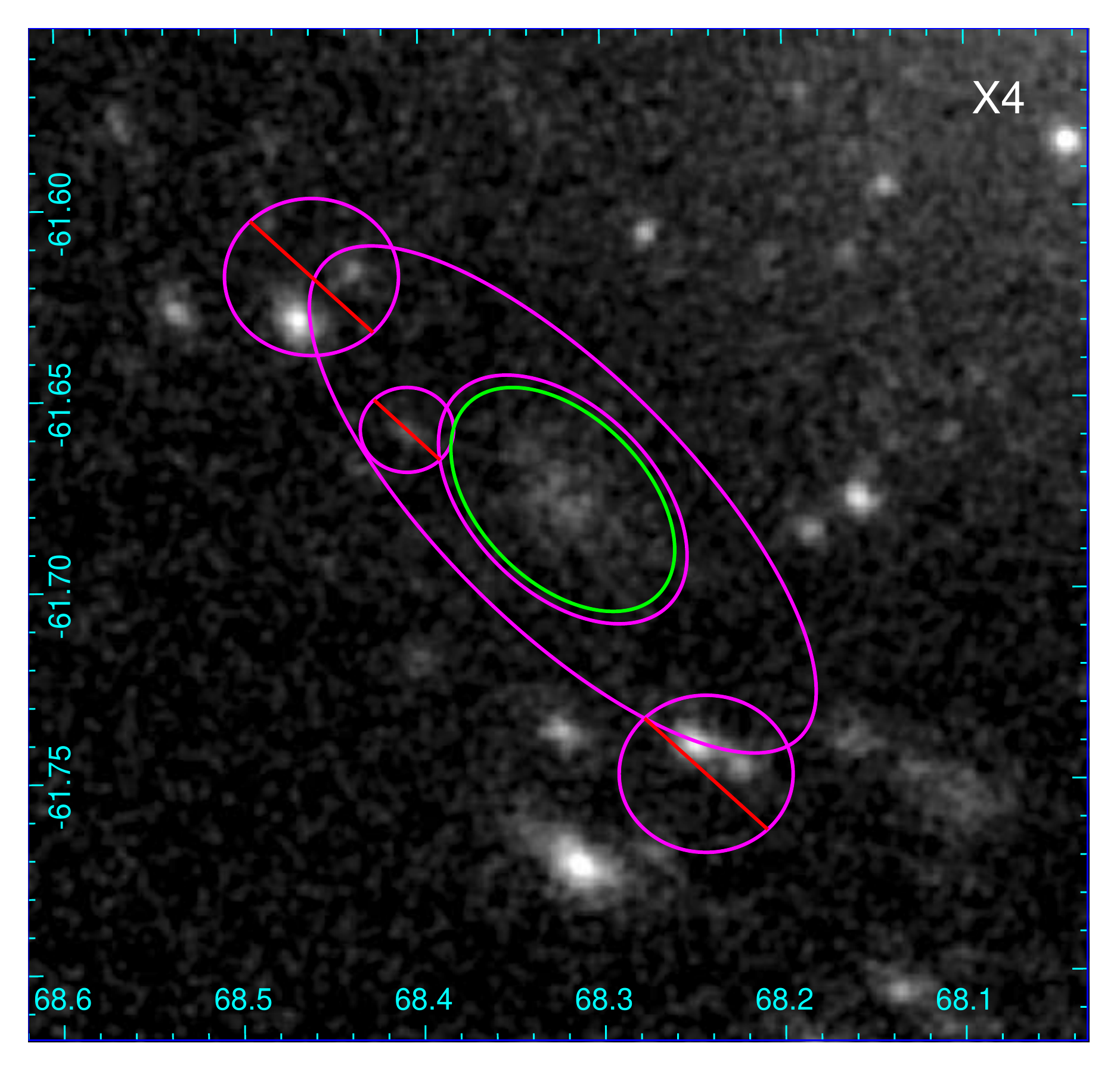}
    \includegraphics[width=0.32\textwidth]{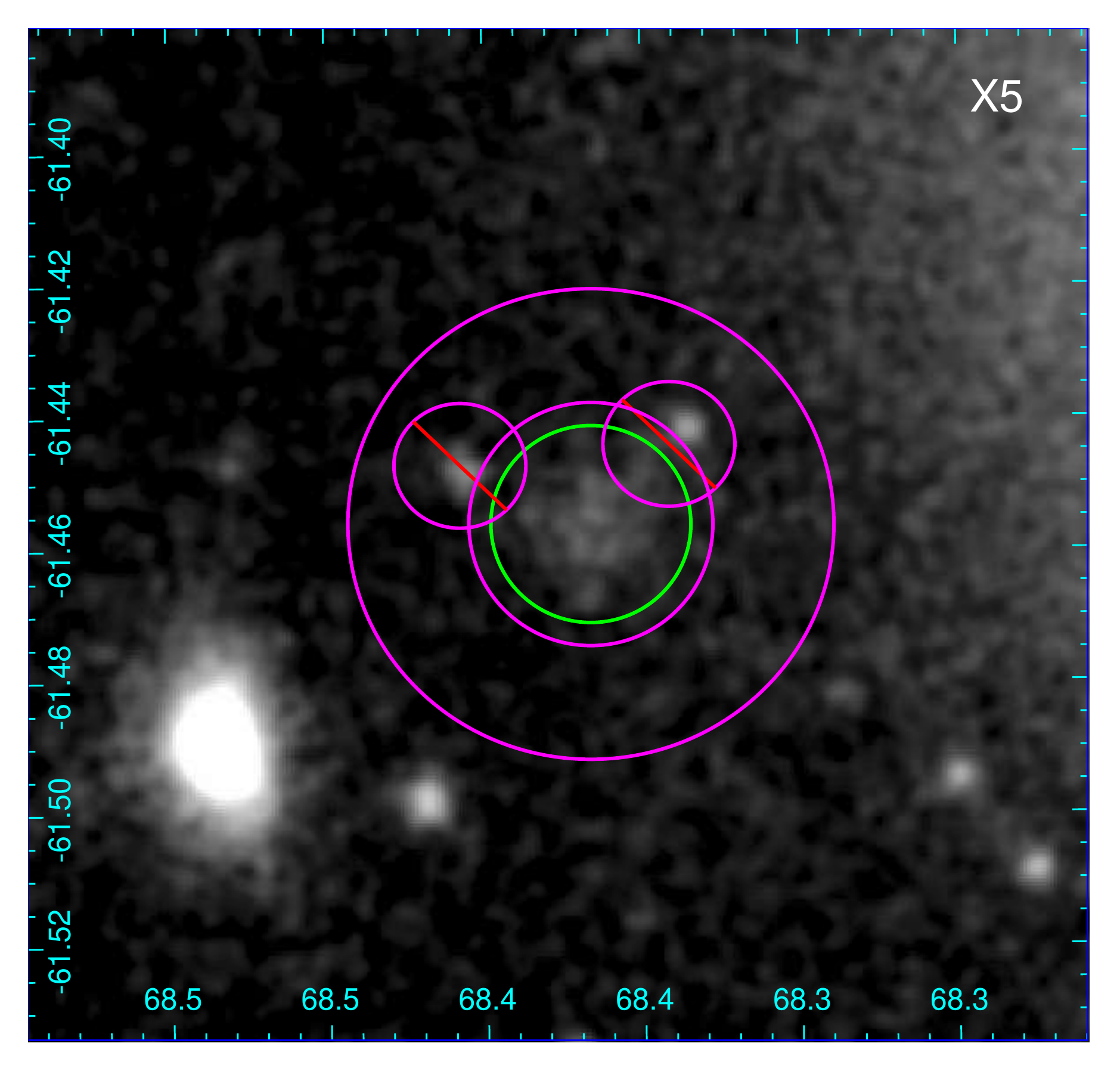}
    \includegraphics[width=0.32\textwidth]{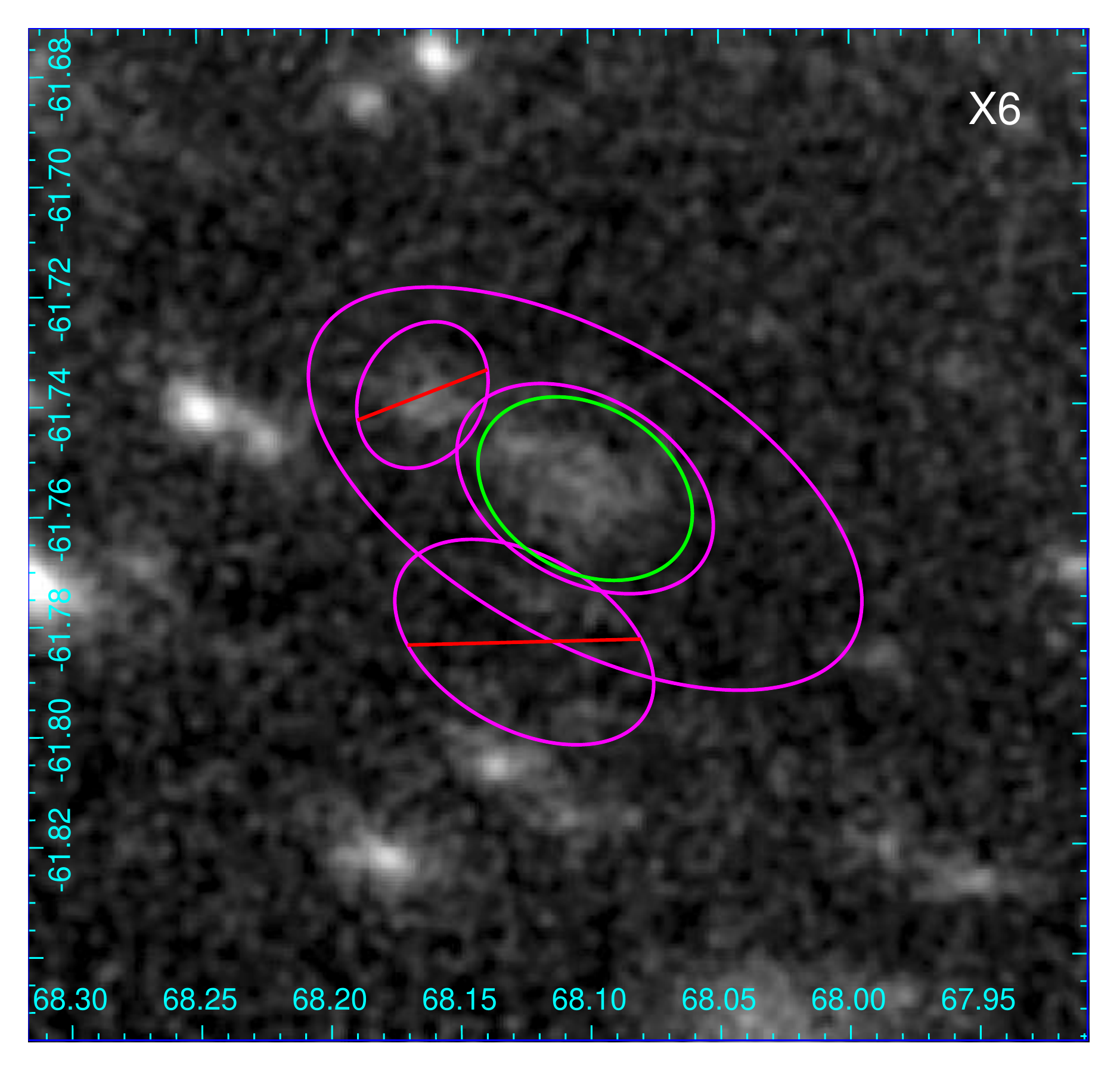}\\
    \includegraphics[width=0.32\textwidth]{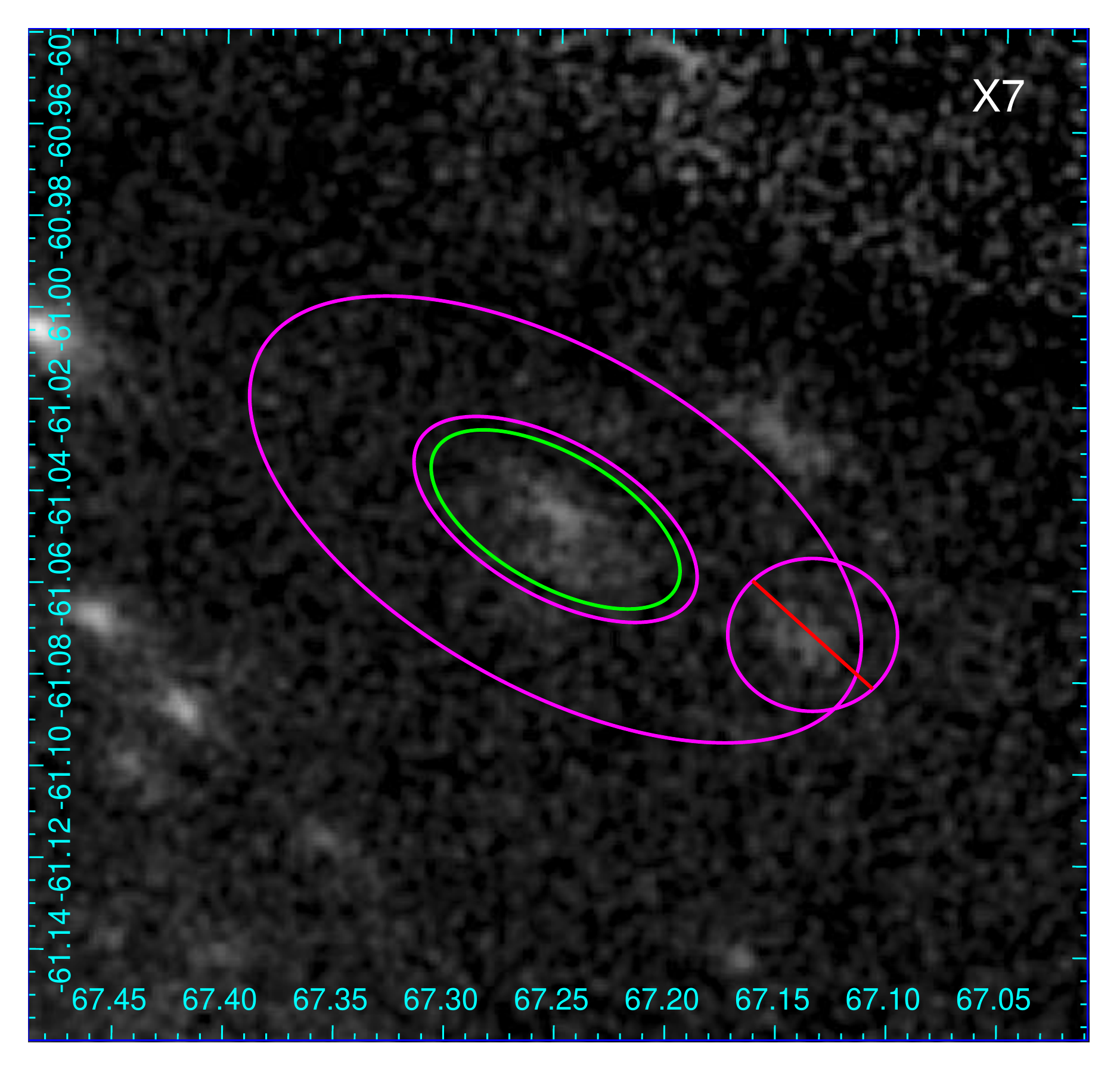}
    \includegraphics[width=0.32\textwidth]{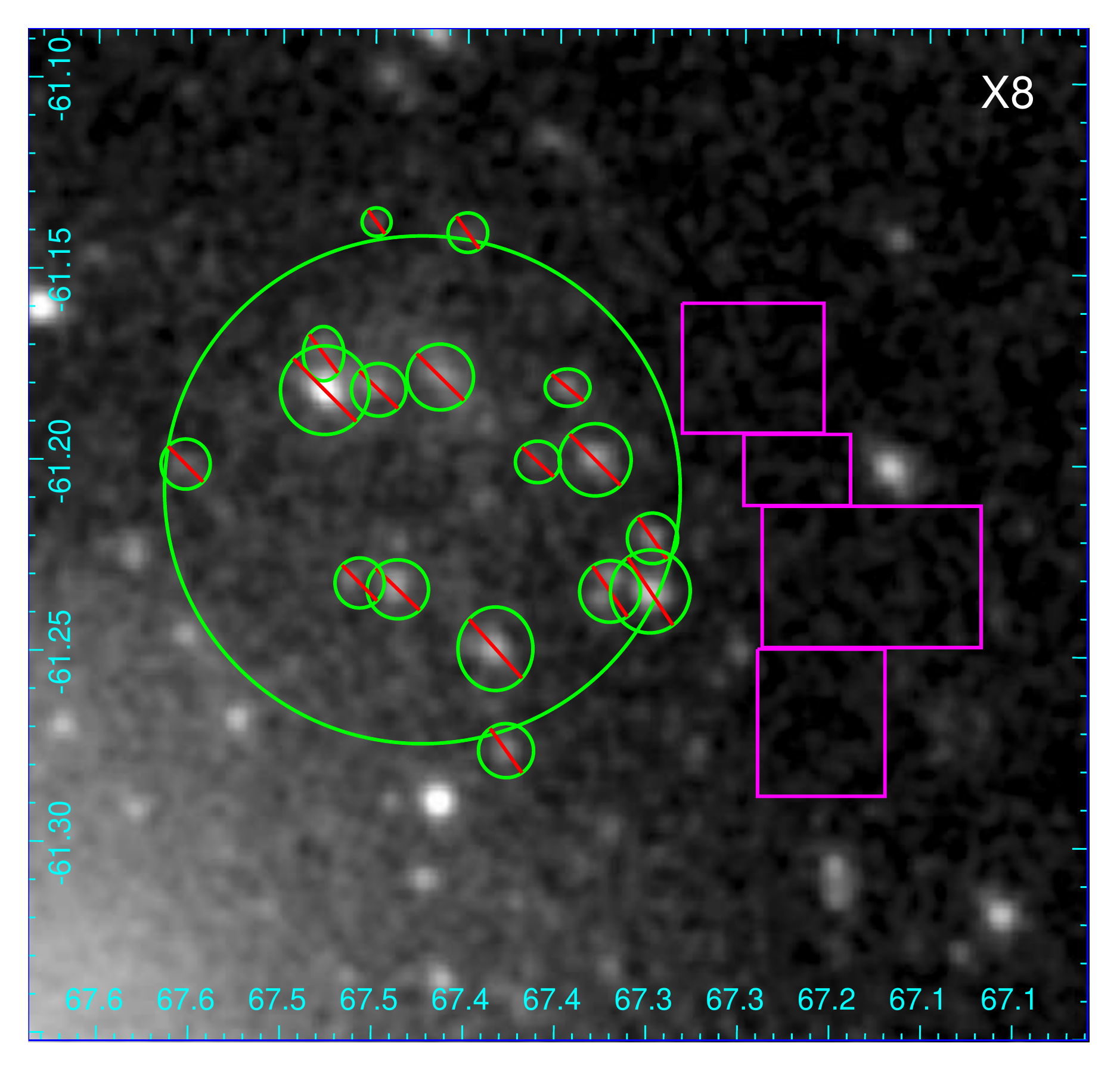}
    \includegraphics[width=0.32\textwidth]{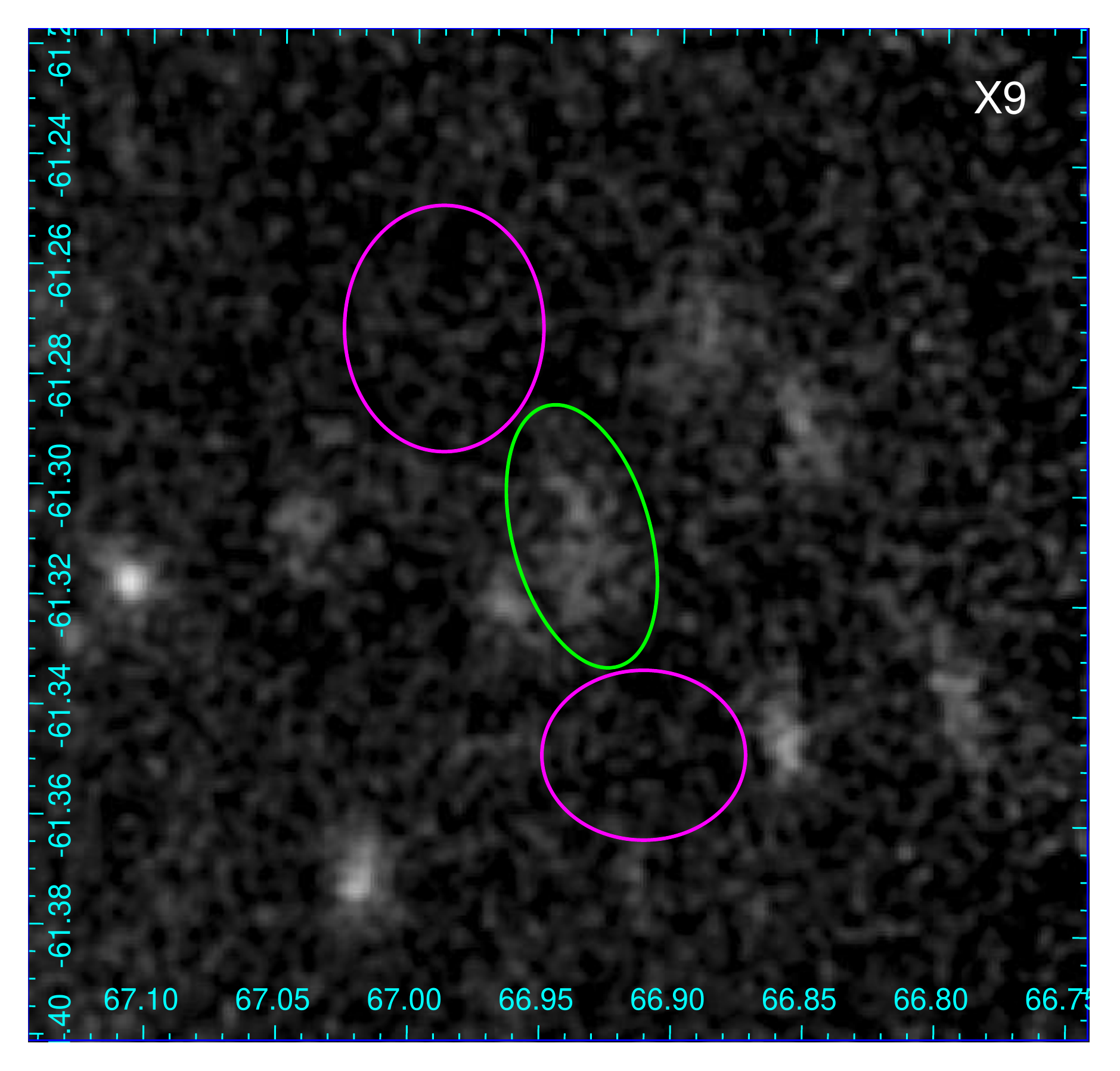}\\
    \includegraphics[width=0.32\textwidth]{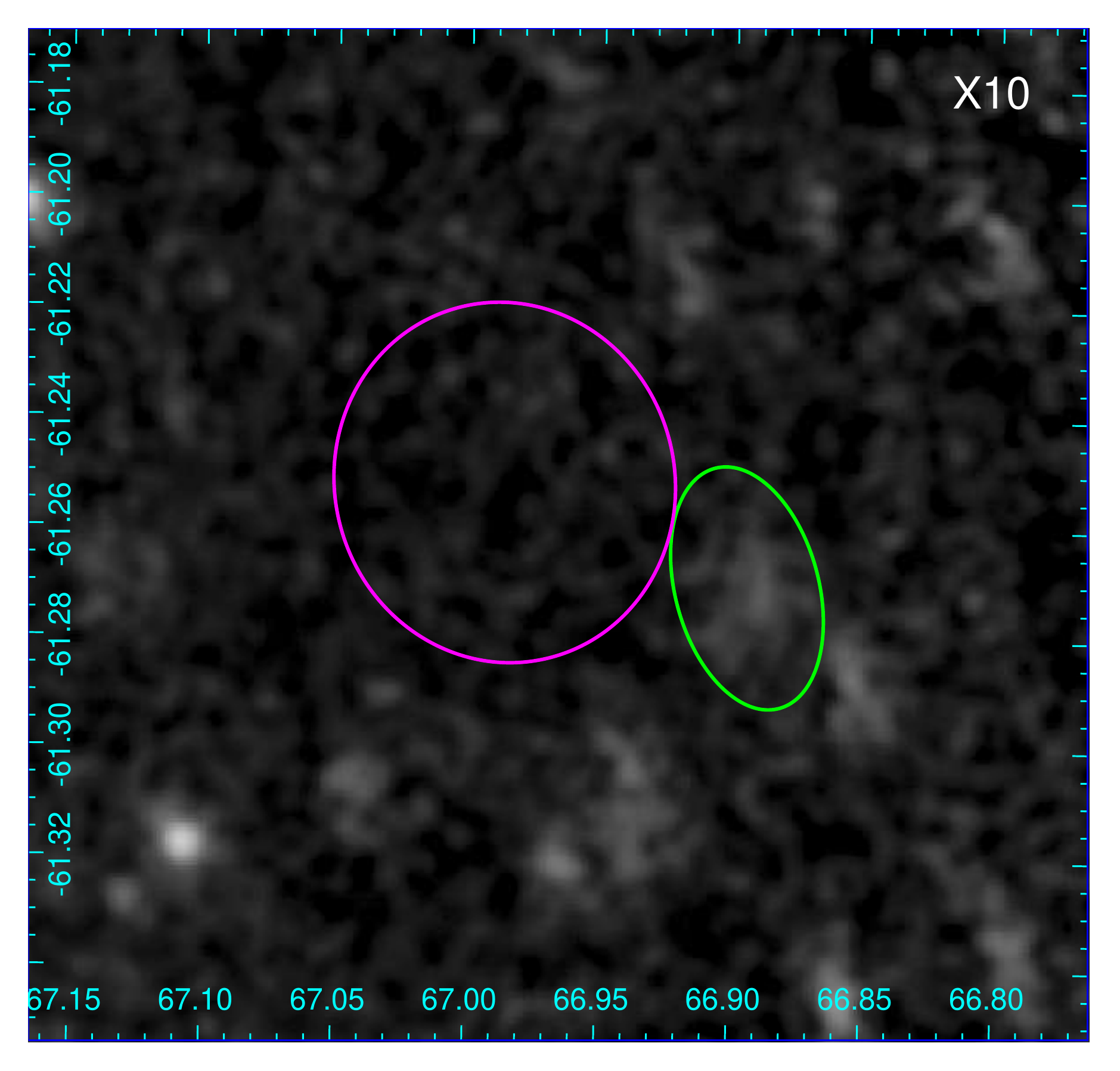}
    \includegraphics[width=0.32\textwidth]{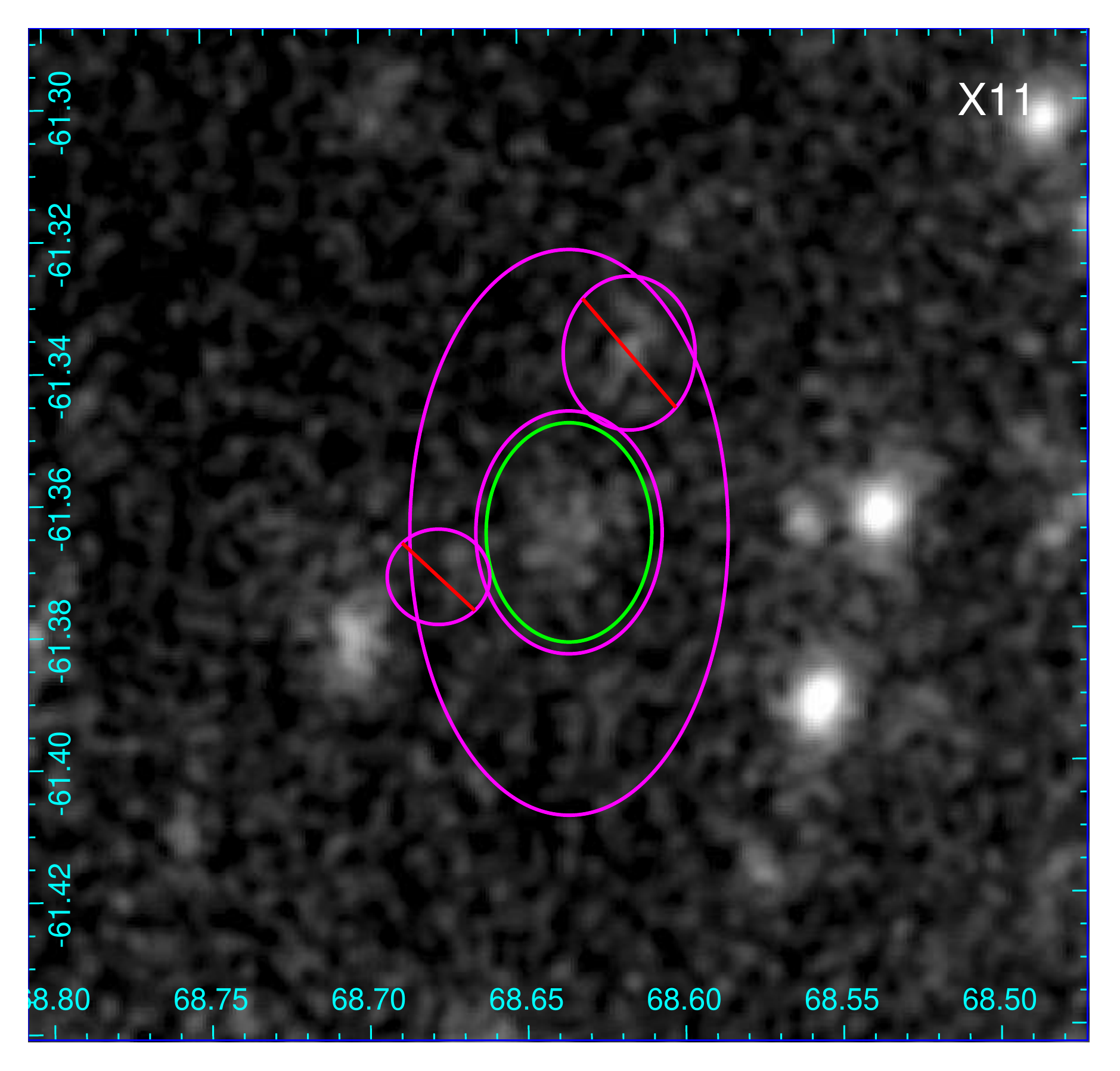}
    \caption{
        Extraction regions used for the spectral analysis of the extended sources in the field.
        The source extraction radius is marked in green, while the background regions are in magenta.
        Shown are 0.3 to 2.3 keV exposure-corrected background-subtracted images, with Gaussian smoothing.
        Excluded regions are marked by a diagonal line.
    }
    \label{fig:specregions}
\end{figure*}

Shown in Fig.~\ref{fig:specregions} are the extraction regions used for the analysis of the external extended sources.

\end{document}